\documentclass[11pt]{article}
\usepackage{amsmath, amssymb}
\usepackage{graphicx}
\newcommand{\mathsym}[1]{{}}

\DeclareMathOperator{\diag}{diag}
\DeclareMathOperator{\Li}{Li}

\renewcommand{\title}[1]{\vbox{\center\LARGE{#1}}\vspace{5mm}}
\renewcommand{\author}[1]{\vbox{\center#1}\vspace{5mm}}

\def\ci{\cite}
\def \V {v}

\def \S {{\rm S}}

\def \td {\tilde}

\def \N {{\mathcal N}}

\def \m {\mu}

\def \bi{\bibitem}
\def \la {\label}

\def \l {\lambda}
\def\foot{\footnote}

\def \sql {{\sqrt \l}}
\def \adss {$AdS_5 \times S^5$\ }
\newcommand{\rf}[1]{(\ref{#1})}
\def \ov {\over}

\def\th{\theta}

\def\ttheta{{{\tilde\theta}}}

\newcommand{\hw}{\hat w}
\newcommand{\hn}{\hat \nu}
\newcommand{\hk}{\hat kappa}

\def\N{{\cal N}}
\def\F{{\cal F}}

\def \ha{{\textstyle{1\ov 2}}}

\def\r{{\rm r}}

\def \J {\mathcal{K}}
\def \del {\partial}

\def \E {{\cal E}}

\def \S {{\cal S}}
\def \J {{\cal J}}

\def \om {\omega}

\def \bi{\bibitem}
\def \la {\label}

\def \l {\lambda}
\def\foot{\footnote}

\def \sql {{\sqrt \l}}

\def \adss {$AdS_5 \times S^5$\ }

 \def \t {\tau}
 
 \def \r {\rho}

\def \ov {\over}

\def \varpi {{\rm w}}

\def \no {\nonumber }

\def \adss {$AdS_5 \times S^5$\ }

\def \N {{\cal N}}

  \def \te {\theta}

\def \k {\kappa}
\def \ci {\cite}

\def\tr{{\rm tr}}

\newcommand{\id}{\mathbf{1}}
\renewcommand{\imath}{i}

\renewcommand{\geq}{\,{\geqslant}\,}
\renewcommand{\leq}{\,{\leqslant}\,}

\newcommand{\binner}[2]{%
  {\langle}\kern-4.15pt{\langle}#1{,}\,#2{\rangle}\kern-4.15pt{\rangle}}

\newcommand{\ffrac}[2]{\raisebox{.5pt}%
  {\footnotesize$\displaystyle\frac{#1}{#2}$}\kern1pt}

\def\id{\protect{{1 \kern-.28em {\rm l}}}}
\def \rK {{\rm K}}
\def \dn {{\rm dn}}

\makeatletter
\renewcommand\section{\@startsection {section}{1}{\z@}%
                                   {-3.5ex \@plus -1ex \@minus -.2ex}%
                                   {2.3ex \@plus.2ex}%
                                   {\normalfont\large\bfseries}}
\renewcommand\subsection{\@startsection{subsection}{2}{\z@}%
                                   {-3.25ex\@plus -1ex \@minus -.2ex}%
                                   {1.5ex \@plus .2ex}%
                                   {\normalfont\normalsize\bfseries}}
\makeatother

\def\b{{\rm b}} %$\beta$ in KRTT
\def\Tr{{\rm Tr}}

\def \vp {\varphi}
\def \ri {{\rm i}}

\def \M {{\cal M}}
\def \F {{\cal F}}  

\numberwithin{equation}{section} \makeatletter
\@addtoreset{equation}{section}

  \def \te {\theta}
  \def \vp {\varphi}
  \def \a {\alpha}
  \def \b {\beta}

\def \s {\sigma}

\def \n {\nu}

\newcommand{\be}{\begin{eqnarray}}
\newcommand{\ee}{\end{eqnarray}}

 \def \om {\omega}

\def\ttheta{{\tilde\theta}}
\def\teta{{\tilde\eta}}

\def\kahat{{\hat\kappa}}
\def\what{{\hat w}}
\def\nuhat{{\hat\nu}}
\def \bmu {\bar \mu}

\def \ri {{\rm i}}
\begin{document}

\textwidth 170mm 
\textheight 230mm 
\topmargin -1cm
\oddsidemargin-0.8cm \evensidemargin -0.8cm 
\topskip 9mm 
\headsep9pt

\overfullrule=0pt
\parskip=2pt
\parindent=12pt
\headheight=0in \headsep=0in \topmargin=0in \oddsidemargin=0in

\vspace{ -3cm} \thispagestyle{empty} \vspace{-1cm}
\begin{flushright} 
\end{flushright}
 \vspace{-1cm}
\begin{flushright} Imperial-TP-RR-01-2010
\end{flushright}
\begin{center}
 %\vspace{1.2cm}

{\Large\bf
%Strong-coupling corrections to 
Generalized scaling function \\ 
%from 2-loop \\
\vskip  0.2cm
from light-cone gauge $AdS_5 \times S^5$ superstring}
 
 \vspace{0.8cm} {
  S.~Giombi,$^{a,}$\footnote{giombi@physics.harvard.edu} 
  R.~Ricci,$^{b,}$\footnote{r.ricci@imperial.ac.uk } 
  R.~Roiban,$^{c,}$\footnote{radu@phys.psu.edu}
  A.A.~Tseytlin$^{b,}$\footnote{Also at
   Lebedev  Institute, Moscow.\ \ 
   tseytlin@imperial.ac.uk }  
   and  
   C.~Vergu$^{d,}$\footnote{Cristian\_Vergu@brown.edu} 
 }\\
 \vskip  0.2cm

\small
{\em
$^{a}$
Center for the Fundamental Laws of Nature,
Jefferson Physical Laboratory, \\
Harvard University, Cambridge, MA 02138 USA

% \vskip 0.05cm
 $^{b}$  The Blackett Laboratory, Imperial College,
London SW7 2AZ, U.K. 

 %\vskip 0.05cm
    $^{c}$ Department of Physics, The Pennsylvania State University,
University Park, PA 16802, USA

%\vskip 0.05cm
$^{d}$ Physics Department, Brown University, Providence, RI 02912, USA
}

\normalsize
\end{center}

 \vskip 0.4cm

\begin{abstract}
We revisit  the computation of the 2-loop correction to the energy 
of a folded spinning  string in $AdS_5$ with an angular momentum $J$  in $S^5$ 
in the scaling limit $\ln S \gg 1$, \ 
 %in which  the  density  of 
 $J \ov \sql \ln S $=fixed. This  correction  gives   the third term in the
%subleading
strong-coupling 
%term  in  
expansion of the generalized scaling function. 
The computation, using the  AdS light-cone gauge  approach developed in our 
previous paper, is done by expanding the $AdS_5 \times S^5$ superstring
partition function  near  the generalized null cusp  world surface 
associated to the spinning string solution. 
%We 
%use the AdS light-cone gauge  approach developed in our previous paper.   
The result  
corrects and extends the previous  conformal gauge  result of arXiv:0712.2479
and is  found to be  in complete agreement  with 
the  corresponding terms in the generalized scaling function as  obtained  from 
the asymptotic Bethe ansatz   in arXiv:0805.4615
(and also partially 
from the quantum  $O(6)$ model and the Bethe ansatz  data   in arXiv:0809.4952). 
This provides  a highly  nontrivial strong coupling comparison
%test
of the    Bethe ansatz proposal 
%against 
with the  quantum  $AdS_5 \times S^5$  superstring theory, 
which  goes beyond the leading  semiclassical term effectively controlled by the 
underlying algebraic curve.  
The 2-loop computation we perform  involves all the 
structures  in the AdS light-cone gauge 
superstring action of hep-th/0009171   and 
thus 
tests its ultraviolet finiteness  and, through the agreement  with the 
Bethe ansatz, its quantum integrability. 
We do most of the  computations for a  generalized  
spinning string solution or the corresponding null cusp surface 
that involves both the orbital momentum and  the 
winding  in a  large circle of $S^5$. 
  
\end{abstract}

\def \g {\gamma}
\def \F {{\cal F}} \def \V   {{ V}} 

\def \el {\ell} \def \jj {\ell} \def \ff {{\rm f}}
\def \bea{\be}
\def \eea{\ee} \def \re  {\rf} \def \K {{\rm K} }
\def \edo {\end{document}}

\def \hn {\hat \nu}
\def \hm {\hat m} 
\def \hk {\hat \k} \def \hw  {\hat m}

\def \nua  {\nu_{_e}  }
\def \hna {\hat \nu_{_e} }
\def \ge  {\gamma_{_e}}

\newpage 

\section{Introduction \label{intro}}
\setcounter{footnote}{0} 

%%%%%%%%%%%%%%%%%%%%%%%%%%%%%%%%%%%%%%%%%%%%%%%%%%%%

The correspondence between   the  fast-spinning  ($S \gg 1$) 
folded closed strings 
in $AdS_5\times  S^5$   and    twist  operators in  the $\N=4$ SYM theory
is a remarkable tool  for  uncovering  and checking 
the detailed structure of the  AdS/CFT correspondence 
(see e.g. \ci{gkp,ft1,bgk,bes,ftt,rtt,krj,am2,benna,bkk,rt1,rt2,frs,grom,bass,rej}). 
In particular,  matching   the expressions   for the 
 2-loop term in the  universal scaling function 
(cusp anomaly)  computed directly from the quantum superstring \ci{rtt,rt1}
and extracted \ci{benna,bkk} from the strong coupling  expansion of the asymptotic Bethe 
ansatz \ci{bes}  provided a non-trivial test  of the latter. 

New lessons are  learned  and new 
%interesting  structures come into play and  new
 detailed checks  are   possible 
when one considers $(S,J)$ strings    dual to  large twist  operators 
Tr$(D^S \Phi^J)$  in the special  strong-coupling  
scaling limit \ci{bgk,ftt}  
\be \la{ho}  \l \gg 1, \ \ \ \ \ \ \  \S \equiv { S \ov \sql}  \gg1\ ,\ \ \ \ \ \ 
 \ \J \equiv  { J \ov \sql}  \gg1   \ , \ \ \ \ \ \ \ 
\el \equiv { \pi J \ov  \sql   \ln S }={\rm fixed}  \ . \ee 
The  folded spinning string solution with spin $S$ in $AdS_5$ \ci{gkp} 
and orbital momentum $J$ in
$S^5$ \ci{ft1} 
simplifies 
in the scaling limit \rf{ho}  becoming  the following 
{ ``homogeneous''} configuration in  $AdS_3\times S^1$ \ci{ftt}
\be
 ds^2=-\cosh^2 \rho\ dt^2+d\rho^2+\sinh^2 \rho\  d\theta^2+d\vp^2\ , 
 \ \ \ \ \  \ \ \ \ \  \ \ \ \ \ \ \ \ 
 \la{ada} \\
t=\kappa\tau~,~~~~~~~~~\rho=\mu\sigma \ , \ \ \ \ \ \     \    \
\theta=\kappa\tau~,~~~~~~\vp=\nu
\tau~,\ \ \ \ \ \ \   \k,\mu,\nu  \gg 1   \ , \ \ \ \  \ \ 
\label{so} \\ 
\kappa^2=\mu^2+\nu^2              \ , \ \ \ \ \ \  
\mu\approx  \frac{1}{\pi}\ln \S   \ , \ \ \ \ \  \
\nu = \J                                          \ , \ \ \ \ \   \    
\hn \equiv { \nu \ov \mu }= {\rm fixed}  \ , \  \  \ \ \
\hk  \equiv { \k \ov \mu } = \sqrt{1 + \hn^2} ~~.
\label{Vi}
\ee
Originally  $0 \leq \s < 2 \pi $; 
rescaling  $\s\to \bar \s = \mu \s $ with  $\mu \to \infty$ we get 
$\rho = \bar \sigma  \in (0, \infty)$.  
% $\mu$ is thus proportional 
The  effective string  length,  
$L= 2 \pi \mu = 2 \ln \S  \to \infty $,  
 scales out of the classical action and  quantum corrections. 
For $L \to \infty$ the closed  folded string 
becomes effectively a combination of two infinite open 
strings  and  
%this solution  becomes  
is  related to a $J \not=0$ 
generalization of  the  null cusp of \ci{kru}  (see  \ci{krtt,rt2} and below).
At the classical  level $\hn = \el$.\foot{In general, $\el$  in \rf{ho}  defined in terms of  
expectation value of the quantum operator $J$ will be 
different   from  its classical value $\hn$, see \ci{rt2} and  section \ref{E_vs_Z}.}
The classical  energy in this limit is   ($\E\equiv {E \ov \sql}$) 
\def \te {\textstyle}
\be \la{cla}
\E_0 - \S = \k  = {\te{ 1 \ov \pi}}  \ff_0( \el  ) \ \ln \S   \ , \ \ \  \ \ \ \ \ \ \ \ \ \ \ 
\ff_0( \el  )= \sqrt{ 1 + \el^2}  \ . \ee
Quantum string corrections  $\sim {1 \ov (\sql)^n} $
change  the   coefficient of $\ln S$ in $E$, i.e. 
\bea 
 E-S = \  {\te {{\sql\ov \pi}}} \   \ff ( \jj, \l) \  \ln S  \ , \ \ \ \ \ \ \ \ \ \ \ \ \ \ 
  \ff ( \jj, \l)= {\te  \ff_0(\jj) + { 1 \ov \sql} \ff_1(\jj) + { 1 \ov (\sql)^2}  
   \ff_2(\jj)}  + ...\ . \la{fef}
  \eea
 The small $\el$ expansion   of  the {\it generalized scaling function }
 $\ff ( \jj, \l)$
   can be organized as follows \ci{am2,rt2} \foot{The formal expansion in \rf{hjk} is to be understood 
 in the sense that  higher powers of $\ln \jj$ are suppressed
 compared to lower  ones at each given order of the  strong   coupling expansion
 in which we {  first}
  take $\l\gg 1$  and { then}  take $\jj$ to be small.} 
 \be 
 && \ff ( \jj, \l)
 = f(\l) + \jj^2 \big[ q_0 (\l)  + q_1(\l) \ln \jj  
 + q_2(\l) \ln^2 \jj   + ...\big]\no \\
&&  \ \ \ \ \ \ \ \ \ \ \ \ \ \ \ \ \ \
 +\  \jj^4 \big[ p_0 (\l)  + p_1(\l) \ln \jj  + p_2(\l) \ln^2  \jj+ ...\big] \ +\   O(\el^6) \ . 
  \la{hjk}
 \ee
 As was  argued in \ci{am2},  the relation to $O(6)$ sigma model at small $\ell$ 
 determines the coefficients $c_n,d_n$  of the two leading  logarithmic terms 
 in the  large $\lambda $ expansion of the  $\ell^2$  part  of 
 \rf{hjk}
 % The  input is  the expression  of free energy of the $O(6)$ model \ci{has,EH} 
% and the string-theory  value of the 1-loop coefficient in $q_0$ in  \rf{exc}
% that determines
\be \la{ggff}
 \ff ( \jj, \l)- f(\l)  = \  \ell^2 \sum^\infty_{n=1} {\te { {1 \ov (\sql)^{n-1}} \ 
  \big( c_n \ln^n \ell + d_n \ln^{n-1} \ell
 + ... \big) +   O(\el^4) }}  \ . \ee
%   \la{pred}
%&& c_1 = -2 \ , \ \ \   d_1= { 3\ov 2} \ , \ \ \ \ \ \ \ 
%  c_2 = 8  \ , \ \ \   d_2= - { 6} \ , \ \ \ \ \ \ \ 
 %  c_3 = -32  \ , \ \ \   d_3= 12 \ ,  ...   \ee 
% Here $c_1,d_1$ follow from \rf{fa} and $c_n,\ d_n$ with $n>1$ are the values 
% determined \ci{am2} by the $O(6)$ model in the limit \rf{amm}.\foot{The values of $c_n$ 
% are essentially determined by the value of the 1-loop coefficient $c_1$, while 
% the values of $d_n$ are sensitive to the value of the ``string'' 1-loop 
% constant $d_1$  or  $\kk$ in \rf{scm}.}
  %according to \ci{am2}.
  The  exact 1-loop result  \ci{ftt} 
   \be 
\ff_1( \jj ) &=& {\te \frac{1}{ \sqrt{ 1 + \jj^2} } \left[ {
  \sqrt{ 1 + \jj^2}  -  1
 + 2 ( 1 + \jj^2 ) \ln (1 + \jj^2)  -   \jj^2   \ln \jj^2  
% \no\\ && ~~~~~~~~~~~~~~~~~~~~~~~~~~~~~~~~~~~~~~~~~~~~~~~
  - \ 2( 1 + \ha  \jj^2) \ln [   \sqrt{ 2 +   \jj^2}
    (1 + \sqrt{1 + \jj^2})] 
    %\ov   \sqrt{ 1 + \jj^2}  
    }\right] } \ \no \\
 &=& {\te {-{3 \ln 2}  - 2 \jj^2 ( \ln \jj   - { 3 \ov 4}) 
 + \jj^4 ( \ln \jj   - { 3 \ov 8}   \ln 2   - { 1 \ov 16} )
 + O(\jj^6)}} \la{fa} \eea
 implies  that $c_1 = -2  , \ \   d_1= { 3\ov 2}$ which together with the $O(6)$ model data 
 determine \ci{am2} the values of $c_n,d_n$ in \rf{ggff}, e.g.
 $ c_2 = 8   , \ \   d_2= - { 6} $. 
 
The  explicit 2-loop  string  result for the two leading     terms in the 
$\el\to 0  $ expansion of  $\ff(\ell, \l) $
  found in \ci{rt2} was 
 \be\la{tww}
 \ff_2(\ell) &=& - K   + \el^2  ( 8  \ln^2  \el - 6 \ln \el  + q_{_{02}} ) +  O(\el^4)  \ , \\
 q_{_{02}}{}_{_{\rm  string ?}}  &=& {\te {  - { 3 \ov 2} \ln 2 +   { 7 \ov 4} -   2 K  }}
    \ . \la{qq} \ee
Here the $\ell^0$ 
%$\ell$-independent 
term $K$ is the Catalan's constant    found earlier in \ci{rt1}; 
 the  coefficients of the $\el^2 \ln^2  \ell$ and $\el^2 \ln  \ell$ 
 matched  the prediction  based on the proposed 
 $O(6)$ sigma model related  \ci{am2}. 
 
The 1-loop  string result \rf{fa} was reproduced in \ci{krj}  by considering 
 the corresponding strong coupling  scaling limit of the asymptotic Bethe ansatz of 
 \ci{afs} with the 1-loop phase of  \ci{bt,hl}  (which itself was extracted  from other 
 string 1-loop corrections). 
 The  derivation \ci{grom} of the next-order   term 
 $ \ff_2(\ell)$ from the asymptotic  Bethe ansatz  with the all-order   phase of
   \ci{bes}  produced  
 the same  logarithmic terms as in  \rf{tww} but   a somewhat
  different value of  the constant $q_{_{02}}$, 
 \be 
  q_{_{02}}  = {\te {  - { 3 \ov 2} \ln 2 +   { 11 \ov 4} }}   \ . \la{qw} \ee
 The same    value \rf{qw} of  $q_{_{02}}$  
 %as  coming out  of  the  Bethe ansatz \rf{qw}
 was  found also  in  \ci{bass}   by extending the suggestion   of \ci{am2} -- 
 i.e. by using the relation   of   the energy
density of the quantum $O(6)$  sigma  model to the generalized scaling function  
  while  identifying  the $O(6)$ model  mass gap 
  with a  dynamical scale extracted from the integral 
FRS equation \ci{frs} which itself follows from the  asymptotic Bethe ansatz.
A disagreement  with the apparent string result \rf{qq} of \ci{rt2}
  then again  implies a  disagreement with
the  Bethe ansatz, i.e. that  it does not   capture the 
string   mass scale  correctly.\foot{An apparent  disagreement 
of the BA result of \ci{grom}   with yet another
result for $\ff_2$    found in   \ci{volin} directly  from the 
FRS equation  disappears  if  one drops  terms singular 
in $\ell \to 0$ in the   expression of \ci{volin}; it 
should be   due to a 
non-commutativity of limit  taken to arrive at  the 
FRS equation from the full  BA system and 
 the scaling limit.} 

Ref. \ci{grom}  also found   the  $\ell^4$ (and $\ell^6$) terms in $
 \ff_2(\ell)$,
 \be \la{ffq}
 (\ff_2)_{\ell^4}
 = {\te { 
 \ell^4\Big[ -6\ln^2 \ell-  (\frac{7}{6} - 3 \ln 2 )  \ln \ell
 -{9 \ov 8}  {\ln^2 2}    +\frac{11}{8}  \ln2  
   -\frac{233}{576}+\frac{3}{32}K  \Big]  }} \  ,  \ee
which could be compared to  superstring theory  provided 
the corresponding computation is extended to $\ell^4$ order (which would be very
 challenging in the
approach   of \ci{rt2}). 

\

The aim of the present paper is to resolve this annoying disagreement 
between  the string theory and the Bethe ansatz  results 
by redoing the 2-loop  string  computation in a  different  and much simpler 
 way than in \ci{rt2}. In \ci{rt2}  the conformal gauge  was  used  in which the 
 propagator  of bosonic modes is  complicated  making the evaluation 
 of the 2-loop graphs very tedious.   
 % and is largely done using  symbolic manipulation program.
 Here we shall use the  AdS light-cone gauge approach   explained in our previous
  paper \ci{grrtv} (where we considered  the 2-loop correction  for $\ell=0$). In this  approach  
   the computation becomes much more   transparent.  
 % We shall show that  the  new  superstring computation does 
 
 We shall find  that the superstring result is, in fact,  given not by \rf{qq} 
 but  by \rf{qw}, i.e. it matches  
   the  Bethe ansatz    value of  \ci{grom}. 
  We will also find the $\ell^4$ term in $\ff_2$ which is   again exactly  the same as the 
  $\el^4£$ term  \rf{ffq}   found in  \ci{grom}.
  This  provides a very   non-trivial  test of the quantum  integrability of
   the   \adss   superstring
   by  demonstrating  that 
     quantum string corrections are 
    described,  in the scaling 
    limit when the finite size   the   world sheet  cylinder can be ignored, 
    by the  asymptotic Bethe  ansatz with the BES \ci{bes} phase.

 The reason  why 
 the conformal gauge computation  of \ci{rt2} failed to produce the 
 same result  \rf{qw} for $ q_{_{02}}$   is probably related to the contribution 
 of the non-1PI  2-loop diagrams that were not  analyzed in detail in \ci{rt2}
 and also to the implementation   of the Virasoro  condition at the quantum level 
 (in particular, the assumption that $\langle H_{2d} \rangle =0$). 
 %R
 Evidence in this direction may be identified in the light-cone gauge calculation. Indeed,
 the fluctuation action contains a term linear in fluctuation fields and proportional
 to the Virasoro constraint. At the one-loop level a tadpole contribution, linear in fluctuation
 fields, is nonvanishing. From the standpoint of the effective action for fluctuations,
 this tadpole contribution leads to a (divergent) correction to the relation between the 
 parameters of the classical solution. It therefore follows that the classical Virasoro relation
 is nontrivialy modified at the quantum level.
 %AT
 \foot{In a  general  gauge theory,  the partition function 
 computed in perturbation theory by expanding  near a classical  
  solution should be non-trivial and gauge-independent
 provided the fluctuation fields satisfy the ``vacuum''  
 asymptotic conditions (decay at infinity) 
 and the same is true also for the gauge-transformed    fluctuation fields. 
 It is possible   that the  presence of massless   fluctuation fields in
  the conformal gauge computation may 
  lead to a violation of these assumptions. We thank  I.Tyutin   for a clarifying
 discussion of this point.
 }
 
 \

 Here  will  be able to  compute the 2-loop  corrections 
 to the  generalized scaling  function   for a  more general asymptotic
  solution than \rf{so}, \rf{Vi}  which contains one  additional  parameter -- 
   the  winding number $m$  of the string  around the $S^1$ in $S^5$. The corresponding
    string background  that
   generalizes \rf{so}, \rf{Vi} is (in conformal gauge)
   \be 
   t&=&\kappa\tau~,~~~~~ \r= \r(\s) \ , \ \ \ \ \ \theta=\kappa\tau  + \vartheta(\s) \ , \ \ \ \ \ \ 
   \vp=\nu \tau  + m \s  \ ,
   % \ \ \ \ \     \ \k,\mu,\nu, m   \gg 1 
   \la{p} \\
    \cosh \rho(\s)& =&{
   \te { \sqrt{1 + \g^2 } }}     \cosh (  \mu\sigma)  \ , \ \ \ \ \ \  
 \tan \vartheta (\s)  =  \g  \coth (  \mu\sigma) \ , \ \ \ \ \ \ 
 \g \equiv {\nu m \ov \k \m} \ , 
\label{ooo} \\  
\kappa^2&=&\mu^2+\nu^2 + m^2  \ ,   \ \ \ \ \ \ \ \ \ 
\mu\approx  \frac{1}{\pi}\ln \S\gg 1    \ , \ \ \ \ \  \nu = \J \gg 1 \  \ ,    
\label{i}\\  \la{jq}
\k,\mu,\nu, m   \gg 1  \ ,  \ && \ 
\hn\equiv { \nu \ov \mu} = { \rm fixed} \ ,\ \ \ \      
\hm \equiv { m  \ov \mu} = { \rm fixed}   \ ,\ \ \ \ 
\hk \equiv {\k \ov \mu}  = \sqrt{ 1 + \hn^2 + \hm^2} \ . %\pi \J \ov \ln S}
% \ , \ \ \ \ \ \ \ \ 
% \hm\equiv { m  \ov \mu} = {\rm fixed} % \pi  m  \ov \ln S} 
\ee
The background  \rf{p}, \rf{ooo}  is an exact solution for finite values of parameters 
assuming 
%$\sigma$ takes values in an infinite interval (i.e. 
one  does not impose the periodicity condition in $\s$, i.e. 
if one formally considers an open string.
Alternatively,  it can be  viewed as  a large spin  limit of a  closed
 string spinning  in $AdS_3$  and wound on $S^1
  \subset S^5$, 
 in  which case
 % $\s\in (0,{\pi\ov 2})$ (for a half-arc or one quarter of the full string) 
%should take  values in a finite $(0,2\pi)$ interval  but
%while 
 the parameters  should scale as in \rf{i}, \rf{jq}.
The presence of non-zero  winding ($\nu m \not=0$)  ``blows up'' the folded 
string -- its shape in $AdS_3$   becomes that of an ellipse (see section
\ref{action_and_solutions}).\foot{A similar 
background  appeared \ci{tst} as a limit of two-spin $S_1=S_2$ solution \ci{tst,rha}
in 
$AdS_5$.  
%solution does not appear to  be   discussed explicitly   in the literature.
 A generalization of the spiky string solution \ci{krusp} to the case of non-zero 
 momentum $J$ and winding
 $m$ in $S^1$ of $S^5$  was  considered in \ci{krut}  but   the corresponding large spin 
 $(\ln \S \gg \J)$ 
 asymptotic solution
 (generalizing the one in \ci{krutt})   that  should, in fact,  be equivalent 
  to the one in \rf{p}, \rf{ooo}   was not explicitly written down  there.
  } 
 It has classical energy 
 %  higher energy   than a folded string which  is point-like 
 %in $S^5$  
 (cf. \rf{cla}; here  $ \ell= \hn $) 
 \be \la{ca}
\E_0 - \S = \k  = {\te{ 1 \ov \pi}}  \ff_0( \el, \hm  ) \ \ln \S   \ , \ \ \  \ \ \ \ \ \ \ 
\ff_0( \el,\hm  )=\hk= \sqrt{ 1 + \el^2 + \hm^2} 
 = {\te \sqrt{ 1 + {\pi^2 \J^2 \ov \ln^2 \S}  + { \pi^2  m^2  \ov \ln^2 \S}  } } \ .
% \el\equiv { \nu \ov \mu} = { \pi \J \ov \ln S} \ , \ \ \ \ \ \ \ \ 
% \hm\equiv { m  \ov \mu} = { \pi  m  \ov \ln S} \ . 
 \ee
 On the gauge theory side the corresponding dual  operator 
 should represent  a higher anomalous dimension state 
 in the sl(2) sector which should lie high  above the ``ground state'' 
 in the band of twist operators (cf. \ci{bgk,kz}).\foot{In the scaling limit
 $\ln S \gg 1$, ${ m \ov \ln S}=\;$fixed the winding number is very large
 but the way this parameter is encoded in the structure  of the 
 dual operator  is not obvious.} 
 
 %While  this   solution (representing   a string wrapped on a big circle of a sphere) 
 %is  unstable for $\nu^2  < m^2$ 
 % certain values of  the parameters, its study  is  still  
 While this solution has higher  energy than folded string, 
  its study   is of interest as in this case the string is again stretched to the boundary of $AdS_3$ 
 and  is thus  related, as we shall see below, 
  to an (euclidean) open string world surface having  a  null  cusp Wilson loop interpretation. 
 Also, despite its appearance, the solution \rf{p}, \rf{ooo}
  turns out to be  equivalent to a homogeneous 
 one, 
 i.e. the quantum  corrections to its energy are explicitly computable by standard diagrammatic 
 methods 
 as we shall 
 demonstrate below.  
  
 \

 As in \ci{rt2} our strategy will be to  start  with  an  equivalent solution 
 in Poincar\'e coordinates  (related  by a world-sheet  euclidean rotation 
 and an $SO(2,4)$ transformation as in \ci{krtt}) 
   that can be  interpreted as on open string world surface ending on a
  null cusp at  the $z=0$ boundary  and  extended also in the $S^5$. 
  We shall then choose the AdS light-cone gauge as in \ci{mt2,mtt}   and 
  compute   the corresponding quantum \adss superstring  partition 
  function $Z$ in the 2-loop approximation 
  by expanding near this classical solution. Since 
   this  solution turns out to be  effectively homogeneous in the $\m \to
  \infty$ limit,   $W=-\ln Z$ will  be given, up to  an  (infinite) world-sheet volume 
  factor, by   a non-trivial function 
  of $\hn,\hm$  and the  inverse string tension ${2\pi  \ov \sql}$, 
  \be 
&&W=-\ln Z =W_0 + W_1 + W_2 + ... \ , \ \ \ \ \ \ \ \ \ \ 
W=  {\te \frac{\sqrt{\lambda}}{2\pi}} V   {\cal F}(\hn,\hm, \l) \ ,  \la{we} \\ 
%\frac{\sqrt{\lambda}}{2\pi}\frac{V_2}{4} {\cal F}(\hat \nu,\hat w)\\
 &&{\cal F}(\hn,\hm, \l) = \sum_{n=0}^{\infty} {\te  \frac{1}{(\sqrt{\lambda})^n}} \,
  {\cal F}_n(\hn,\hm)\,. \la{fe}
\ee
In contrast  to 
%what happened   in 
the case of $\hn,\hm=0$ (i.e.  $J=0,\ m=0$)
  when  $W$  was directly proportional to the cusp
anomaly $f(\l)$ 
% or    $E-S \ov \ln S$ 
 (cf. \rf{fef}),  for $\nu\not=0$,  as explained in  \ci{rt2}, 
and as  reviewed and adapted to light-cone gauge  in section~\ref{E_vs_Z} below,  
a further  transformation is required  to  obtain  the 
corresponding  quantum correction to the  closed-string energy $E-S$  
 and thus the 
generalized scaling  function $\ff(\el,\hm,\l)$ 
%from  ${\cal F}(\hn,\hm, \l)$
 (i.e.  the analog of  
 %the one 
 $\ff(\el,\l)$ in \rf{fef}
in  the case of $\hm\not=0$ with $\el= \hn + O( { 1 \ov \sql})$) 
 from  ${\cal F}(\hn,\hm, \l)$.
 That implies, in particular,  that 
 even if one might  try to relate the $J,m\not=0$ null cusp surface 
 to a generalized  cusp  Wilson loop, its anomaly given by the logarithm $\ln Z$ 
 of the string partition function will not give directly the generalized scaling function.

As we are interested in comparison to the Bethe ansatz results \rf{qw}, \rf{ffq} of \ci{grom} 
which were found only for  $\hm=0$,
in this paper  we shall compute  the 1-loop and 2-loop partition  
function (or   ${\cal F}_1(\hn,\hm)
$ and $ {\cal F}_2(\hn,\hm)$ in \rf{fe})   for
   generic arguments  but will 
%for  simplicity we shall 
extract the  generalized scaling function only for $\hm=0$
(for this reason we shall ignore the  dependence on $\hm$ in section \ref{E_vs_Z}). 
%\foot{ {\bf  we know the recipe for $\hm=0$... how does it generalize to $\hm\not=0$ ?! 
%should be similar} } {\bf do we really want to do this? We do have $F_2$ in terms of $w$ after 
%all... In sec 2 $w$ is on the same footing as $\mu$, that is, it is one extra parameter that enters 
%the definition of the vacuum once $\nu$ and $\kappa$ chemical potentials are present.}

In section \ref{action_and_solutions} we shall   introduce 
 the  AdS light-cone gauge action for the  \adss superstring
 and  find the classical open string  solution that  represents
   a generalized null cusp 
 extended also along $S^1$ in $S^5$. We shall then
 introduce  a closed string  solution  that generalizes  the folded spinning string 
 of \ci{gkp,ft1}  
 to the case of non-zero winding in  $S^5$ 
 %(in which case the string is no longer folded aga but has circular topology  
  (which  should belong to the family of ``rounded''  spiky strings   discussed in \ci{krut})
  and   explain  why  its large spin  asymptotics \rf{p}, \rf{ooo}
  is equivalent to the generalized null cusp solution.

 In section \ref{expansion_and_1loop} we shall  discuss  the expansion   
 of the  light-cone gauge action  near
  the generalized null cusp  
 solution  of section \ref{action_and_solutions} 
 and compute the 1-loop  correction to  its partition function. In the limit 
  $\hm=0$ the result will of course  agree with \rf{fa} as was already 
  found in \ci{krtt} (for $\hn=0$) 
   and in \ci{rt2} (for $\hn\not=0$).
 
 In  section \ref{2loopZ} we shall describe  the Feynman diagram 
  computation of the 2-loop correction to $Z$ or
  $\F_2(\hn,\hm)$
 to fourth order in  small $\hn=\el + O({1\ov \sql})$ and small 
 $\hm$ expansion.
 %order $\hn^4$ in small $\hn$  (i.e. small $\el$)  expansion. 
 Then   in section \ref{generalized_scaling_function}
  we shall show that the $\el^2$ and $\el^4$  terms in the 
 corresponding generalized scaling   function 
 $\ff_2(\el)\equiv  \ff_2(\el,\hm=0)$   
 reconstructed according to
 the  rules of  \ci{rt2} and  section \ref{E_vs_Z}   are  in full  agreement  with the Bethe ansatz 
 results \rf{tww}, \rf{qw}, \rf{ffq} 
 of \ci{grom}. We also extract the exact expressions of the coefficients of the two leading logarithms,
 $(\ln \ell)^2$ and $\ln \ell$. The former agrees with the Bethe ansatz result \ci{grom}. The latter 
 is new; its series expansion reproduces the similar terms obtained from the Bethe ansatz.
Section \ref {remarks} will contain some concluding remarks. 

Few useful two-dimensional momentum 
integrals are collected in Appendices~\ref{sec:1loop_int}, 
\ref{sec:int_computation} and \ref{sec:2loop_int}. 
The reduction of tensor integrals to scalar integrals is discussed in Appendix~\ref{sec:tensor_ints}.
Appendix~\ref{FermProp} presents  the details of the  fermionic propagator.
 Appendix~\ref{sec:thermodynamics}  reviews a thermodynamics relation used in section \ref{E_vs_Z}.
 Appendix~\ref{sec:one_loop_E_J} contains 
  the explicit  computation of the  one-loop expectation values of
   the current $J$  and of the energy $E-S$
  that tests the general relations derived in section \ref{E_vs_Z}.

%that   checks   the general  expression  for it discussed in section 2 ???

%%%%%%%%%%%%%%%%%%%%%%%%%%%%%%%%%%%%%%%%%%%%%%%%%
%%%%%%%%%%%%%%%%%%%%%%%%%%%%%%%%%%%%%%%%%%%%%%%%%

 %\section{Target space energy from string partition function in light-cone gauge}
\section{Generalized  scaling function  from string partition function \label{E_vs_Z}}

Here we shall  follow, with some clarifications,  ref.~\ci{rt2} and  discuss  
 the relation between   the  string partition function 
computed by expanding near a classical solution and  the corresponding 
quantum-corrected  AdS energy $E$. 
For a  generic 2-d  sigma model  the 
expectation values of 2-d  conserved quantities $Q_i$ (e.g.  spins)
% or windings {\bf ? what 
%would be the conserved charge whose eigenvalue is the winding?}) 
 computed   in a semiclassical approximation 
 %(and  thus expected to be large) 
can be found using a thermodynamical approach, i.e. by 
 adding  chemical potentials  to the 2-d 
world sheet   Hamiltonian 
and   obtaining the expectation values as derivatives of the partition  function, 
%and the expectation values are found as derivatives of the partition  function. 
\be
{\widetilde H}_{2d}=H_{2d}+\sum_i h_i Q_i~ \ , \ \ \ \ \ \ \ \ \ \ \ \ 
Z= e^{- \beta \Sigma (h_i) } =  \tr\  e^{-\b  {\widetilde H}_{2d}}   \ . 
\ee
Such a strategy relies on the fact that $Q_i$ are conserved  and mutually commuting, 
 $[H_{2d}, Q_i] = 0$,\  $[Q_i,Q_j]=0$.
 %AT
 %Here $\beta$ is the large  time interval. 
%From the standpoint of the undeformed theory, 
  $Z(h_i) $ %in the presence of chemical potentials
 may be also interpreted as the  sigma model 
partition function in  a nontrivial homogeneous background parametrised  by $h_i$.

The case of  string   theory is similar, but  it is necessary to take into account 
 the Virasoro  constraints.
% additional arguments are needed for the construction of the expectation 
%values of charges. 
From the point of view of the path integral computation of the 
string sigma model partition function in which the chemical potentials 
appear as semiclassical background field parameters,  
% in which 
%of the  background interpretation of the sigma model partition function 
%calculation  implies that 
the Virasoro constraints should relate the chemical potentials 
and thus should modify the expressions   for the expectation values of the 
charges as derivatives of the  partition function.

We will be interested in the expectation values of $E-S$ ($E$ and $S$ are  the 
$AdS_5$  energy  and  spin) and $J$ (the $S^5$ orbital momentum). 
%\iffalse
%It is an interesting question 
%whether their expressions in terms of the independent chemical potential (say $\nu$)
%\be
%E-S=(E-S)(\nu) \,\qquad J=J(\nu)
%\label{ESJvsnu}
%\ee
%depends on world sheet gauge choices. While $E$, $S$ and $J$ are conserved quantities 
%labeling (from a target space standpoint) the states of the theory, the chemical potential 
%$\nu$ is entirely a world sheet quantity with no direct target space physical interpretation. 
%This leads to the {\bf (tentative?)} conclusion that, separately, the equations (\ref{ESJvsnu})
%need not be gauge-choice independent; however, the relation $E-S=(E-S)(J)$ obtained by 
%eliminating $\nu$ between these two equations is gauge-choice independent.
%\fi
%Since world sheet Hamiltonians have a strong dependence on world sheet gauges, 
%it is not immediately clear which undeformed world sheet Hamiltonian $H_{2d}$ 
%we should use in
As in \ci{rt2}, it is   natural  to view $\k$ and $\nu$   in \rf{so} 
as the corresponding chemical potentials, i.e. consider 
\be
{\widetilde H}_{2d}=H_{2d}+\kappa (E-S)-\nu J \ , 
\ee
%The main constraint however arises from the requirement that 
where we should  require that $H_{2d}$ is such that 
$
[H_{2d},(E-S)]=0\ ,  \  [H_{2d},J]=0, 
$
and $\k$ is a function of $\nu$ according to the  classical Virasoro
%AT
 condition \rf{Vi}.\foot{In general, 
$\k$  may depend  on other parameters not corresponding to Noether charges.}
% {\bf Perhaps
%mention here that the homogeneous background interpretation of the calculation
%allows in principle other parameters to appear in $\k(\nu)$, such as parameters 
%of the solution not corresponding to an additional conserved quantity.}
%To allow for a comparison between different gauges it is perhaps useful to carry out a 
%gauge-invariant discussion before committing to any particular gauge. Thus, at least 
%temporarily, we will take 
%For example, we  may  define   $H_{2d}$  to 
%be the world sheet Hamiltonian defined schematically 
%as 
% $ H=p{\dot q}-L~.$  This  expression,  
 %is  determined by the 
%This is also proportional to the (phase space)
% world sheet stress tensor and, 
%in general, will depend on the world sheet metric.
Then the    partition function is 
\be
Z[\kappa(\nu),\nu]=\Tr\  e^{-\beta{\widetilde H}_{2d} (\k(\nu),\n) } 
= e^{-\beta\Sigma(\nu)}  ~, 
\ee
where the trace involves  a sum over all the states of the theory.\foot{Note that 
in this formulation the states   carry no chemical potential dependence.} 
Then \be
\frac{d\Sigma(\nu)}{d\nu}=\frac{d\kappa(\nu)}{d\nu}
\langle E-S\rangle-\langle J\rangle \ .  
\label{dSdn}
\ee
A further equation,  
 which  is the analog of the usual relation between the free energy 
and the internal energy for 
 statistical mechanical systems, 
 %; see the appendix E for a derivation)
is found  in the limit $\beta \to \infty$ 
on a world sheet of  infinite spatial extent (see Appendix \ref{sec:thermodynamics}):
\be
\Sigma(\nu)=\langle H_{2d}\rangle +\kappa(\nu)\langle E-S\rangle-\nu\langle J\rangle\ . 
\label{UFrel}
\ee 
While the  expressions for $\Sigma(\nu)$ and  the 
resulting   expectation values  may  depend
on the choice of gauge, the existence of the relations (\ref{dSdn}) and (\ref{UFrel})
should be gauge-independent.
% For certain gauges however it may be difficult to argue 
%directly for their existence. It is also somewhat
%AT
%Since  it may be  difficult to argue directly for the existence 
%of these relations in a Lagrangian/path integral formalism defined 
% order by order in perturbation theory, 
% here  we 
%We will adopt the point of view that adopted below equation (\ref{ESJvsnu})
% that the only 
We shall assume that the only invariant 
information contained in the equations  (\ref{dSdn}) and (\ref{UFrel}) is the relation
\be
\langle E-S\rangle= \langle E-S\rangle  (\langle J\rangle)
\label{desired}
\ee
obtained by eliminating $\nu$  from  these two equations.
 This expresses 
$\langle E-S\rangle$ in terms of $\Sigma$ and $\langle H_{2d}\rangle$ and 
their  derivatives evaluated as 
 functions of $ \langle J\rangle $.\footnote{The latter quantity is also a function of $\nu$: 
 while $H_{2d}$ 
carries no chemical potential dependence, the probability measure used to compute the average is
 $\exp{({-\beta{\widetilde H}_{2d}})}$.}
 
% An important quantity that needs further discussion, in particular its relation to the partition 
% function, is $\langle H_{2d}\rangle$. 
 
 Below   we will  use the AdS light-cone gauge to  compute $Z$ and thus 
 obtain  (\ref{desired}).
% We will be interested in using the AdS light-cone gauge to 
 %find the relation (\ref{desired}).  It this context 
 It is important to stress  that  $H_{2d}$ is not the light-cone Hamiltonian 
  $H_{lc}=-P^-$  but  the usual world sheet 
 Hamiltonian evaluated in the 
 light-cone gauge. Since the latter Hamiltonian is nothing but a linear combination of the Virasoro 
 constraints which are solved in the light-cone gauge, the operator $H_{2d}$  should 
 %(and thus its expectation  values)
  vanish identically, i.e. 
  \be  \langle H_{2d}\rangle =0 \ . \la{heh} \ee 
Indeed, we may relate   $ H_{2d}$ to  the Lagrangian by 
%the relation between Lagrangian and Hamiltonian formalism is given by the
% Legendre transform
\be
{\cal L}={\dot x}^-P^++{\dot x}^+ P^- + {\dot x}_i P^i - H_{2d}~~, 
\ee
where $x_i$  labels  all the fields transverse to the light-cone directions.
$H_{2d}$ is a sum of Virasoro constraints with  coefficients which are 
components of the world sheet metric.
 In the light-cone gauge one sets 
 $
 x^+=\tau , \  ~P^+=p^+={\rm fixed}
 $
 and 
%and one integrates out the world sheet metric, i.e. one
 solves the Virasoro  constraints. Then  the light-cone 
Lagrangian, which 
 is used to evaluate the partition function in the  path integral formalism, is 
 (modulo a total  derivative term) 
 \be
{\cal L}
%&=&{\dot x}^-p^+ +
= {\dot x}_i P^i - (-P^-)={\dot x}_i P^i - H_{lc}~~.
\ee
Using \rf{heh} we   
conclude  from  \rf{dSdn}, \rf{UFrel}
   that %in light-cone gauge 
the equations 
determining the target space energy in 
terms of the $AdS_5$ and $S^5$ spins on a world sheet 
of infinite spatial extent (i.e. in the large  spin limit) are
\be
\label{thermo}
\frac{d\Sigma(\nu)}{d\nu}=\frac{d\kappa(\nu)}{d\nu}
\langle E-S\rangle-\langle J\rangle \ ,  \ \ \ \ \ \ \ \
\Sigma(\nu)= \kappa(\nu)\langle E-S\rangle-\nu\langle J\rangle \ .  \la{ahh} 
\ee
With these clarifications, it then follows that the relation (\ref{desired}) is exactly the one 
derived in \cite{rt2}, i.e.  the one following, upon solving for $\nu$,  from 
\be
 E -S\equiv \langle E-S\rangle
&=&-\Big[ \nu\frac{d\kappa(\nu)}{d\nu}-\kappa(\nu)\Big]^{-1}
\Big[\Sigma(\nu)-\nu\frac{d\Sigma(\nu)}{d\nu}\Big] \ , 
\la{ej} \\
J\equiv \langle J\rangle  &=&-\Big[\nu\frac{d\kappa(\nu)}{d\nu}-\kappa(\nu)\Big]^{-1}
\Big[\Sigma(\nu)\frac{d\kappa(\nu)}{d\nu}-\kappa(\nu)\frac{d\Sigma(\nu)}{d\nu}\Big] \  . 
\la{pq}
\ee
In the specific  case of the solution in \rf{so}, \rf{Vi}
where $\k= \mu \sqrt{ 1 + \hn ^2} $  we get as in \rf{we}, \rf{fe}\foot{Here  $\beta$   plays the role of the
(infinite)  time interval.}
%For the specific relation $\kappa({\hat\nu})$ relevant for us we have:
\be
W= \beta\Sigma(\hn)&=&\frac{\sqrt{\lambda}}{2\pi}V  {\cal F}(\hn)\ , 
~~~~~~~ V=2\pi\mu\beta~,  \ \ \ \ \  \   {\cal F}= 1 + { 1 \ov \sql } \F_1
 +  { 1 \ov (\sql)^2 } \F_2+ ...    \la{pqr}  \\
%~~~~~\kappa=\sqrt{1+{\hat \nu}^2}
%\cr
%{\cal F}({\hat\nu})&=&1+\sum_{n=1}^\infty { 1 \ov  (\sql)^{n} }
% {\cal F}_n({\hat\nu})
%\ee
%\be
E-S&=& \  \M \sqrt{1+{\hat\nu}{}^2 }\ 
\big[ {\cal F}({\hat\nu})-{\hat\nu}\frac{d{\cal F}({\hat\nu})}{d{\hat\nu}}\big] \ , 
\la{q} \\
J 
&=&  \  \M \  \big[ {\hat\nu}{\cal F}({\hat\nu})-
(1+{\hat\nu}{}^2) \frac{d{\cal F}({\hat\nu})}{d{\hat\nu}}\big] \ , \la{o}
\ee
where $\M = {\sql \ov 2 \pi} L= \sql  \mu $ is the ``string mass'' (tension 
$\times$  length)
%,\foot{Recall that we are in 
%,the scaling  limit $\sql \ln S >> J , \ \ S >> \sql,\  \lambda \gg 1, $
%, makes $\ln \S$, $\ln S$ and $\ln S/J$ indistinguishable.}
\be \la{my}
{\cal M} \equiv    { \sql \ov 2 \pi} { V \ov \beta}    
=\sql \mu =  {\sql  \ov \pi} \ln S  \ \gg\  1 
\ . \ee
We shall  check the consistency  of the relations \rf{ej}, \rf{pq}  or  \rf{q}, \rf{o}
in Appendix~\ref{sec:one_loop_E_J}
by directly evaluating the  expectation values of $E-S$ and $J$ in the 1-loop
 approximation. 

Defining 
\be 
%\la{defr}
&&{\rm f} (\ell) 
%= \hat {\rm f} (\hn ) 
\equiv  { E-S \ov {\cal M}}  \ , \  \ \ \ \ \ \ \ \ \ 
\ell\equiv \frac{ J}{ {\cal M}  } = \hn + { 1 \ov \sql } \ell_1 (\hn) 
 +  { 1 \ov (\sql)^2 } \ell_2(\hn) + ...    \ , \la{ln} 
\ee
we find from \rf{q},\rf{o}
\be
 {\rm f} (\ell) &=& \sqrt{1+{\hat\nu}{}^2 }\ 
\big[ {\cal F}({\hat\nu})-{\hat\nu}\frac{d{\cal F}({\hat\nu})}{d{\hat\nu}}\big] \ , 
\la{quu} \\
\ell 
&=&  \  {\hat\nu}{\cal F}({\hat\nu})-
(1+{\hat\nu}{}^2) \frac{d{\cal F}({\hat\nu})}{d{\hat\nu}} \ , \la{ouuu}
\ee
allowing one to  compute  $ {\rm f} (\ell)$  from  a given expression  for 
${\cal F}({\hat\nu})$  by  solving for $\hn$. 
Note that differentiating over $\hn$ one finds from  \rf{quu},\rf{ouuu} 
\be 
\la{gert}
{ d {\rm f} (\ell)  \ov d \ell} =  { \hn \ov  \sqrt{1+\hn^2} }  \ , \ \ \ \ \ \ \ \ \ \ \ \ 
{  {\cal F} ( \hn) \ov \sqrt{1 + \hn^2} }   = {\rm f} (\ell)  - \ell   { d {\rm f} (\ell)  \ov d \ell}
\ . 
 \la{impl}
\ee
These equations suggest that it might be possible to interpret the construction described 
here as minimizing the difference between the target space energy and the AdS spin 
for fixed angular momentum $J$.
%In this way  we find 
%Eliminating $\hat\nu$  from \rf{q}, \rf{o}  
The relations  \rf{quu}, \rf{ouuu} lead to  
the following expression for the  
quantum   corrections to  the  generalized scaling function $\ff(\ell)$ in  \rf{fef} 
in terms of $\F$  (here ${\cal F}_0=1$)\foot{$\ell_n (\hn) $ in \rf{ln} are  determined by $\F_n$ and their derivatives 
according to  the  relations \rf{ej}, \rf{pq} 
\ci{rt2}.}
\be
%E-S &=&\frac{\sqrt{\lambda}}{\pi}\ln S\ \ \Big[
%{\rm f}_0+\frac{1}{\sql }{\rm f}_1+\frac{1}{(\sql)^2}{\rm f}_2+\dots\Big]\ , \la{oo} \\
{\rm f}_0 &=& \sqrt{1+\ell^2}\vphantom{\Big|} \ ,  \ \ \ \ \ \ \ \ \ 
{\rm f}_1 = \frac{{\cal F}_1(\ell)}{\sqrt{1+\ell^2}} \ ,  \la{gg} \\ 
{\rm f}_2&=&\frac{1}{\sqrt{1+\ell^2}}\Big[{\cal F}_2(\ell)
+\frac{1}{2}\Big(\frac{\ell}{\sqrt{1+\ell^2}}\,{\cal F}_1(\ell)
-\sqrt{1+\ell^2}\,\frac{d {\cal F}_1(\ell)}{d\ell}\Big)^2\Big] \nonumber\\[2pt]
&=&\frac{{\cal F}_2(\ell)}{\sqrt{1+\ell^2}}  +\frac{1}{2}(1+\ell^2)^{3/2}
\Big(\frac{d{\rm f}_1}{d\ell}\Big)^2 \ . \la{pp}
\ee
%Here 
%\be
%\ell\equiv \frac{\pi J}{\sql \ln S} = \hn + { 1 \ov \sql } \ell_1 (\hn) 
% +  { 1 \ov (\sql)^2 } \ell_2(\hn) + ...    \ , \la{ln}
%\ee
%with $\ell_n (\hn) $ determined by $\F_n$ and their derivatives 
%according to  the  relations \rf{ej}, \rf{pq} 
%\ci{rt2}. 
%As follows  from from \rf{quu},\rf{ouuu} 
%\be 
%\la{gert}
%{ d {\rm f} (\ell)  \ov d \ell} =  { \hn \ov  \sqrt{1+\hn^2} } \ , \ \ \ \ \ \ 
%\ee
%implying that  $\rm f$ is the generating function  of transformation from (a function of) 
%variable $\hn$  to $\ell$. 

What remains then is  to compute  the  partition function or $\F(\hat \nu)$, 
use it  to determine  $\ff_2(\el)$ and compare the result with \rf{tww}, \rf{ffq}. 
This  appears to be technically most straightforward 
 in the light-cone gauge  ``open string'' (null cusp)  picture
 discussed in the next section. 

%\end{document}

%\newpage

%%%%%%%%%%%%%%%%%%%%%%%%%%%%%%%%%%%%%%%%%%%%%%%%%%%%%%%%%%%%%%%%%%%%%%%%%%%%%%%%%%%%%%%%%%%%%%%%%%%%%%%%%%%%%%%%%%%%%%%%%%%%%%%%
\section{String action in AdS light-cone gauge,  generalized null cusp   solution
and closed spinning string with winding in $S^5$ \label{action_and_solutions}}
%%%%%%%%%%%%%%%%%%%%%%%%%%%%%%%%%%%%%%%%%%%%%%%%%%%%%%%%%%%%%%%%%%%%%%%%%%%%%%%%%%%%%%%%%%%%%%

Our starting point will be   the  \adss   superstring action in AdS 
light-cone gauge  \ci{mt2,mtt}.
Using this  action below we  shall  discuss a classical solution   representing a generalized null 
cusp  and then describe  its relation to a closed spinning  string solution 
in conformal gauge.

\subsection{Action}

The AdS 
light-cone gauge is defined  in the Poincar\'e coordinates in $AdS_5$  in which the 
%This string theory in Poincar\'e patch in light-cone gauge was
%described in \ci{mt2,mtt}.
 10d metric  may be written as ($m=0,1,2,3; \ M=1,...,6$)
\be
ds^2 = z^{-2} (dx^m dx_m + d z^M d z^M) = z^{-2} (d x^m d x_m +
d z^2) + du^M du^M\ , \ \ \ \ \ \ \la{mei}\\
x^m x_m =
x^+ x^- + x^* x \ , \ \ \ \ \ x^\pm = x^3 \pm x^0 \ , \ \ \ \ \ \
x,x^* = x^1 \pm  \mathrm{i} x^2\ , \ \ \ \ \ \ u^Mu^M=1\ . \la{mee} 
\ee 
We shall later  use the following parametrization  of $S^5$\
 ($a=1,2,3,4$): 
\be%gin{gather}\label{upar}
u^{a} = \frac{y^{a}}{1+{\te {\te {\te \frac{1}{4}}}}y^2}\ , \qquad \
u^{5} = \frac{1-{\te {\te {\te \frac{1}{4}}}}y^2}{1+{\te {\te {\te \frac{1}{4}}}}y^2} \cos\varphi\ , \qquad\
u^{6} = \frac{1-{\te {\te {\te \frac{1}{4}}}}y^2}{1+{\te {\te {\te \frac{1}{4}}}}y^2} \sin\varphi\ . 
\la{nee}
\ee
%\\\text{with $y^2\equiv \sum_{a=1}^4 (y^a)^2$}, \quad \text{and $a = 1,\dotsc,4$.}\end{gather} 
 The angle $\varphi$
parameterizes a large circle $S^1 \subset  S^5$ at $y^a=0$.

%where $u^M u^M =1$ with $M=1, \dotsc ,6$, $x^m x_m =
%x^+ x^- + x^* x$, with $x^\pm = x^3 \pm x^0$ and $x = x^1 + \mathrm{i}
%x^2$.
The  AdS light-cone gauge is defined by imposing $ \Gamma^+ \theta^I=0$ 
on the two  Majorana-Weyl  fermions  in the superstring  action 
 as well as 
%\footnote{As in the standard conformal gauge, there are residual
%diffeomorphisms allowing one to choose also $x^+ = p^+ \tau$.}
\begin{equation}
  \label{ga}
  \sqrt{-g} g^{\a\b} = \diag(-z^2, z^{-2})\ , \qquad \qquad x^+ = p^+ \tau \ .
\end{equation}
 Then $x^-$ is determined from  the equations of motion
for $g_{\a\b}$, i.e. from the  analog of the Virasoro constraints.
 
The  resulting \adss   superstring  action  can be written as \ci{mt2,mtt} ($z^M=z \, u^M$)
\be
\label{s}
I &=& \ha T \int d \tau \int%_0^{2\pi \ell}
 d \sigma \; \mathcal{L}\ , \quad  \quad  \quad T =
\frac{R^2}{2 \pi \alpha'} = \frac {\sqrt{\lambda}}{2 \pi} \ , \\
\mathcal{L} &=& \dot{x}^* \dot{x} + (\dot z^M  + \mathrm{i}  p^+ z^{-2} z^N 
\eta_i {\rho^{MN}}^i{}_j \eta^j)^2  + \mathrm{i} p^+ (\theta^i \dot{\theta}_i +
       \eta^i\dot{\eta}_i - h.c.) - (p^+)^2 z^{-2} (\eta^2)^2 \no \\
&&
  \quad - z^{-4} ( x'^*x'  + {z'}^M {z'}^M) - 2 \Big[\ p^+ 
       z^{-3}\eta^i \rho_{ij}^M z^M (\theta'^j - \mathrm{i}
       z^{-1} \eta^j  x') + h.c.\Big]\;.  \label{la}
\ee
This  action has manifest $SO(6)\simeq SU(4)$ symmetry.  The fermions are
complex $\th^i = (\th_i)^\dagger,$ $\eta^i = (\eta_i)^\dagger\ $
$(i=1,2,3,4)$ transforming in fundamental representation of $SU(4)$. 
 $\rho^{M}_{ij} $ are  off-diagonal blocks of
six-dimensional gamma matrices in chiral representation and 
$(\rho^{MN})_i^{\hphantom{i} j} = (\rho^{[M}
  \rho^{\dagger N]})_i^{\hphantom{i} j}$ and
$(\rho^{MN})^i_{\hphantom{i} j} = ( \rho^{\dagger [M}
  \rho^{N]})^i_{\hphantom{i} j}$ are  the $SO(6)$ generators.
  % in the $\mathbf{4}$ and $\mathbf{\bar{4}}$ representations.

In  what follows  we  shall consider  the euclidean world sheet 
version \ci{grrtv} of this  action that may be  obtained 
by $\tau \to  -\mathrm{i} \tau, \ p^+ \to  \mathrm{i} p^+$.\foot{More precisely, 
one should start with a euclidean world sheet action before fixing the light-cone gauge.  
Equivalently, the transformation to the euclidean action can be  done by 
 $\s \to \mathrm{i} \sigma$  as in \ci{grrtv}.} 
$p^+$   can be set to 1 by  rescaling   the 
 string length  and  the fermions;  we shall assume  this in what
follows (see also \ci{mtt,grrtv}). 
The resulting  euclidean Lagrangian is then  
%{\bf  check consistency with  sect 3} 
\be 
\mathcal{L}_E = \dot{x}^* \dot{x} + (\dot z^M  + \mathrm{i}  z^{-2} z_N 
\eta_i {\rho^{MN}}^i{}_j \eta^j)^2  + \mathrm{i}  (\theta^i \dot{\theta}_i +
       \eta^i\dot{\eta}_i - h.c.) -  z^{-2} (\eta^2)^2 \no \\
   \quad  +  z^{-4} ( x'^*x'  + {z'}^M {z'}^M) + 2 \mathrm{i} \Big[\ 
       z^{-3}\eta^i \rho_{ij}^M z^M (\theta'^j - \mathrm{i}
       z^{-1} \eta^j  x') + h.c.\Big]\;.  \label{euc}
\ee

\subsection{Generalized null cusp  solution}

Let us now construct a  bosonic solution of this euclidean action 
 for which  
only the radial coordinate $z$ and one isometric angle $\varphi$ of 
$S^5$   are nontrivial (i.e. $x = x^* = 0$, $y^a = 0$). 
The relevant  part of \rf{euc} is then 
\begin{equation}
\mathcal{L}_E = \dot{z}^2 + z^2 \dot{\varphi}^2 +
  \frac{1}{z^4} \big({z'}^2 + z^2 {\varphi'}^2\big).
\end{equation} 
% Here, as usual, the prime stands for derivative with
%respect to $\sigma$, while dot stands for derivative with respect to
%$\tau$.
The  corresponding equations of motion are:
\be
-\ddot{z} - \partial_\sigma \big(\frac{z'}{z^4}\big) -
2\frac{{z'}^2}{z^5} - \frac{1}{z^3} {\varphi'}^2+z \dot{\varphi}^2 = 0,\ \ \ \ \ \ \
\partial_\tau \left(z^2 \dot{\varphi} \right) + \partial_\sigma
\big(\frac{1}{z^2} {\varphi'}^2\big) = 0 \ .  \la{ja}
\ee
The euclidean  analog  of the Virasoro constraints  determine the derivatives of $x^-$:
%The Wick rotated Virasoro constraints, evaluated on the solutions for
%which $x = x^* = 0$ and $y^a = 0$, are given by
\be
  \dot{x}^- + \dot{z}^2 + z^2 \dot{\varphi}^2 - z^{-2} (z^{-2} {z'}^2 +
  {\varphi'}^2) = 0,\ \ \ \ \ \ \ 
  \ha  {x'}^- + \dot{z} z' + z^2 \dot{\varphi} \varphi' = 0 \ . \la{ch}
\ee
A simple solution of the second equation (\ref{ja}) is $\varphi =0$;
the equations for $z$ and $x^-$  are then
solved by 
%the method of separation of variables,  $z(\tau, \sigma) = g(\tau) h(\sigma)$, 
%leading finally to 
%\begin{equation}
%g(\tau)=\sqrt{\tau}, \quad
%h(\sigma)=\frac{1}{\sqrt{\sigma}}.
%\end{equation}  By solving for $x^-$ from the Virasoro constraints we
%obtain the following solution
\begin{equation}
  z = \sqrt{\frac \tau \sigma},  \qquad x^+
  = \tau\ , \qquad x^- = - \frac 1 {2 \sigma} \ , \ \ \ \ \ \ 
  \quad \varphi = 0 \ . 
\end{equation} 
 This is the well-known null cusp solution of
\cite{kru} written in this light-cone gauge ~\cite{grrtv}:
since $z= \sqrt{ - 2 x^+ x^-}$ this open-string euclidean world sheet 
surface ends on  null cusp  at the 
 boundary $z=0$.
%we have $x^+ x^- = 0$ so the cusp solution intersects the boundary
%along two intersecting light-like lines.

More generally,  making a separation of variables  ansatz 
\be  \la{sepa} \varphi=a(\tau)+b(\sigma) \ , \ \ \ \ \ \ \ \ \ \ \ 
z(\tau, \sigma) = g(\tau) h(\sigma) \ , \ee
% The equations become
%(after dividing by $h$ and multiplying by $g^3$ the equation for $z$ and
%multiplying by $g^2$ and dividing by $h^2$ the equation for $\varphi$):
%\begin{align}
%0 &= -g^3 \ddot{g} - \left(\frac{h''}{h^5}-2\frac{{h'}^2}{h^6}\right)
%- \frac{{\varphi'}^2}{h^4} + g^4 \dot{\varphi}^2,\\
%0 &= g^2 \partial_\tau (g^2 \dot{\varphi}) +
%\frac{1}{h^2} \partial_\sigma \left(\frac{\varphi'}{h^2}\right).
%\end{align}
%Restricting to separated variable ans\"atze we search for $\varphi$ of
%the form $\varphi=a(\tau)+b(\sigma)$.  Then,
%\begin{align}
%0 &= -g^3 \ddot{g} - \left(\frac{h''}{h^5} -
%  2 \frac{{h'}^2}{h^6}\right) - \left(\frac{b'}{h^2}\right)^2 + (g^2 \dot{a})^2,\\
%0 &= g^2 \partial_\tau(g^2 \dot{a})+\frac{1}{h^2} \partial_\sigma
%\left(\frac{b'}{h^2} \right).
%\end{align}
one finds a particular solution of the equation of $\vp$ in \rf{ja} if 
%Some simple solutions of the $\varphi$ equation (certainly not the
%general factorized solution) are given by:
\begin{equation}
g^2 \dot{a} = \nua\ , \qquad \ \ \ \ \   h^{-2} {b'} = m \ , \ \ \ \ \ \ \ \ 
\nua,\, m={\rm const} \ . 
\end{equation}
Then the equation for $z$  is solved by 
\be
&&g(\tau) = \sqrt{2\kappa}\sqrt{\tau}\ ,\ \  \qquad
h(\sigma) = \frac{1}{\sqrt{2\mu}\sqrt{\sigma}} \ ,\\
  \label{eq:constraint}
&& \ \ \   \kappa^2 = \mu^2 - \nua^2 + m^2 \ . \la{cow}
\ee
As a result, we obtain the following solution
\begin{equation}
  \label{SJ}
  z = \sqrt{\frac{\kappa}{\mu}} \sqrt{ \frac{\tau}{\sigma}}\ , \ \ \ \ \ \ \ 
   x^+ = \tau\ , \qquad x^- = -\frac{\kappa \mu - \nua m}{2 \mu^2}\ \frac 1
  \sigma \ , 
   \qquad\ \ \ \ 
  \varphi = \frac{\nua}{2\kappa} \ln\tau + \frac{m}{2\mu} \ln\sigma %\ , \qquad\ \ \ 
 % x^+ = \tau\ , \qquad x^- = -\frac{\kappa \mu - \nua w}{2 \mu^2} \frac 1 \sigma 
 \ .
\end{equation}
% where the fields $x^{\pm}$ are obtained by solving the Virasoro constraints.  
%We should note here that it is precisely when
%the constraint~\eqref{cow} holds that $\dot{x}^- =0$, so
%$x^-$ only depends on $\sigma$.  
 Except in the degenerate case
$\kappa \mu - \nua m = 0$, this surface also ends on two intersecting
light-like lines at  the boundary $z=0$.

%Because we will often encounter ratios of the quantities $\mu$, $\nua$,
%$\kappa$ and $w$ it proves convenient to define the following
%quantities
%\begin{equation}
%  \hat{w} = \frac w \mu, \quad
%  \hat{\kappa} = \frac \kappa \mu, \quad
%  \hat{\nua} = \frac \nua \mu
%\end{equation} Note that what we call $\hat{\nua}$ here was called
%$\ell$ in~\cite{rt2}.

The induced two-dimensional metric for the
solution~\eqref{SJ}  is found to be (cf. \rf{ga}) 
\begin{equation}
  d s^2 = z^{-2} (d x^+ d x^- + d z^2) + d \varphi^2 = {\te {\te {\te \frac{1}{4}}}} ({\mu^2 +   m^2})
    \big(\frac {d \tau^2}{\kappa^2 \tau^2} + \frac {d
      \sigma^2}{\mu^2 \sigma^2}\big) \ .
\end{equation} 
This  can be  transformed to the  conformal gauge form by an obvious coordinate redefinition:\foot{The 
world-sheet coordinate $t$ here should not be confused with the   AdS time coordinate in 
\rf{so}, \rf{p}.}
%If we redefine the world sheet coordinates to read
\begin{equation}
  d s^2 = \frac 1 4 \left(1 + \hat{m}^2\right) \left(d t^2 +
    d s^2\right)\ ,\qquad  t = \frac \mu \kappa \ln \tau\ , \ \ \ \qquad s = \ln \sigma\ . \la{sts}
\end{equation}
As in \rf{Vi}, \rf{jq}   we shall often use the rescaled parameters 
\begin{equation}
\hna \equiv  \frac \nua \mu \  , \ \ \ \ \ \ \    \hat{m} \equiv  \frac m \mu\ , \qquad
  \hat{\kappa} \equiv  \frac \kappa \mu = \sqrt{ 1 - \hna^2 + \hm^2 } \ 
  . \la{resa} 
\end{equation}
Then the conformal gauge form of the solution  in \rf{SJ} is 
\be 
&& z=\sqrt{ \hk} \ e^{ { {1 \ov 2}} ( \hk t -s)} \ , \ \ \ \ 
x^+ = e^{  \hk t}  \ , \ \ \ \ \ x^- = - \ha ( \hk - \hna \hw)\  e^{-s}  \  , \ \ \ \ 
\vp =  \ha ( \hna t + \hw s )    \ ,  \la{tt} \\
&& x^+ x^-  = - \ha ( 1- \ge) z^2  \ , \ \ \ \ \ \  \ \ \ \ 
\ge \equiv   \frac {\nua m}{ \mu \kappa} = {  \hna \hw \ov \hk }  \ . \la{yty}
\ee 
The value of the euclidean  string 
action on this  classical solution   is
\begin{equation}
  I_{E}  %\frac T 2 \frac1 4 (1 + \hat{w}^2) \int d t d s = 
 = \frac {\sqrt{\lambda}}{2 \pi} (1 + \hat{m}^2) V \ ,
 \ \ \ \ \ \ \ \ \ \ \ 
 V \equiv \frac 1 4 \int d t d s \equiv \frac 1 4  V_2 \ . \la{vvv}
\end{equation} 
%where $V \equiv \frac 1 4 \int d t d s \equiv \frac 1 4 V_2$.
%Following \ci{kru,amm,krtt}
It  is useful to   write the solution~\eqref{SJ} in the   $R^{2,4}$ embedding coordinates
of $AdS_5$  
%(for which $X_0^2 + X_5^2 - X_1^2 -   X_2^2- X_3^2 -   X_4^2=1$) 
\be
  \label{eqg}
&&  X_0 = \frac {x_0} z\ , \quad X_i = \frac {x_i} z\ , 
  \quad X_4 = \frac 1 {2 z} (-1 + z^2 + x_m x^m)\ ,
   \quad X_5 = \frac 1 {2 z} (1 + z^2 + x_m x^m)\ , \\
&& \ \ \ \ \ \   X_0^2 + X_5^2 - X_1^2 -   X_2^2- X_3^2 -   X_4^2=1  \ .    
\ee
 %It is easy to see that in these coordinates the
%surface is defined by
We find that  the surface is described by 
\be  X_0^2 - X_3^2 = \frac 1 2 (1 - \ge) \ , \ \ \ \ \ \ \ 
  X_5^2 - X_4^2 = \frac 1 2 (1 + \ge) \ , \ \ \ \ \ \ \ X_1 = X_2 = 0 \ . \la{homo}
  % \\
%&&  \ \ \ \ \ \ \ \g \equiv \frac {\nua w}{ \mu \kappa} \ . 
  \ee
  %  - \frac {\nua w}{2 \mu \kappa}, \quad X_5^2 - X_4^2 = \frac 1 2 + \frac {\nua w}{2 \mu \kappa},
 %    \quad X_1 = X_2 = 0. \end{equation}  
 If we perform a formal  coordinate transformation
\be\la{coy}
  X_0 = \frac 1 {\sqrt{2}} (Y_2 - Y_0),\quad   X_5 = \frac 1 {\sqrt{2}} (Y_0 + Y_2),
  \quad  X_3 = \frac 1 {\sqrt{2}} (Y_5 - Y_1),  
  \quad X_4 = \frac 1 {\sqrt{2}} (Y_1 + Y_5),
\ee
we find that the surface becomes 
\begin{equation}
  \label{een}
  Y_0 Y_2 - Y_1 Y_5 = \ha \ge \ ,\ \ \  \qquad Y_0^2 - Y_1^2 +  Y_2^2 - Y_5^2 = 1.
\end{equation} 
This can be  put into the ``canonical''  form  by further swapping 
  $Y_2$ and $Y_5$. 
When  $\nua m = 0$, i.e. $\ge=0$, this is  the  familiar form of the null cusp
solution  in the embedding coordinates \ci{kru,amm,krtt}. 
%Again, the standard null cusp surface of \ci{kru} 
%is recovered in the limit $\g=0$. 
 % In order to put the equations in a more familiar form
 %we could swap $Y_2$ and $Y_5$, but the form presented above will be
 %more convenient for comparison with a closed string solution we will
 %discuss below.

Since  the null  cusp solution  can be related (by an analytic continuation and $SO(2,4)$
transformation)  to the asymptotic form \rf{so} 
of the large spin closed
 string solution~\cite{krtt}  it is natural to ask if its $m\not=0$   generalization 
 \rf{SJ}  has a similar closed string counterpart.

 \subsection{Closed string spinning in $AdS_3$  with
  momentum and winding in $S^1 \subset S^5$
    } 
 
%In the conformal gauge, 
%To be completed ....
To find a generalization of  the folded closed string  spinning in $AdS_3$ and 
orbiting in $S^1 \subset S^5$ \ci{ft1}  to the case of  non-zero winding number $m$ 
it is useful first to consider the corresponding solution in flat $R_t \times R^2 \times S^1$
space with coordinates $(t,x_1,x_2,\vp)$. The   folded string  spinning in $R^2$ 
   is described (in conformal gauge) 
  by\foot{In this subsection  we  consider the  Minkowski signature string  action  and use the
 conformal gauge.}
  $t= \k \tau, \ \   x_1 + {\rm i} x_2   =   \k  \sin k\s\  e^{ {\rm i} k \tau}  ,$  
  where $k=1,2,...$ is the number of folds ($k=1$ for standard folded string). 
  Its generalization   with  momentum  and  winding in $S^1$ is 
  \be 
 && t= \k \tau\ , \ \ \  \ \ \  \ \ \  \vp = \nu \tau + m \s \ , \ \ \ \ \la{flt} \\
 &&   x_1 = \ha \big[ \k_+  \sin k(\tau + \s)   - \k_- \sin k(\tau - \s)\big] \ , \ \ \ 
 x_2= \ha \big[-  \k_+  \cos k (\tau + \s)   + \k_- \cos k (\tau - \s)\big]  \ , \no \\
 &&\k_+ = k^{-1} \sqrt{ \k^2   - ( \nu +  m)^2} \ , \ \ \ \ \ \ \ \ \ 
  \k_- = k^{-1}  \sqrt{ \k^2   - ( \nu -  m)^2}  \ .   \la{jj}
  \ee
Then  the spin, angular momentum and energy are (the string tension is $T= 2 \pi \a'$):

\noindent
$S= { 1 \ov 2 \a'}  k^{-1}  ( \k^2 - \nu^2 -m^2), \ \  J = { 1 \ov  \a'} \nu, \ \
E=  { 1 \ov  \a'} \k$, i.e. 
\be   E= \sqrt{2\a'^{-1} k  S  +   J^2 +\a'^{-2} m^2 }   \ . \la{enq} \ee 
There are thus  4 independent parameters: $S,k,J,m$.  For $\nu m\not=0$ the 
string is no longer folded  but has a fixed-time profile of an ellipse;  e.g.
at $\tau=0$ we get   $  x_1 = \ha ( \k_- + \k_+) \sin k\s, \ \ 
 x_2 = \ha ( \k_- - \k_+) \cos k\s, $  i.e. 
 $ ({ 2x_1 \ov   \k_+ + \k_-})^2 + ({ 2x_2 \ov  \k_+ - \k_-})^2 =1 $, 
 with the smaller axis  going to zero in the limit $\nu m\to 0$. 
 Writing  $x_1 + {\rm i} x_2  \equiv\  \r \ e^{{\rm i} \theta}, \ \  \theta = \tau + \vartheta(\s) 
  $ we conclude that 
\be 
\rho = \ha \sqrt{  \k_+^2 + \k_-^2 - 2 \k_+ \k_- \cos 2 k\s } \ , \ \ \ \ \ \ 
\tan \vartheta = { \k_- - \k_+ \ov \k_- + \k_+ } \cot k\s    \ . \la{tqp} \ee 
Here $\vartheta$ changes from ${\pi \ov 2} $ to 0 as $\s$ changes from $0$ to $\pi \ov 2k$
so that $k$ is also the winding number of  $\theta$,  \ $\theta(\s + 2 \pi) = \theta(\s) + 2 \pi k$.  
This  elliptic string is thus 
% constructed out of 4 segments,
 representing a ``blown-up'' folded string. 

Since for small enough spin  the string in $AdS_3$  
should be small and thus moving in an approximately flat space 
this suggests to 
consider the following ansatz for the coordinates of the winding generalization of the 
folded spinning  string 
in the $AdS_3 \times S^1$ 
metric \rf{ada} \foot{We assume that the string wraps 
 big circle of $S^5$. 
 One may also consider a case when  the string is wrapped  on an arbitrary (e.g. small)  circle
 of $S^2 \subset S^5$, i.e. 
 $ds^2 = d \psi^2 + \sin^2 \psi \ d \vp^2 $, \ \ 
 $ \vp = m ( \tau + \sigma), \  \psi=\psi_0$=const. 
 In this case the relations  below are still  valid with the replacement 
 $\nu^2 + m^2 \to   2 \sin^2 \psi_0\ m^2 , \ \ 
 \nu m \to   \sin^2 \psi_0\  m^2$.}
\be 
   t=\kappa\tau~,~~~~~\ \   \vp=\nu \tau  + m \s  \ , \ \ \ \ \ \ \ \ 
    \r= \r(\s) \ , \ \ \ \ \ 
   \theta=\om\tau  + \vartheta(\s) 
   \ . 
   \la{pe} \ee
As in flat space, the   dependence of
   $ \theta$  on $\sigma$  (implying that the string will no longer be
straight in $AdS_3$) is required to satisfy the Virasoro conditions
 (which  are first integrals of  equations of motion)  
\be
 && \rho'^2 - \k^2 \cosh^2 \r   +  \om^2 \sinh^2 \r 
   + \sinh^2 \r \ \vartheta'^2     + \n^2 + m^2 
   = 0\ ,  \la{kot1} \\ 
&&    \ \ \ 
  \  \    \om\ \sinh^2 \rho \ \vartheta'  + \nu m = 0\ ,  \la{kot2}
\ee
%Here the parameters  $\k,\m,\n,w$   refer to the   closed  string solution 
%and will be  related to the  parameters in \rf{SJ} below. 
implying 
\be 
&&\rho'^2  =  \mu^2  \cosh^2 \r   -   \bar \mu^2  \sinh^2 \r  - 
 { c^2  \ov \sinh^2 \rho}  \ ,\la{ut} \\ 
&& 
\mu^2 \equiv \k^2 -  \nu^2 - m^2 \ , \ \ \ \  \ \ 
\bar \mu^2 \equiv \om^2 -  \nu^2 - m^2 \ , \ \ \ \  \ \ 
 \ \ \ c\equiv  { \nu m \ov \om}   \ . \la{gj}
\ee
The equation for $\r$  \rf{ut} is the same  
%as appears 
as in the case of the 2-spin 
($S_1=S_2$)   solution in $AdS_5$ \ci{tst,rha}. It can be rewritten as 
\be 
x'^2 = \mu^2 x^2 (x^2-1)   - \bar \mu^2 (x^2-1)^2 - c^2\equiv
 (\bmu^2-\m^2) (x^2-a_{-})  (a_{+}-x^2)
\ ,
\ \ \ \ \  \ \  x\equiv \cosh \r  \ ,
\label{yyy}
\ee
where\foot{Since  $x=\cosh \rho \geq 1$ we
   have  $\bmu \geq  \mu$ and
    $1 \leq a_{-} < a_{+}$.}
\begin{equation}
a_{\pm}= \frac{2 \bmu^2 - \m^2 \pm \sqrt{\m^4-4 c^2 (\bmu^2-\m^2)}}{2 (\bmu^2- \m^2)} \ . \la{aaq}
\end{equation}
The solution to the eq. \rf{yyy}  for $\rho$  is 
\begin{equation}
\cosh \rho= 
\frac{\sqrt{a_{-}}}{{\rm dn}[\  {\rm p}\,  \sigma,\ {\rm q} \ ]} \ ,  
\ \ \ \ \ \ \ \ \ 
{\rm p}\equiv   \frac{c\ \sqrt{a_{+}}}{\sqrt{(a_{+}-1)(a_{-}-1)}} \ ,   \ \ \ \ \ \ 
{\rm q} \equiv   1 - { a_- \ov a_+}   \ ,  
\label{sol1}
\end{equation}
where $\rm dn$ is the Jacobi  elliptic function.
Thus  the 
radial string coordinate  $\rho$ changes in the interval  $
 \r_- \leq \r \leq \r_+  , \ \   \cosh \rho_{\pm}= \sqrt{ a_\pm}    , 
$
where  $\r_-=0$  ($a_-=1$) when $ \nu m =0$  ($c=0$). 
As in \ci{tst,rha},
we will assume
that  $\rho$  starts at its minimum $\rho_{-}$ at $\sigma=0$ and goes to its  maximum
 $\rho_{+}$ at $\sigma=\frac{\pi}{n}$ where $n$ is an integer.
  To get a closed   string   defined on  $0 \leq \sigma \leq 2\pi$ we need to glue together
   $2 n$ such segments (or  $n$  string ``arcs'')  
    imposing the  periodicity condition
   $\rho(\sigma+2 \pi)=\rho(\sigma)$.
    Since the  period of   $\dn[z, u]$ function is $2 \rK[u ]$  (where $\rK[u ]$ is an elliptic integral) 
     we obtain the periodicity condition 
\begin{equation}
  \frac{2 \pi}{n}   \frac{c\ \sqrt{a_{+}}}{\sqrt{(a_{+}-1)(a_{-}-1)}}  =
  2 \rK[ 1 - { a_+ \ov a_-}  ] \
  .  \label{opi}  
\end{equation}
Again as  in \ci{tst,rha}, the  solution of the equation for $\vartheta$ 
in \rf{kot2}  can be written in terms of the
elliptic integral of the  third kind $\Pi[x,y,z]$ (and $\Pi[x,z]\equiv \Pi[x,{\pi \ov 2},z]$). 
 %we may  follow \ci{tst}
 To get a closed  string we are to 
    glue together several arcs to cover the whole 
 $2 \pi$  range of $\theta$, so that 
 \be  \vartheta(\sigma + 2 \pi) = \vartheta(\sigma) +  2\pi k  \ , \ \ \ \la{jkl} \ee
 where $k$ is  an  integer winding  number (generalizing the number of folds).
 Then the condition that the string is closed in $\theta $ is  
 $\vartheta(2 \pi)= 2 \pi k = 2 n \vartheta(\frac{\pi}{n}) $, i.e.   \ci{tst} 
 \begin{equation}
 \frac{\pi k}{n} {\frac{\sqrt{a_{+}(a_{+}-1)}} {\sqrt{a_{-}-1}}}  = \Pi[ { a_+ - a_- \ov a_+ -1}, 
1 - {a_+ \ov a_-} ]  ~~.
 \label{ond}
\end{equation}
As in flat   space, the presence of winding in $S^1$ 
 appears to change  the  topology  of the string profile 
  from a folded one into a  circular one. 
The  periodicity  conditions \rf{opi} and \rf{ond} 
 put constraints on the parameters.  One special  solution of \rf{kot1}, \rf{kot2} is 
 the $(S,J)$ circular string of  \ci{park}: \ $\r=$const, \ $\theta= w \tau +   k \s$ 
 for which \rf{kot2} implies  $k S + m J =0$. Then for $k=1$  the
  winding number $m$  is
 determined in terms of $S,J$.  This is also  a feature of some other special  solutions. 
 For example, for $k=1$  one finds the ``3-arc''  solution \ci{tst} 
 for which $n=3$. These solutions should  be  a special subclass of 
 generalized ``rounded'' spiky strings  with  momentum and winding in $S^1$ 
 considered in \ci{krut}. 
 
 The energy and the spin  of these   solutions 
 can be expressed in terms of elliptic integrals (cf. \ci{tst}) and one  may study various limits. 
 Here  we will not go into a detailed  analysis  of the moduli space of such solutions and concentrate 
 on the asymptotic large spin solution.

%%%%%%%%%%%%%%%%%%%%%%%%%%%%%%%%%%%%%%%%%%%%%%%%%%%%%%%%%%%%%%%%%%%%%%%%%%%%%%%%%%%%%%%%%%%%%%%%%%%%%%
  To obtain  the  asymptotic   solution 
  for  which the farmost  points of the string   reach the boundary, 
  i.e. $\rho_+ \to \infty$ or $a_+\to \infty,\  \bmu \to \mu,\  \om \to \k$, 
  it is sufficient to go back to the equations 
\rf{kot1}, \rf{kot2}  and solve them for $\om=\k$. One then finds for $\r$ and $\vartheta$ in \rf{pe} 
\foot{Here   $\k,\m,\n,m,\g$ 
  refer to the   closed  string solution; they  
 will be  related to the  parameters in \rf{SJ} below.}
\be 
 && {\rm cosh}\ \rho=   {\te { \sqrt{1 + \g^2 } }}     \cosh (  \mu\sigma)   \ , \ \ \ \ \ \  \ \ \  
{\rm tan}\ \vartheta   =   [  \g  \coth (  \mu\sigma)] \ ,\label{ooy} \\
 &&  \g \equiv {\nu m \ov \k \m} = { c \ov \mu} \  , \ \ \ \ \  \ \ \ \ \ \ \  
 \mu^2 = \k^2 -  \nu^2 - m^2  \ , 
\label{oiy}  \ee 
i.e. the solution \rf{p}, \rf{ooo}  already  mentioned in the Introduction. 
If we relax the condition of periodicity  in $\s$ 
(and assume that $\s$ takes values  in an infinite line) 
 then this is  an exact solution of \rf{kot1}, \rf{kot2}  with $\om=\k$ 
for arbitrary values of the parameters. 
To view 
\rf{ooy} as an asymptotic  limit of a closed string with finite spin, finite orbital momentum and 
finite winding  
 discussed above  we need  to assume  the scaling limit  \rf{jq}
 of large parameters with their ratios  being fixed, i.e. 
 \be 
 \k=\om,\ \mu,\ \nu,\ m\  \gg 1\  ,\ \ \ \ 
 \hat{\nu} \equiv  \frac \nu \mu , \ \    \hat{m} \equiv  \frac m \mu , \ \ 
  \hat{\kappa} \equiv  \frac \kappa \mu = \sqrt{ 1  + \hn^2 + \hm^2 }, \ \
  \g={  \hn \hm \ov  \sqrt{ 1  + \hn^2 + \hm^2 }}
    % \ \ \  \ = \ {\rm fixed} 
 \ .  \la{scaa} \ee
%In any case, 
 For $\bar \s = \m \s$ changing   from 0 to $\infty$   the radial 
$AdS_3$ coordinate changes from its minimal value  $ \rho_-={\rm arccosh}  {
    { \sqrt{1 + \g^2 } }}$    to  infinity  while 
   $\vartheta$ changes from $ { \pi \ov 2}$ to 
   $ {\rm arctan} \  \g $.
   
   %\foot{
  %%%%%%%%%%%%%%%%%%%%%%%%%%%%%%%%%%%%%%%%%%%%%%%%%%%%%%%%%%%%%%%%%%%%%%%%%%%%%%%%%%%%%%%%%%%%% 
   This  asymptotic solution  describes  just one  half-arc stretch of the string
   in $AdS_5$ with $ 0 \leq \s \leq { \pi \ov n}$.
   Eq. \rf{jkl}  then implies 
   the condition:  \  $ 2 n\ {\rm arccot}\ \g= 2 \pi k$, 
   i.e. $\g = \cot { \pi k \ov n}$.  
   %, i.e. one quarter of an ellipse for $n=2$. 
     The energy and the spin    of the corresponding closed-string 
      solution  are then given by 
     \be 
     \E_0= 2n  \k \int^{ \pi \ov n}_0    { d \s \ov 2 \pi }    \cosh^2 \r\  , 
   \ \ \ \ \   \ \ \S= 2n  \k \int^{ \pi \ov n}_0  {  d \s \ov 2 \pi }    \sinh^2 \r ~~. \ee
    % where $n=2$ for a folded string. 
     Assuming that $\mu  \to \infty$   we get 
     \be  \E_0-\S=  \k \  ,\ \  \ \ \ \S=  2n   \k \int^{ \pi \ov n}_0 
      {  d \s \ov 2 \pi }\ \big[ (1+ \g^2)    \cosh^2 (\m \s)  -1 \big] 
      \approx   {  n \k (1+ \g^2)  \ov 8  \pi \m  } e^{ 2 \pi \m \ov n}   \ee 
      so that 
     $ \mu \approx { n \ov 2 \pi}  \ln S $, i.e.  
    % To have a smooth limit to the 
    %  folded string case  for  $\g\to 0$   we need to set $n=2$, getting (cf. \rf{ca}) 
     \be  \E_0-\S \approx    \sqrt{ { n^2 \ov 4\pi^2} \ln^2 S + \nu^2  + m^2 }  \ . 
     \la{ener} \ee
     %i.e. the expression in \rf{ca}.
     %%%%%%%%%%%%%%%%%%%%%%%%%%%%%%%%%%%%%%%%%%%%%%%%%%%%%%%%%%%%%%%%%%%%%%%%%%%%%%%%%%%%      
   When $\g=0$ we recover
   %go back to the case of 
   the asymptotic  solution 
   \rf{so} describing a straight stretch of a  spinning   string with $\r$  changing from 
    0 to $\infty$
   (AdS boundary). 
   For $\g\not=0$ the  string is  bent, i.e. 
   $\r$ depends on $\vartheta= \theta(\tau=0,\sigma)$
   as  
   %[{\bf check this and above  ...}] 
    \be 
   \cosh^2 \rho = { 1 + \g^2 \ov   1 - \g^2 \cot^2 \vartheta } 
  \ . \la{varr} \ee 
  We sketch the profile  of one arc (two joined half-arcs each described by the above asymptotic
  solution)  of the  string 
  %at fixed  $\tau$ 
  % (an ellipse  in $(\rho,\theta)$  plane of 
 located in  $AdS_3$  which also  circles  $S^1$
%  which indeed  looks like   an arc of a spiky string) 
  in Figure 1. 
\begin{figure}
\begin{center}
\includegraphics[width=.5\textwidth]{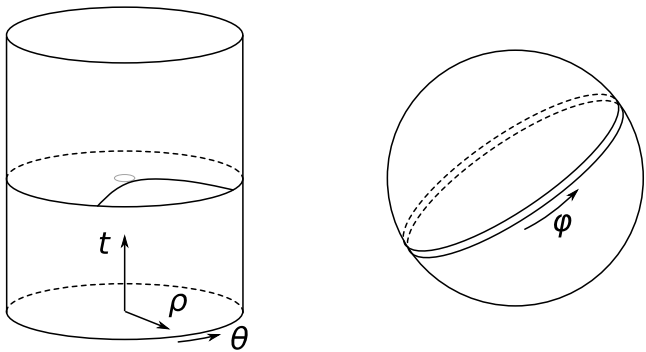}
\parbox{13cm}{\caption{Profile  of an arc of   fast-spinning
   string     in $AdS_3$   wound   around  a
big circle of $S^5$. 
%The size of the smaller axis of the ellipse 
%is determined by $\g$, i.e.  in   the case of zero winding number
% the string becomes straight and folded 
%and  passes through  the center of $AdS_5$.
}
\label{string}}
\end{center}
\end{figure}

%AAT
One may  of course construct a closed  string solution 
out of 4 half-arcs   each described by \rf{ooy}   by considering 
one arc in Figure 1 folded on itself.  In this case $n$   will be equal to 2 and  
the  energy  \rf{ener} will have a  smooth  limit $m\to 0$  matching  
the energy of the straight  folded  string non-winding case.
It is not clear, however, if such asymptotic closed string  solution can be considered 
as a large  spin limit of a finite spin  closed string solution
(see also sect.  3 of  \ci{tst} for similar remarks).
If the latter  exists, it should 
probably be described by a more general ansatz than \rf{pe}; also, in that case there would 
be, in contrast to what happens in the straight folded string case,  a 
 discontinuity between the small spin limit (described  by an approximately elliptic string) 
 and the large spin limit (described by that bent folded string).

As was already mentioned, the above  asymptotic  solution \rf{ooy} 
should be  equivalent 
   to  a particular   asymptotic 
    limit of the ``rounded''  spiky string  solution 
   % (the one with ``inverted'' spikes \ci{mos})
       with rotation and
     winding  in a big circle 
     of $S^5$   considered   in  ~\cite{krut}.
Indeed,  the  large spin limit  of the spiky string 
is conformally equivalent to the large spin limit of the  straight string 
 \ci{krutt}   and the same should be true for  $\nu m \not=0$.

Let us also note that a background similar to \rf{ooy} appeared in \ci{tst} as a large spin limit 
of the $S_1=S_2$  solution in $AdS_5$ \ci{tst,rha}.  
Its interpretation, however, was  different: there the angles 
of the $S^3\subset AdS_5$ ($ds^2 = d\theta^2 + \cos^2 \theta\ d\phi_1^2 + 
\sin^2 \theta\ d\phi_1^2$)  
were $\theta=\vartheta(\s), \ \phi_1=\phi_2= \k \tau$, while here 
$\theta=\k \tau    + \vartheta(\s), \ \phi_1=\phi_2= 0$.
The asymptotic  large spin 
solution in \ci{tst} was shown there to be $SO(2,4)$ equivalent to the 
asymptotic limit of  the folded spinning string. This is no longer so 
for the present  solution   \rf{ooy}  as we shall discuss below. 

%%\iffalse 
%%X1 + i X2 =   cos theta exp i phi1
%%X3 + i X4 =   sin  theta exp i phi2
%%then in S1=S2 case we had 
%%X1 + i X2 =   cos theta(s)   [ cos k t  + i sin k t ] 
%%X3 + i X4 =   sin  theta(s)   [ cos k t  + i sin k t ] 
%%while   here 
%%X'1 + i X'2 =   cos (k t + theta(s)) 
%%X'3 + i X'4 =   sin  ( k t +  theta(s) ) 
%%that seems   to suggest that the two  are related by 
%%a rotation -- indeed -- 
%%X'1= X1 - X4 ,    X'_2=0 
%%X'3=  X2  + X3 ,   X'4=0  
%%but this is   not a rotation so  the solutions are not really equivalent 
%%\fi  
%%%%%%%%%%%%%%%%%%%%%%%%%%%%%%%%%%%%%%%%%%%%%%%%%%%%%%%%%%%%%%%%%%%%%%%%%%%%%%%%%%%%%%%%%%%%%%
%%%%%%%%%%%%%%%%%%%%%%%%%%%%%%%%%%%%%%%%%%%%%%%%%%%%%%%%%%%%%%%%%%%%%%%%%%%%%%%%%%%%%%
%In order to compare with the previous solution, let us write this
%closed string solution in global coordinates

\subsection{Equivalence between the  generalized  null cusp 
and  the large spin   limit of the closed string solution} 
%%%%%%%%%%%%%%%%%%%%%%%%%%%%%%%%%%%%%%%%

Writing the solution \rf{ooy} in the  embedding $AdS_5$  coordinates 
\be   X_0 + \mathrm{i} X_5 = \cosh \rho\ e^{\mathrm{i} t}, \ \ \ \ \ \ \ \  X_1 +
  \mathrm{i} X_2 = \sinh \rho\ e^{\mathrm{i} \theta}, \ \ \ \ \ \  X_3=X_4=0 \la{emr}  \ee 
  we get  
%\begin{equation}
%\  X_0 + \mathrm{i} X_5 = \cosh \rho e^{\mathrm{i} t}, \quad X_1 +
%\  \mathrm{i} X_2 = \sinh \rho e^{\mathrm{i} \theta}.
%\\end{equation}  After some computation we find that
\begin{align}
  X_0 + \mathrm{i} X_5 &= \sqrt{ 1 + \g^2}\  \cosh (\mu \sigma)\ e^{\mathrm{i} \kappa
    \tau}\ ,\   \ \ \ \ \ \ \ \ 
  X_1 + \mathrm{i} X_2 &= \big[\sinh (\mu \sigma) + \mathrm{i}
    \g \cosh (\mu \sigma)\big] \ e^{\mathrm{i} \kappa\tau}  \  ,   \la{tak}
\end{align} 
implying that 
% Equivalently this can be written as
\begin{equation}
  X_0 X_2 - X_1 X_5 =   {\te {  \g  \ov {\sqrt{1 +  \g^2} } } }(X_0^2 + X_5^2)\ , \  \qquad\ \ 
  X_0^2 + X_5^2 - X_1^2 - X_2^2  = 1 \ .  \la{uu}
\end{equation}  
By applying two $SO(2,2)$ boosts in the planes $(02)$ and $(15)$
\be
&&  X_0 = \cosh v Y_0 + \sinh v Y_2,\ \  \ \ \ \ \   X_2 = \hphantom{-}\sinh v Y_0 + \cosh v Y_2,  \\
&&   X_1 = \cosh v Y_1 - \sinh v Y_5,  \ \  \ \ \ \ \  X_5  = -\sinh v Y_1 +
  \cosh v Y_5   \la{k} \  , 
 \ \ \  \ \  
 \tanh (2 v) = {\te {  \g  \ov {\sqrt{1 +  \g^2} } } }   \la{ye} 
   \ee
%  =\left(1 + \tfrac {{\kappa'}^2 {\mu'}^2}{{\nu'}^2
   % {w'}^2}\right)^{-\frac 1 2}$ 
%    the resulting equations simplify and
%we obtain that the surface in the $Y$ coordinates is given by
we can transform the equation for world surface \rf{uu} into the form
\begin{equation}
  \label{ed}
  Y_0 Y_2 - Y_1 Y_5 = \ha \g \ , \qquad\ \ \ \ 
  Y_0^2+ Y_5^2  - Y_1^2 - Y_2^2  = 1 \ .
\end{equation}
Comparing  this  to \rf{een}  we observe
that the   euclidean world sheet solution  for the null cusp 
 is related to the  Minkowski world sheet solution for the asymptotic limit 
 of the closed string 
 by the following analytic continuation\foot{The continuation of $Y_2,Y_5$ is 
 induced by  continuation of the time-like world sheet coordinate.} 
% (induced by  continuation of the time-like world sheet coordinate) 
%The open string solution in~\eqref{een} can be mapped
%to the closed string solution in~\eqref{eq:sol_global_closed} by the
%following transformations
\be
 && Y_0 \to Y_0, \qquad Y_1 \to Y_1, \quad Y_2 \to \mathrm{i} Y_2, \quad
  Y_5 \to \mathrm{i} Y_5,\la{cuc} \\
&& \nu_{_e} \to \mathrm{i} \nu,\ \  \  \mu \to \mu ,\  \ \  \kappa \to
 \kappa ,\  \ \  m \to m ,\ \  \g_{_e} \to \mathrm{i} \g,\ 
  \quad \ha \hk  t \to -\mathrm{i} \k \tau,\ \ \  \ha 
  s  \to \mu \sigma \la{cu}
\ee
Here  the transformations of the world sheet coordinates can be found, e.g.
%are  the ones that are required  to match
by comparing  the expressions for the $S^5$  angle 
 $\vp$  in \rf{tt} and in \rf{p}, \rf{pe}   ($\vp$ remains  real under the 
 euclidean rotation, with $\nu$  and $\tau$  rotating in ``opposite'' directions). 

%The last two transformations in the equation above are
%needed to make the solutions for the angle $\varphi$ agree as well.
%Note that under these transformations the constraint $\mu^2 = \kappa^2
%+ \nu^2 - m^2$ from the open string solution is mapped to the
%constraint ${\mu'}^2 \equiv {\kappa'}^2 - {\nu'}^2 - {w'}^2$, which we
%encountered while studying the closed string solution.  Also, this
%transformation converts between euclidean and lorentzian world sheet
%signature.

\

Following  the same logic as in \ci{rt1} (where the null cusp with $\nu=0$ was considered) 
and in 
\ci{rt2} (where  the $\nu\not=0$ generalization was considered)
below we shall use the euclidean  null cusp form of the 
solution \rf{SJ} or \rf{tt}  to 
compute the string partition function $Z(\hn,\hw)$ in the 
2-loop approximation, defining it in terms of  the path integral 
 with the string action \rf{euc}  in the AdS  light-cone gauge. 
We shall then  extract from $W= - \ln Z$  the expression for 
the  generalized scaling function  by applying  the relations \rf{pqr}--\rf{pp}
from  section 2. 

As we are ultimately   interested in the dependence on the parameters of the 
 closed-string  solution \rf{p}, \rf{ooo}, we shall  
  rotate the parameters of the 
 open string solution  \rf{SJ}, \rf{tt} as  in \rf{cu}, i.e. $\nu \to \mathrm{i} \nu$.   
 It turns out to be useful to keep 
 a symmetry between $\nu$ and $m$, so  we shall  use 
 the solution  \rf{SJ}, \rf{tt}  with the {imaginary} $\nu_{_e}=\mathrm{i}  \nu 
  $ {and} the  {\it imaginary}\  
 $m \equiv  \mathrm{i}  w  $
 (thus keeping $\nu_{_e} m = -\nu w   $ in  $x^-$ in \rf{tt} real), i.e. we shall
 start with   (cf. \rf{vvv})
 \be 
 &&\vp =  \ha \mathrm{i} ( \hn t + \hat w s )   \ , \ \ \ \ \  \ \ \ \ 
\hat w \equiv  -\mathrm{i} \hat m  \  , \ \ \ \ \ \ \ \ 
 \hat{\kappa}  = \sqrt{ 1 +  \hn^2 - \hat w^2 } \  ,\la{esa}  \\
 &&
    I_{E} 
 = \frac {\sqrt{\lambda}}{2 \pi} (1 - \hat{w}^2) V \  .  \la{F0}
\ee
We should  of course  set    $w=  -\mathrm{i}  m $, i.e. replace $w$
 by  the  physical closed-string  winding number  $m$ 
  in \rf{p}
in the final expression for the partition function. 

\

The  remarkable  property of  the generalized  null cusp solution \rf{SJ}, \rf{tt}
or the equivalent Minkowski world sheet  solution \rf{p}, \rf{ooo}, \rf{ooy}
is 
that, as in the zero-winding  $m=0$ case  \ci{krtt},  it is still  a homogeneous  solution, 
i.e. it can be put into the form where 
only the isometric angles of the  $AdS_5$ metric   
are non-zero and   linear in $(\t,\s)$. 
In this case  the  fluctuation Lagrangian can be put into a 
form where  it will  have constant ($(\t,\s)$-independent) 
coefficients  and thus the computation of  quantum corrections  can
be done by standard diagrammatic methods. 
 Starting with the form of the solution in \rf{homo}  the  argument is essentially the same 
 as in \ci{krtt}. Namely,  let us  introduce  the new global  $AdS_5$ coordinates 
  $(\b,r,p,q,h)$  
 by setting 
%   To see this explicitly 
%let us group $X_M$  as $
%(X_0^2 - X_1^2) + (X_5^2 - X_2^2)   - (X_3^2 + X_4^2 )=1$
 %and introduce  new   global  $AdS_5$ 
 %coordinates as 
\be &&X_0\pm X_3 = \ \cosh r \ \sin \b \ e^{\pm  p} , \ \ \ 
X_5\pm X_4 =\  \cosh r\   \cos \b  \  e^{\pm  q} \ ,\ \ \ 
X_1 \pm {\rm i} X_2 =\ \sinh r \    e^{\pm {\rm i}  h} \no
 \\
&&ds^2 = - \cosh^2 r\ d\b ^2  + dr^2
+  \cosh^2 r\ (\sin^2 \b\   dp^2 + \cos^2 \b\  dq^2)   + \sinh^2 r\ dh^2   \la{hqq}
\ee 
 As follows from \rf{tt}, \rf{yty}, \rf{homo}, in the case of the 
 generalized null cusp 
\be  \cos  (2 \b)  =  \g_{_e}  \ , \ \ \ \ 
p = \ha ( \hat \k  t +   s)  \ , \ \ \ q= \ha (\hat \k  t  -  s )  \ , \ \ \ \ \ \ \ 
 r=0\ ,\ \  \ \   h=0\ .  \la{ryy} \ee 
It is  easy to check that  conformal-gauge conditions  are satisfied if this 
background is supplemented by $\vp= \ha ( \hat \n_{_e} t+ \hat m s)$  in \rf{tt}. 
Since only the  isometric  directions $p$ and $q$  are non-constant and are
only   linear 
in world-sheet coordinates this is a homogeneous background.

%%%%%%%%%%%%%%%%%%%%%%%%%%%%%%%%%%%%%%%%%%%%%%%%%%%%%%%%%%%%%%%%%%%%%%%%%%%%%%%%%%%%%%%%%%%
\section{Expansion of the  string  action near
 the classical solution \\ and the 1-loop 
partition function \label{expansion_and_1loop}}

Let us now  expand the euclidean light-cone gauge Lagrangian \rf{euc}
near the classical solution \rf{SJ} or \rf{tt},  with  $\nu_{_e}= \mathrm{i} \nu, \ 
 m = \mathrm{i} w$  as in   \rf{esa}, and 
 explicitly identify  a field redefinition 
 that puts the  fluctuation Lagrangian into
  the form where it has constant
 coefficients. This will enable  us   to  compute the corresponding partition
 function  in semiclassical expansion by evaluating standard  Feynman diagrams. 
  The 1-loop partition function will be determined by the quadratic 
  part of the fluctuation  action. The 2-loop corrections will
   be  discussed in the next section. 
 
%  construct the all-order action 
%(since
% just the $z$ background is not a solution, 
% this action has a tadpole which we of course keep) and 
%then introduce the $\phi$ background and construct the all order action.

\subsection{Expansion of the action}
 It is useful to 
 construct the expansion of the 
 action 
two steps:  we  shall first  consider only the expansion near $z$
 background in \rf{SJ} and will include the $\vp$  background in \rf{esa}   later. 
 
The first step is essentially identical to the expansion  of the
 action around the standard  ($\nu,w =0$)  null cusp  solution 
\cite{grrtv}. 
It is useful to define the 
 fluctuations around the $z$  solution   with extra rescalings 
%to be  defined as follows (some of the constant 
%factors are introduced for later convenience)
\be
&& z=\sqrt{\kahat}\, \sqrt{\frac{\tau}{\sigma}}\ {\tilde z}
%=\sqrt{\frac{\tau}{\sigma}}e^{\tilde \phi}
~,~~~z^M=
\sqrt{\kahat}\,\sqrt{\frac{\tau}{\sigma}}\ {\tilde z}^M~
%~~~\alpha=\sqrt{\frac{\kappa}{\mu}}
\ , \ \ \ \ 
x = \sqrt{\kahat}\,\sqrt{\frac{\tau}{\sigma}} {\tilde x}
~,~~~~
\theta=\frac{1}{\sqrt{\sigma}}{\td\theta}
~,~~~~
\eta=\frac{1}{\sqrt{\sigma}}{\td\eta}\ ,  \la{fla}\\
&& 
{\tilde x}={\tilde x}_1+{\rm i} {\tilde x}_2
\ , \ \ \ \ ~~~~
{\tilde z}\equiv e^{\tilde\phi}=1 +\tilde\phi+ ...
\ , \ \ \ \  ~~~~
{\tilde z}^M={\tilde z} u^M 
\ , \ \ \ \  ~~~~
u^M u^M=1~~.\la{flf}
 %\cr
 %x=\frac{1}{\sqrt{2}}(x_1+{\rm i} x_2)~~~~~~~x^*=\frac{1}{\sqrt{2}}(x_1-{\rm i} x_2)
 %~~~~~~~~x_\perp=(x_1,x_2)
\ee
Then the Lagrangian written in terms of the coordinates $(s,t)$ in \rf{sts} 
becomes (we rescale overall factor of $\hk^2$) 
%It is also necessary to do a 2d coordinate transformation 
%(\ref{redefs})\be  t={\kahat}\,\ln \tau
%~,~~~~~~s=\ln \sigma
%%~,~~~~~~dt ds=\kahat\, \frac{d\tau d\sigma}{\tau\sigma}
%~,~~~~~~\tau\partial_\tau=\kahat\,\partial_t
%~,~~~~~~\sigma\partial_\sigma=\partial_s\ . 
%\ee
%Then, the action for fluctuations is
%(we set $p^+=1$ by rescaling the fermions) \foot{
%To get the Lagrangian below from its Minkowski counterpart 
%one may formally replace $ s \to i s$.
%Note also that here  we  assume we consider the open string case,
%i.e. relax periodicity condition  in  $\sigma$.  If we start with  $ 0  <  \sigma \leq 2
%\pi\ell $ 
%then the range of $s$ is  $ -\infty   <   s \leq  \ln ( 2 \pi \ell)$.} 
%The action, obtained by writing out explicitly the hermitian conjugation  
%in equation (\ref{la}) is and cancelling an overall factor of $\alpha^2$ against 
%the measure:
\be
{\cal L}  &=&
 \big|\partial_t {\tilde x}+\frac{1}{2}\kahat\,{\tilde x}\big|^2
 + ( \partial_t {\tilde z}^M +\frac{1}{2} \kahat\, {\tilde z}^M  
 + \frac{{\rm i} }{{\tilde z}^2} 
{\td \eta}_i  \rho^{MN}{}^i{}_j {\td \eta}^j  {\tilde z}^N )^2 \cr
&&~~~~~~~~~~~~~~~~~~~~~~
+ \  \frac{1}{{\tilde z}^{4}} \Big[
\big|\partial_s {\tilde x} -\frac{1}{2}{\tilde x}\big|^2 
+ (\partial_s{\tilde z}^M -\frac{1}{2} {\tilde z}^M)^2\Big]
 \cr
&&~~~~~~~~~~~~~~~~~~~~~+  {\rm i} 
({\td \theta}^i \partial_t{\td \theta}_i
+{\td \eta}^i\partial_t{\td \eta}_i + {\td \theta}_i
\partial_t{\td \theta}^i
+{\td \eta}_i\partial_t{\td \eta}^i)
-  \ \frac{1}{{\tilde z}^{2}} ({\td \eta}^2)^2  
\cr
&& 
+ 2\ {\rm i}\Bigl[\ \frac{1}{{\tilde z}^{3}}{\td \eta}^i \rho_{ij}^M {\tilde z}^M
(\partial_s{\td \theta}^j -\frac{1}{2}{\td \theta}^j 
- \frac{1}{{\tilde z}} {\td \eta}^j
 (\partial_s {\tilde x}-\frac{1}{2}{\tilde x}))\cr
&&~~~~~~~~~~~~~~~~~~~~~
           +\frac 1 {{\tilde z}^{3}}{\td \eta}_i (\rho^\dagger_M)^{ij} {\tilde z}^M
\big(\partial_s {\td \theta}_j -\frac{1}{2} {\td \theta}_j
+ \frac{{\rm i}}{{\tilde z}} {\td \eta}_j(\partial_s{\tilde x}^* -\frac{1}{2}{\tilde x}^*)\big)\Bigr] \ . 
\label{action_step_1}
\ee 
Let us now  discuss the  dependence on the $S^5$  angle  $\vp$ in \rf{nee}. 
Since  $\varphi$ represents an 
isometry direction,  it should  be possible to find a coordinate transformation 
such that only the derivatives of $\varphi$ appear in the action.
\iffalse
to put the action
into a form where  $\vp$  never appears without a derivative acting on it.
\fi
% The structure  of equation~\eqref{???upar} implies that 
According to \rf{nee}  the angle  $\varphi$ may be
shifted by  a  rotation in the $(56)$ plane.
Let us make such a rotation with an arbitrary angle $\check{\vp}$,
%$(R \tilde{z})^A = R^A_{\hphantom{A} B} \tilde{z}^B$, where the rotation
(the rotation matrix $R(\check{\vp})$  acts in  $(56)$ plane)
% restricted to lines and columns $5$ and $6$, reads
\begin{equation}
 (R \tilde{z})^A = R^A_{\hphantom{A} B} \tilde{z}^B \ , \ \ \ \ \ \ \ \   R =
 \begin{pmatrix}
  \hphantom{-} \cos \check{\varphi} & \sin \check{\varphi}\\
   -\sin \check{\varphi} & \cos \check{\varphi}
 \end{pmatrix} \ , \ \ \ \ \ \ 
 (\partial R)R^{-1}=\partial {\check\varphi}
\begin{pmatrix}
0 & 1\cr -1 & 0
\end{pmatrix}
~~. \la{yti}
\end{equation}
To define  the action of this rotation on the fermions, let us 
introduce the following  matrices
\begin{align}
 M^i_{\hphantom{i} j}(\check{\varphi}) &= \exp (-\frac 1 2
   \rho^{\dagger[5}\rho^{6]} \check{\varphi})^i_{\hphantom{i} j} = \delta_j^i \cos \frac
 {\check{\varphi}} 2  - (\rho^{\dagger [5}\rho^{6]})^i_{\hphantom{i}
   j}\ \sin \frac {\check{\varphi}} 2,\\
  M_i^{\hphantom{i} j}(\check{\varphi}) &= \exp (-\frac 1 2
    \rho^{[5}\rho^{\dagger6]} \check{\varphi})_i^{\hphantom{i} j} =  \delta_i^j \cos \frac
  {\check{\varphi}} 2 -
  (\rho^{[5}\rho^{\dagger 6]})_i^{\hphantom{i} j}\ \sin \frac {\check{\varphi}}2.
\end{align}  
Then  the action of the rotation $R$ on
the fermions $\theta$ is (the action on $\eta$  is the same) 
%we only write the action on the fields $\theta$; the
%action on the fields $\eta$ is identical to the action on the fields
%$\theta$ with similar indices)
\begin{align}
 \theta^i  \to M^i_{\hphantom{i} j}(\check{\vp}) \theta^j =
 \theta^j M_j^{\hphantom{j} i}(-\check{\vp}),\  \ \ \ \ \ \ \ \ 
 \theta_i  \to M_i^{\hphantom{i} j}(\check{\vp}) \theta_j =
\theta_j M^j_{\hphantom{j} i}(-\check{\vp}).
\end{align}
In the representation we are using the matrices $\rho^{\dagger[5}\rho^{6]}$ and $\rho^{[5}\rho^{\dagger 6]}$ 
are diagonal, purely imaginary and complex conjugate of each other. 
If $\check{\varphi}$ were a constant,  this would be a symmetry of
the action.  
%But in the following we will use $\check{\varphi}$ to
We may then use  such rotation to eliminate the dependence of the action on the
constant part of $\varphi$, i.e. to make the action depend only on derivatives 
of $\varphi$; in this case the  action will have constant  coefficients since 
$\vp$ in \rf{esa} is linear function of the world sheet coordinates. 
The rotated form of the action  can be written as  
%classical
% background for $\varphi$.
%Since this classical solution for $\varphi$ depends on the world sheet
%coordinates, applying the transformations above on $\tilde{z}^A$ and on
%the fermions will produce extra contributions containing derivatives of
%$\varphi$.  Since these derivatives are constant, we see that we
%obtain a homogeneous background.
%It is not hard to construct the transformation of the various components of the fluctuation 
%action (\ref{action_step_1}). Putting them all together leads to:
%%%%%%%%%%%%
%%%%%%%%%%%%%%%
\be
{\cal L}  &=&
 ~~\big|\partial_t {\tilde x} +\frac{1}{2}\kahat\,{\tilde x}\big|^2
\cr
&&
 + \Big( 
\partial_t (R{\tilde z})^M -[((\partial_t R)R^{-1})(R{\tilde z})]^M
+\frac{1}{2}\kahat\, (R{\tilde z})^M   
+ \frac{{\rm i} }{{\tilde z}^2} 
{\teta}_i  \rho^{MN}{}^i{}_j {\teta}^j  (R{\tilde z})^N \Big)^2 \cr
&&
+ \  \frac{1}{{\tilde z}^{4}} \Big[
\big|\partial_s {\tilde x} -\frac{1}{2}{\tilde x}\big|^2 
+ \Big(\partial_s R{\tilde z}^M 
-[((\partial_s R)R^{-1})(R{\tilde z})]^M -\frac{1}{2} R{\tilde z}^M\Big)^2\Big]
 \nonumber\\[3pt]
&&~~~~~~~~~~~+  {\rm i} 
\Big(\,
\ttheta^k(\partial_t\delta^k_l-\frac{1}{2}\partial_t{\check\varphi}
(\rho^{[5}\rho^{\dagger 6]} ){}_{k}{}^{l})\ttheta_l
+\teta^k(\partial_t\delta^k_l-\frac{1}{2}\partial_t{\check\varphi} 
(\rho^{[5}\rho^{\dagger 6]} ){}_{k}{}^{l})\teta_l
\cr
&&~~~~~~~~~~~~~~~
+ \ttheta_k(\partial_t\delta^k_l-\frac{1}{2}\partial_t{\check\varphi}
(\rho^{\dagger[ 5} \rho^{6]}){}^{k}{}_{l})\ttheta^l
+\teta_k(\partial_t\delta^k_l-\frac{1}{2}\partial_t{\check\varphi}
(\rho^{\dagger[ 5} \rho^{6]}){}^{k}{}_{l})\teta^l\Big)
-  \ \frac{1}{{\tilde z}^{2}} ({\teta}^2)^2 \bigg] 
\cr
&& 
+ \ \frac{2{\rm i }}{{\tilde z}^3}\ \Bigl[\ 
{\teta}^k\rho_{kl}^M (R{\tilde z})^M 
\Big(\partial_s{\ttheta}^l-\frac{1}{2}\partial_s{\check\varphi}
(\rho^{\dagger[ 5} \rho^{6]}){}^{l}{}_{u}\ttheta^u 
-\frac{1}{2}{\ttheta}^l - \frac{{\rm i}}{{\tilde z}} \,{\teta}^l
 (\partial_s {\tilde x} -\frac{1}{2}{\tilde x} )\Big)\cr
&&~~~~~
           +
{\teta}_k (\rho^\dagger_M)^{kl} (R{\tilde z})^M
\Big(\partial_s {\ttheta}_l-\frac{1}{2}\partial_s{\check\varphi}
(\rho^{[5}\rho^{\dagger 6]} ){}_{l}{}^{u}\ttheta_u
-\frac{1}{2} {\ttheta}_l+ \frac{{\rm i}}{{\tilde z}}\, {\teta}_l
(\partial_s {\tilde x}^* -\frac{1}{2} {\tilde x}^*)\Big)\Bigr] \ . 
\label{Action}
\ee 
%%%%%%%%%
Here  bosonic connection term
$(\partial_t R)R^{-1}$ vanishes in the directions $1,2,3,4$; in the $(56)$ plane 
it is given by \rf{yti}.

A family of choices for the rotation angle ${\check\varphi}$ in $R$ which renders 
 all coefficients in  the action constant  is 
\begin{equation}
\check\varphi=\frac{\rm i}{2}(\hat\nu \,t + \hat w \,s) +\delta \, \tilde\varphi (t,s)\ ,
\label{rot-par}
\end{equation}
where the first term represents our  classical background \rf{esa}
and $\tilde\varphi (t,s)$ is the   fluctuation field. 
The parameter $\delta$ 
may be further changed arbitrarily by background-independent rotations.
 It should 
disappear from all $SO(6)$-invariant quantities; 
we shall   keep it at intermediate stages as this  provides %observing this represents 
a test of 
the calculations.\footnote{The bosonic part of the Lagrangian is actually 
independent of $\delta$, but  interaction terms involving fermions have a non-trivial 
dependence on it. Also, for $\delta\ne 1$, the action contains terms with no derivatives on 
${\tilde\varphi}$.} 
%{\it This $\nuhat$ is the closed string real quantity; 
%$\what$ however is imaginary from a closed string perspective. This is is because 
%in these variables Virasoro looks like $\kahat^2=1+\nuhat^2-\what^2$. To be fixed later --R}.
%
Using \rf{nee} the   rotated coordinates are then ($a=1,\ldots,4$)
\begin{equation}
\begin{aligned}
(R\tilde z)^a= e^{\tilde\phi} \,\frac{y^a}{1+{\te  \frac{1}{4}}y^2}\,,~~~~  \ \ \ \  \ \ 
(R\tilde z)^5 +{\rm  i}(R\tilde z)^6 =e^{\tilde\phi}\,\frac{1-{\te \frac{1}{4}}y^2}
{1+{\te {\te {\te \frac{1}{4}}}}y^2}\, \ e^{{\rm i} (1-\delta)\tilde\varphi } 
%\cos[(1-\delta)\tilde \varphi]\,\,,~~~~
%(R\tilde z)^6=e^{\tilde\phi}\,\frac{1-{\te {\te \frac{1}{4}}}y^2}{1+{\te {\te \frac{1}{4}}}y^2}\,
%\sin[(1-\delta)\tilde\varphi] \,.
\end{aligned}
\end{equation}

\subsection{Quadratic part of the action}

It is easy to extract the quadratic bosonic 
part of (\ref{Action})  (we ignore  total derivatives and also drop a term 
linear in fluctuations and 
proportional to $(1+{\hat\nu}^2-{\hat w}^2-\hat\kappa^2)$; here $\a=(t,s)$)
\begin{equation}\la{qqw}
\begin{aligned}
{\cal L}_{B}^{(2)} &= \partial_{\alpha} \tilde x \partial_{\alpha} \tilde x^* 
    + {\te {\te \frac{1}{4}}}(1+\hat \kappa^2) \tilde x\tilde x^* 
    + %\sum_{a=1}^4 \big[
    \partial_{\alpha} y^a \partial_{\alpha} y^a + {\te {\te \frac{1}{4}}}(\hat\nu^2+\hat w^2) y^a y^a
    %\big]
    \\
&+ \partial_{\alpha} \tilde \phi \partial_{\alpha} \tilde \phi +\partial_{\alpha} \tilde\varphi \partial_{\alpha} \tilde\varphi
+2 \ri  (\hat \nu \partial_t \tilde\varphi-\hat w \partial_s \tilde\varphi) \tilde \phi+(\hat \k^2-\hat \nu^2)\tilde\phi^2
\cr
&\equiv ({\tilde x},{\tilde x}^*,{\tilde \phi},{\tilde \varphi},{y}_a)K_B
({\tilde x},{\tilde x}^*,{\tilde \phi},{\tilde \varphi},{y}_a)^T~~. 
\end{aligned}
\end{equation}
%%%%%%%%%%%%%%%%%%
%\iffalse
%The determinants of the corresponding kinetic operators are given by
%\begin{equation}
%\begin{aligned}
%&\tilde x,\tilde x^{*}:~~ \left( p^2+ {\te {\te \frac{1}{4}}}(1+\hat \kappa^2)\right)^2 \\
%&y^a:~~ \left(p^2+{\te {\te \frac{1}{4}}} (\hat \nu^2+\hat w^2)\right)^4 \,,~~~~a=1,\ldots,4\\
%&\tilde\phi,\tilde\varphi:~~ p^4 + \hat \kappa^2 p_0^2 + p_1^2 -2 \hat \nu \hat w \, p_0 p_1\,.
%\end{aligned}
%\end{equation}
%For $w=0$, this is exactly the same spectrum as in conformal gauge\footnote{To be more precise, in conformal gauge one finds two additional massless modes whose contribution is cancelled by the conformal gauge ghosts.}. The relevant propagators in momentum space are readily computed to be
%\fi
%%%%%%%%%%%%%%%%%%%
%Taking into account 
%the overall numerical factor of $1/2$ in the action  (\ref{s}), 
The  bosonic 
propagator is then (note the overall factor of $1/2$ in the action  (\ref{s}))
\be
\label{Bo}
G_B(p)\equiv K_B^{-1}(p) &=&
\begin{pmatrix}
0 & \frac{2}{p^2+{\te {\te \frac{1}{4}}}(1+ \kahat^2)} & 0  & 0 & {\bf 0}_{1\times 4} \cr
\frac{2}{p^2+{\te {\te \frac{1}{4}}}(1+ \kahat^2)} & 0& 0  & 0 & {\bf 0}_{1\times 4} \cr
0 & 0 & \frac{p^2}{{\cal D}_B(p)} & \frac{\nuhat p_0-\what p_1}{{\cal D}_B(p)} 
                      & {\bf 0}_{1\times 4} \cr
0 & 0 & \frac{-{\nuhat} p_0+\what p_1}{{\cal D}_B(p)} & \frac{\kahat^2-\nuhat^2+p^2}{{\cal D}_B(p)} & {\bf 0}_{1\times 4} \cr
 {\bf 0}_{4\times 1} & {\bf 0}_{4\times 1} & {\bf 0}_{4\times 1} & {\bf 0}_{4\times 1} & 
 \frac{{\bf 1}_{4\times 4}}{p^2+{\te {\te \frac{1}{4}}}(\nuhat^2+\what^2)}
\end{pmatrix}
\\[5pt] \la{Bob}
{\cal D}_B(p)&\equiv &p^4 + \hat \kappa^2 p_0^2 + p_1^2 -2 \hat \nu \hat w \, p_0 p_1
= p^2(p^2+1)+(\hat\nu^2-\hat w^2) p_0^2-2\hat\nu \hat w  \, p_0 p_1\ .
\ee
One may easily identify the bosonic fluctuation spectrum from the poles of this propagator:
 it consists of eight 
massive fields  and  for $\what=0$ 
 reproduces the massive part of the conformal gauge 
bosonic fluctuation spectrum  \ci{ftt}  around the  solution \rf{so}. 
% string. 
%%%%%%
%\iffalse
%\begin{equation}
%\begin{aligned}
%&G_{\tilde\phi\tilde\phi}=\frac{p^2}{{\cal D}_B(p)}\,,~~G_{\tilde\phi\tilde\varphi}(p)=-G_{\tilde\varphi\tilde\phi}(p)=\frac{\hat \nu p_0-\hat w p_1}{{\cal D}_B(p)}\,,~~G_{\tilde\varphi\tilde\varphi}(p)=\frac{\hat \kappa^2-\hat \nu^2+ p^2}{{\cal D}_B(p)}\,,\\
%&{\cal D}_B(p)=p^4 + \hat \kappa^2 p_0^2 + p_1^2 -2 \hat \nu \hat w \, p_0 p_1\,,\\
%&G_{\tilde x \tilde x^*}(p)=\frac{2}{p^2+{\te {\te \frac{1}{4}}}(1+\hat\kappa^2)}\,,
%~~G_{y^a y^b}(p)=\frac{\delta^{ab}}{p^2+{\te {\te \frac{1}{4}}}(\hat \nu^2+\hat w^2)}\,,~a,b=1,\ldots,4
%\label{Bose-prop}
%\end{aligned}
%\end{equation}
%\fi
%%%%%%
The  terms quadratic in fermions extracted from (\ref{Action}) are:
\begin{equation}\la{fee}
\begin{aligned}
{\cal L}_{F}^{(2)}=&{\rm i} 
\Bigr[\,
\ttheta^k(\partial_t\delta^k_l-\frac{\mathrm{i}\hat\nu}{4} 
(\rho^{[5}\rho^{\dagger 6]}){}_{k}{}^{l})\ttheta_l
+\teta^k(\partial_t\delta^k_l-\frac{3\mathrm{i}\hat\nu}{4} 
(\rho^{[5}\rho^{\dagger 6]}){}_{k}{}^{l})\teta_l\\
&+ \ttheta_k(\partial_t\delta^k_l-\frac{\mathrm{i}\hat\nu}{4}
(\rho^{\dagger [ 5}\rho^{6]}){}^{k}{}_{l})\ttheta^l
+\teta_k(\partial_t\delta^k_l-\frac{3\mathrm{i}\hat\nu}{4}
(\rho^{\dagger [ 5}\rho^{6]}){}^{k}{}_{l})\teta^l\Bigr] 
\\
& +2{\rm i}\Bigl[
{\teta}^k\rho_{kl}^5 
\Big(\partial_s{\ttheta}^l-\frac{\mathrm{i}\hat w}{4}(\rho^{\dagger[5}\rho^{6]}){}^{l}{}_{u}\ttheta^u 
-\frac{1}{2}{\ttheta}^l \Big)+{\teta}_k (\rho^{\dagger}_5)^{kl}
\Big(\partial_s {\ttheta}_l-\frac{\mathrm{i}\hat w}{4}(\rho^{[5}\rho^{\dagger 6]}){}_{l}{}^{u}\ttheta_u
-\frac{1}{2} {\ttheta}_l \Big)\Bigr] \cr
\equiv &(\tilde\theta^i,\tilde\theta_i,\tilde\eta^i,\tilde\eta_i)
K_F(\tilde\theta^i,\tilde\theta_i,\tilde\eta^i,\tilde\eta_i)^T~~.
\end{aligned}
\end{equation}
The corresponding kinetic operator matrix  in the momentum space is 
{\small 
\begin{equation}
K_F={\rm i}
\begin{pmatrix}
0 &  \ri p_0 {\bf 1}_4 -\frac{\ri  \hat \nu}{4} \rho^{[5} \rho^{\dagger\,6]} & 
-(\ri p_1+\textstyle{\frac{1}{2}} ) \rho^5 -\frac{\ri  \hat w}{4} \rho^6 & 0 \cr
\ri p_0 {\bf 1}_4 -\frac{\ri  \hat\nu}{4}  \rho^{\dagger [ 5} \rho^{6]}  & 0 & 0 & 
-(\ri p_1+\textstyle{\frac{1}{2}} ) \rho^{\dagger}_5 -\frac{\ri  \hat w}{4} \rho^{\dagger}_6 \cr
(\ri p_1-\textstyle{\frac{1}{2}} ) \rho^5 - \frac{\ri  \hat w}{4} \rho^6  & 0 & 0 & 
\ri p_0 {\bf 1}_4 -\frac{3\ri \hat \nu}{4}  \rho^{[5} \rho^{\dagger\,6]}  \cr
0 & (\ri p_1-\textstyle{\frac{1}{2}} )\rho^{\dagger}_5 - \frac{\ri  \hat w}{4} \rho^{\dagger}_6  & 
\ri p_0 {\bf 1}_4 -\frac{3\ri \hat \nu}{4} \rho^{\dagger[5} \rho^{6]} & 0
\end{pmatrix}~~.
\la{fw}\end{equation}
}
%The notation here is such that $\rho^M_u = (\rho^M)^{ij}$ and $\rho^M_d=\rho^M_{ij}$. 
Its inverse is somewhat complicated so we present  it in
 Appendix~\ref{FermProp}. 
 It has the following  structure, which exposes its poles
\be
K_F^{-1}(p)=\frac{N_+(p)}{{\cal D}_F(p)} + \frac{N_-(p)}{{\cal D}_F^*(p)} 
~,  \ ~~~~~\ \ \ 
{\cal D}_F(p)=(p_0 - \frac{\ri \nuhat}{4})^2+(p_1+ \frac{\ri  \what}{4})^2 
+ \frac{{\kahat}^2 + {\what}^2}{4} \ .
\la{fep}\ee
%where  the matrices $N_{\pm}(p)$ are complex conjugate of each other.
 All 8 
fermions are thus massive with equal masses. This is consistent
 with the surviving $SO(4)$ symmetry 
and charge conjugation \cite{am2}.

\subsection{One-loop partition function}

Using the determinants of the kinetic operators \rf{Bo} and \rf{fw} 
%constructed in the previous section
\be
\det K_B&=&
\big[ p^2+ {\te {\te \frac{1}{4}}}(1+\hat \kappa^2)\big]^2
\left(p^4 + \hat \kappa^2 p_0^2 + p_1^2 -2 \hat \nu \hat w \, p_0 p_1\right)
\big[p^2+{\te {\te \frac{1}{4}}} (\hat \nu^2+\hat w^2)\big]^4
\\
\det K_F&=&\Big[ \big( p^2+ \frac{4\hat\kappa^2-\hat\nu^2+3\hat w^2}{16}\big)^2 + 
 {\te {\te \frac{1}{4}}} (\hat\nu p_0 - \hat w p_1)^2\Big]^4~~,
\ee
one finds for the 
the one-loop contribution to the partition function in \rf{we}, \rf{fe} 
\be  Z_1= e^{-W_1} \ , \ \ \ \ \ \  \ \ \ \ 
W_1=\frac{V}{2\pi} {\cal F}_1(\hat \nu,\hat w) \ , \ \ \ \ \ \    V= {1 \ov 4} \int dtds= {1 \ov 4} V_2  \ , 
\la{oou} \ee
%%%%%%%
%%%%%%%
\be
&&{\cal F}_1 = 4\pi \int \frac{d^2p}{(2\pi)^2} \Bigr(2 \ln \big[ p^2+ {\te {\te \frac{1}{4}}}(1+\hat \k^2)\big]+ 
\ln (p^4 + \hat \k^2 p_0^2 + p_1^2 -2 \hat \nu \hat w \, p_0 p_1) + 4 \ln\big[p^2+{\te {\te \frac{1}{4}}}
 (\hat \nu^2+\hat w^2)\big]\no \\ 
&&~~~~~~~~~~~~~~~~~~- 4 \ln \big[ \big(p^2+ \frac{4\hat\kappa^2-\hat\nu^2+3\hat w^2}{16} \big)^2 + 
 {\te {\te \frac{1}{4}}} (\hat\nu p_0 - \hat w p_1)^2\big] 
\Bigr)~~. 
\la{qo}
\ee
The  constraint $\kahat^2=1+\nuhat^2-\what^2$ in the solution \rf{esa} 
then implies that 
the momentum  integral converges in the UV.\foot{
%AT
Note that there is an instability coming from $S^5$ fluctuations 
for  small  enough  orbital momentum or large enough winding, 
$\hat m^2= -\hat w^2  > \hat \nu^2$. The  same instability was present 
 in the case of the circular $(S,J)$ string solution \ci{park} (with $\S n + \J m=0$); it is 
 of course absent in large $\J=\nu$ expansion  for finite $m$.}
% for all allowed values of parameters.
 Evaluating  this integral  leads to 
%%%%%%
\iffalse
One can check that the momentum integration is finite for any $\hat\kappa,\hat \nu,\hat w$ 
satisfying the eq. of motion constraint $\hat \kappa^2-\hat\nu^2+\hat w^2=1$. Evaluating the 
integral, we get 
\fi
%%%%%%
\begin{equation}
\begin{aligned}
{\cal F}_1(\hat \nu,\hat w) = &-1+\hat w^2+\sqrt{(1+\hat \nu^2)(1-\hat w^2)}-(\hat \nu^2+\hat w^2) \ln(\hat \nu^2+\hat w^2)+2(1+\hat \nu^2) \ln(1+\hat \nu^2)\\
& - (2+\hat \nu^2-\hat w^2) \ln\big[\sqrt{2 + \hat \nu^2-\hat w^2}( \sqrt{1 + \hat \nu^2} + \sqrt{1 - \hat w^2})\big] \ . 
\label{F1}
\end{aligned}
\end{equation}
For $\what =0$ we thus recover the result for the  asymptotic $(S,J)$ folded string \rf{so} 
 or for the  null cusp with $\nu\not=0$ 
 obtained in 
conformal gauge in  \cite{ftt,rt2}. 
%As a consistency check, one can see that the one-loop
% effective action $W_1=\frac{V_2}{8\pi} {\cal F}_1(\hat \nu,\hat w)$ is 
Note that $W_1$ is  invariant under the simultaneous exchanges 
$
\nu \leftrightarrow w  , 
~~
\kappa \leftrightarrow \mu , 
~~
V_2 \leftrightarrow \kahat^2 V_2~, $
%\la{ujj}
as implied by the symmetry of the classical solution \rf{tt}, \rf{esa}. 
%consistently with the symmetries of the solution.

%\end{document}
%For $\hat \nu=0$, the integration yields instead
%\begin{equation}
%\Gamma_1 = -\frac{\hat \kappa^2}{4\pi} \left( 
%1 -\sqrt{1 + \tilde w^2} + 2 \tilde w^2 \ln\tilde w - 
%   2 (1 + \tilde w^2) \ln(1 + \tilde w^2) + (2 + \tilde w^2) \ln[\sqrt{2 + \tilde w^2}(1 + \sqrt{1 + %\tilde w^2})] \right)
%\end{equation}
%where $\tilde{w} = \hat w /\hat \kappa$. This is related to the $w=0$ result by exchanging $\tilde %w=w/\kappa$ with $\hat \nu = \nu/\mu$ 
%%%%%%%%%%%%%%%%%%%%%%%%%%%%%%%%%%%%%%%%%%%%%%%%%%%%%%%%%%%%%%%%%%%%%

%\end{document}

\section{Two-loop partition function \label{2loopZ}}

\begin{figure}
\begin{center}
\includegraphics[width=90mm]{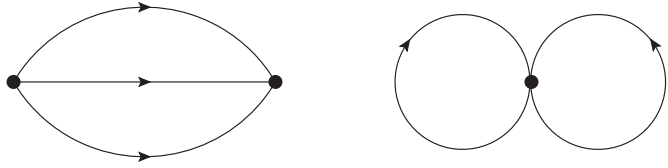}
\parbox{13cm}{\caption{The 2-loop 1PI topologies: ``sunset'' and ``double-bubble''. The propagators 
here are  either bosonic or fermionic.}
\label{1PI}}
\end{center}
\end{figure}

\begin{figure}
\begin{center}
\includegraphics[width=55mm]{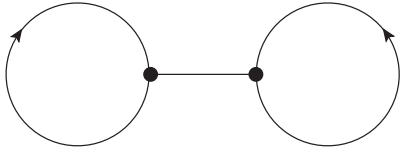}
\parbox{13cm}{\caption{The 2-loop tadpole topology. 
The non-vanishing graphs   have the internal line corresponding  to a $\tilde\phi$-propagator
 while the propagators in the loops   can be either bosonic or fermionic.}
\label{non-1PI}}
\end{center}
\end{figure}

In this section we shall  present the calculation of the 2-loop term in the logarithm of the partition function
in \rf{we}, \rf{fe}, i.e. 
\begin{equation}\la{wv}
W_2 = \frac{V}{2\pi\sqrt{\lambda}}\, {\cal F}_2(\hat \nu,\hat w)\,.
\end{equation}
 $W_2$ receives contributions from all connected vacuum diagrams that have the
  one-particle irreducible (1PI) topologies  in Fig.\ref{1PI}  and   the non-1PI
   ``tadpole" topology in  Fig.\ref{non-1PI}.  As in the case of the cusp without angular
    momentum on $S^5$  discussed in the AdS  light-cone gauge in \ci{grrtv}, here 
    the  non-1PI tadpole graphs 
     are again non-vanishing   and their contribution is important 
   %  play a crucial  role
     for the cancellation of 2-loop UV divergences and for
     reproducing the correct  2-loop term in the  generalized scaling function
     which agrees with the Bethe ansatz result. 
     % matching     the finite part  against the Bethe ansatz result.

For the computation of the 2-loop Feynman diagrams we need to expand the action (\ref{Action}) to fourth
 order in the fluctuations
\begin{equation}
I=I^{(0)}+I^{(2)}+I_{int}^{(3)}+I_{int}^{(4)}+\ldots\ , 
\end{equation}
where the structure of the interaction terms  may be written schematically as
\begin{equation}
\begin{aligned}
&I_{int}^{(3)}=\frac{\sqrt{\lambda}}{2\pi}\int dt ds \Big{(}\frac{1}{3!}V_{i_B j_B k_B}^{(B^3)} 
\Phi^{i_B}\Phi^{j_B}\Phi^{k_B}+\frac{1}{2!}V_{i_B| i_F j_F}^{(BF^2)} 
\Phi^{i_B}\Psi^{i_F}\Psi^{j_F}\Big{)}\\
&I_{int}^{(4)}=\frac{\sqrt{\lambda}}{2\pi}\int dt ds\Big{(}\frac{1}{4!}V_{i_B j_B k_B l_B}^{(B^4)} 
\Phi^{i_B}\Phi^{j_B}\Phi^{k_B}\Phi^{l_B}+\frac{1}{(2!)^2}V_{i_B j_B|i_F j_F}^{(B^2F^2)} 
\Phi^{i_B}\Phi^{j_B}\Psi^{i_F}\Psi^{j_F}\\
&\qquad \qquad \qquad \qquad +\frac{1}{4!}V_{i_F j_F k_F l_F}^{(F^4)} 
\Psi^{i_F}\Psi^{j_F}\Psi^{k_F}\Psi^{l_F}\Big{)}\,.
\end{aligned}
\end{equation}
Here $\Phi^{i_B}=(\tilde\phi,\tilde\varphi,\tilde x,\tilde x^*, y^a)$ includes the 8 bosonic fluctuations and $\Psi^{i_F}=(\tilde\theta^i,\tilde\theta_i,\tilde\eta^i,\tilde\eta_i)$ the 16
 fermionic ones. The vertices carry up to two derivatives or equivalently two momentum factors. 
The sunset and double-bubble contributions to $W=-\ln Z$  correspond to
\begin{equation}
\begin{aligned}
&W_{2~{\rm sunset}}= -\frac{1}{2} \langle I_{int}^{(3)} I_{int}^{(3)}
  \rangle_{_{\mbox{\scriptsize 1PI}}}\ , \ \ \ \ \ \
\ \ \ \ \ \ \ \ W_{2~{\rm bubbles}}= \langle I_{int}^{(4)} \rangle\ ,
\label{1PI-diag}
\end{aligned}
\end{equation}
while the tadpole contribution is obtained from
\begin{equation}
W_{2~{\rm tadpoles}}= -\frac{1}{2} \langle I_{int}^{(3)} I_{int}^{(3)}  
\rangle_{_{\mbox{\scriptsize non-1PI}}}\,.
\label{tad-diag}
\end{equation}
As usual,  the expectation values above are to be computed by inserting the appropriate propagators 
derived from the quadratic part of the action (see \rf{Bo}, \rf{fw}).
%The possible topologies correspond to the ``sunset" and ``double-bubble", see fig.1.  
%These diagrams are one-particle irreducible (1PI). 
%They do not exhaust though all possible contributions to the 2-loops partition function 
%as it turns out that in our background non 1PI contributions (tadpoles) are also present. 

In the following  subsections we  shall present some details of the computation 
of the relevant 2-loop Feynman diagrams. We begin with the analysis of the 
1PI bosonic diagrams and later 
%move to fermions
discuss the fermionic contributions. We end this section with the 
calculation of the tadpole contributions.

\subsection{Bosonic Sunset}

We can arrange the terms entering in the sunset diagram depending on their denominator structure. From the form of the interactions, we see that schematically we have the following possibilities 
\be
&&\int \frac{d^2p\, d^2q\, d^2r}{(2\pi)^4} \delta^{(2)}(p+q+r) \frac{{\cal 
N}(p,q,r)}{{\cal D}_B(p)(q^2+m^2)(r^2+m^2)} \,,~~~~m^2=m_{\tilde x}^2 ~{\rm or}~ m_y^2\nonumber\\
 &&\int \frac{d^2p\, d^2q\, d^2r}{(2\pi)^4}  
  \delta^{(2)}(p+q+r) \frac{{\cal N}(p,q,r)}{{\cal D}_B(p){\cal D}_B(q){\cal D}_B(r)} \,,
\label{sunset-structures}
\ee
%\be
%(p^2+m)(q^2+m)C(r)\,,\qquad  C(p)C(q)C(r)
%\ee 
where the numerator ${\cal N}(p,q,r)$ contains tensors of up to fourth order in momenta, and ${\cal D}_B(p)$ 
is the denominator appearing in the $\tilde\phi$ and $\tilde\varphi$ propagators, see eq.
 (\ref{Bob}).
%\be
%{\cal D}_B(p)= p^4+p^2+(\hat\nu^2-\hat w^2) p_0^2-2 p_0 p_1 \hat w \hat\nu\,.
%\ee
%which comes  from the denominator of the propagators $G_{\tilde\phi \tilde\phi}(p)$ and  $G_{\tilde\phi %\tilde\varphi}(p)$. 
In (\ref{sunset-structures}) we have  introduced the shorthand notation
\be  m_{\tilde x}^2\equiv
 {\te {\te \frac{1}{4}}}(1+\hat \kappa^2)   \ , \ \ \ \ \ \ \ \ \ \ m_{y}^2\equiv {\te {\te \frac{1}{4}}}(\hat \nu^2+\hat w^2) \la{maaa} \ee
 for the masses of the $\tilde x$ and $y^a$ fluctuations in \rf{qq}, \rf{Bo}. 
  
Unfortunately, the presence of the Lorentz non-invariant denominator 
${\cal D}_B(p)$ \ \rf{Bob} makes an exact evaluation of the loop integrals
 technically challenging. We will therefore limit ourselves
 to 
 %obtaining  an approximated answer, by making an 
 expanding  up
  to fourth order in the  parameters $\hat \nu, \hat w$.
%  In this case  
 %Expanding  the Lorentz non-invariant denominator in \rf{Bob}  as
 Since 
\be\label{Cexpansion}
\frac{1}{{\cal D}_B(p)}=\frac{1}{p^2(p^2+1)}+\frac{2 \hat \nu \hat w p_0 p_1  
+(\hat w^2-\hat \nu^2)p_0^2}{p^4(p^2+1)^2}
+\frac{[2  \hat \nu \hat w p_0 p_1 +(\hat w^2-\hat \nu^2)p_0^2]^2}{p^6(p^2+1)^3}+{\cal O}(\hat \nu^6,\hat w^6)\,,\ee
%Note that in this
we find  that 
 %expansion
  the denominators appearing in the expansion 
   are Lorentz invariant, and therefore the evaluation of the loop integrals becomes straightforward.

For the computation of the sunset topology we need the third-order bosonic Lagrangian   
\be
&&{\cal L}_B^{(3)}=-4\,\tilde\phi |\partial_s\tilde x-\ha\tilde x|^2
+2\tilde\phi (\partial_t \tilde\phi^2-\partial_s \tilde\phi^2)+  
2\tilde\phi (\partial_t \tilde\varphi^2-\partial_s 
\tilde\varphi^2)+ 2 {\rm i}\tilde\phi^2 (\hat \nu\partial_t \tilde\varphi
+\hat w\partial_s \tilde\varphi)\nonumber\\
&&\ \ \ \ \ \  +\ \frac{1}{2} (\hat \nu^2-\hat w^2)\, \tilde\phi\, y^a y^a
+2\tilde \phi\left((\partial_t y^a)^2
-(\partial_s y^a)^2\right)-{\rm i} (\hat \nu\partial_t\tilde\varphi+\hat w\partial_s\tilde\varphi)\, y^a y^a \ . \la{cub} 
\ee
A particularly simple contribution, with a single ${\cal D}_B$ in the denominator, 
comes from the $\tilde \phi \tilde x\tilde x^*$ 
interaction. 
%$${\cal L}^{(3)}_{\tilde \phi \tilde x\tilde x^*}=-4\,\tilde\phi |\partial_s\tilde x-\ha\tilde x|^2\,.$$
Evaluation of the corresponding Feynman diagram as obtained from (\ref{1PI-diag})
 yields the integral\footnote{Here and in the following we will omit the overall 
 factor $\frac{2\pi}{\sqrt{\lambda}}V_2$. It will be restored at the end.}
\begin{equation}\label{BosSun1}
 -\frac{1}{2} \int \frac{d^2p\, d^2q\, d^2r}{(2\pi)^4}\,\delta^{(2)}(p+q+r) \frac{p^2 (1+4\,q_1^2)(1+4\,r_1^2)}{{\cal D}_B(p)(q^2+m_{\tilde x}^2)(r^2+m_{\tilde x}^2)}\,.
\end{equation}
After inserting the expansion in (\ref{Cexpansion}) we can perform 
the integral over the momenta by standard techniques, see  Appendix~\ref{sec:int_computation} 
for more details.  
%Before computing the actual integral we need to specify
%our procedure for dealing with integrals over loops momenta.  
As in \cite{rt1,rtt}, all manipulations of tensor structures in the numerators are performed in $d=2$, and the
 resulting scalar integrals are evaluated using an analytic regularization scheme in which power divergent
contribution are set to zero
\be
\int \frac{d^2p}{(2\pi)^2} (p^2)^n=0\,,\  \qquad n \geq 0\,.
\ee
We also use the following notation for the integrals
\be
{\rm I}\mbox{$\left({a\atop m^2}\right)$}&=& \int \frac{d^2p}{(2\pi)^2}\frac{1}{(p^2+m^2)^a}\,,\nonumber\\
{\rm I}{\mbox{$\left({a_1\atop m_1^2}{a_2\atop m_2^2} 
{a_3 \atop m_3^2}\right)$}}&=&\int \frac{d^2p\, d^2q\, d^2r}{(2\pi)^4}\,\frac{ 
\delta^{(2)}(p+q+r)}{(p^2+m_1^2)^{a_1}\, (q^2+m_2^2)^{a_2} \,(r^2+m_3^2)^{a_3}}\,.
\label{int-basis}
\ee
Note that the integrals ${\rm I}{\mbox{$\left({a_1\atop m_1^2}
{a_2\atop m_2^2} {a_3 \atop m_3^2}\right)$}}$ and 
 ${\rm I}\mbox{$\left({a\atop m^2}\right)$}$ with $a>1$ are UV finite,
  while ${\rm I}\mbox{$\left({1\atop m^2}\right)$}$ is UV divergent.
   When any of the masses vanish, both types of integrals have IR divergences. In the 
   following we will also use for convenience the notation 
   ${\rm I}\mbox{$\left({1\atop m^2}\right)$} \equiv {\rm I}[m^2]$.
   We refer to Appendices~\ref{sec:1loop_int}, 
\ref{sec:int_computation} and \ref{sec:2loop_int} 
 for more details on  evaluation of these integrals and their explicit values.  

After reduction to scalar integrals and expansion up to fourth order in $\hat \nu,\hat w$, 
the integral (\ref{BosSun1}) yields the following result 
\begin{eqnarray}
%\begin{aligned}
&&\!\!\!\!\!\!\!\!\!\! \hphantom{-}
{\te {\te \frac{1}{4}}}{\rm I}{\mbox{$\left({1\atop 1}{1\atop \ha} {1\atop \ha}\right)$}}
+{\te {\te \frac{1}{4}}}(\hat \nu^2-\hat w^2)\left[{\rm I}{\mbox{$\left({1\atop 1}{1\atop \ha } 
{1 \atop \ha}\right)$}}-\frac{1}{2} {\rm I}{\mbox{$\left({1\atop
1}{2\atop \ha } {1 \atop \ha}\right)$}} 
- \frac{1}{2} {\rm I}{\mbox{$\left({2\atop 1}{1\atop \ha } {1 \atop
\ha}\right)$}}+ 4 {\rm I}[1] {\rm I}[\ha] \right]\\
&&\!\!\!\!\!\!\!\!\!\!
-\frac{1}{8}\Bigg{[}2\hat \nu^2 \hat w^2 {\rm I}{\mbox{$\left({1\atop 1}{1\atop \ha} {1\atop \ha}\right)$}} 
+(\hat \nu^2-\hat w^2)^2\left({\rm I}{\mbox{$\left({2\atop 1}{1\atop \ha} {1\atop \ha}\right)$}}+{\rm I}{\mbox{$\left({1\atop 1}{2\atop \ha} {1\atop \ha}\right)$}}-{\te {\te \frac{1}{4}}} 
{\rm I}{\mbox{$\left({1\atop 1}{3\atop \ha} {1\atop \ha}\right)$}}
-\frac{1}{2} 
{\rm I}{\mbox{$\left({2\atop 1}{2\atop \ha} {1\atop \ha}\right)$}}
-\frac{1}{8} 
{\rm I}{\mbox{$\left({1\atop 1}{2\atop \ha} {2\atop \ha}\right)$}}
\right)\cr
&&\;
-\frac{1}{8}(5 \hat \nu^4 + 2\hat \nu^2 \hat w^2 + 5 \hat w^4){\rm
I}{\mbox{$\left({3\atop 1}{1\atop \ha} {1\atop \ha}\right)$}} +2 (\hat
\nu^2-\hat w^2)^2 {\rm I}[1] {\rm
I}\mbox{$\left({2\atop \ha}\right)$} +(3 \hat \nu^4-10 \hat \nu^2 \hat
w^2+3 \hat w^4){\rm I}\mbox{$\left({2\atop 1}\right)$} {\rm I}[\ha]\Bigg{]}
\nonumber
%\end{aligned}
\end{eqnarray}
The second contribution to the bosonic sunset with a single ${\cal D}_B(p)$ factor at 
the denominator comes from the cubic pieces in 
the Lagrangian which contain $y^a y^a$ while the cubic
 vertices involving the $\tilde \phi$ and $\tilde\varphi$ fluctuations have a denominator of the type
 ${\cal D}_B(p){\cal D}_B(q){\cal D}_B(r)$. For brevity we do not present 
 their expressions and give 
 only the final result of the bosonic sunset. After expanding to fourth order
  in $\hat \nu,\hat w$ and 
  %plugging in 
  using the explicit values of the integrals 
from the Appendix ~\ref{sec:2loop_int}, we obtain
\begin{equation}\la{su}
\begin{aligned}
&W_{2{\scriptstyle \rm B\ sunset}}=\frac{K}{16\pi^2}+\frac{1}{2} {\rm I}[1]^2
+\frac{1}{32\pi^2}\Big{[}(\hat \nu^2-\hat w^2)K+4 \pi {\rm I}[1] 
\Big{(}\hat \nu^2 (-1+2 \ln 2+24 \pi {\rm I}[1])\\
&+\hat w^2 (1-2 \ln 2+
4 \pi {\rm I}[1])\Big{)}
+(\hat \nu^2+\hat w^2) \ln m_y^2 (\ln m_y^2-16 \pi {\rm I}[1])\Big{]}\\
& -\frac{1}{1024\pi^2}\Big{[}\; \hat\nu^4 (14 K-9+24 \ln 2+464 \pi {\rm I}[1])+2 \hat\nu^2 \hat w^2 (6 K+19-40 \ln 2-304 \pi{\rm I}[1])\\
&~~~~~~~~~~~~
+\hat w^4 (14K-9+24 \ln 2+80 \pi {\rm I}[1])-32 (\hat\nu^4-\hat w^4) \ln m_y^2\Big{]}\,,
\end{aligned}
\end{equation}
where $K$ is the Catalan's constant and  $m_y^2={\te {\te \frac{1}{4}}}(\hat \nu^2+\hat w^2)$ was  defined  in \rf{maaa}. 
%={\te {\te \frac{1}{4}}}(\hat \nu^2+\hat w^2)$.
%Note that 
To obtain the above expression we 
have rewritten all UV divergent 1-loop integrals in
 terms of ${\rm I}[1]$, by using 
 the identity ${\rm I}[m^2]= 
 {\rm I}[1]-\frac{1}{4\pi} \ln m^2$, for $m^2 \neq 0$.

\subsection{Bosonic double-bubble}

Let us now give some details of the evaluation of the bosonic double-bubble diagram, which
 receives contributions from the bosonic quartic interactions 
\be
{\cal L}^{(4)}_B\!\!&=&\!\!
\frac{1}{3}(1-\hat w^2)\tilde \phi^4+2\tilde \phi^2 \partial_\a\tilde\phi\partial_\a\tilde\phi
+ \frac{4}{3} {\rm i} \,\tilde \phi^3(\hat\nu\partial_t \tilde\phi-\hat w\partial_s \tilde\phi)
-2 {\rm i}\,\tilde \phi (\hat\nu\partial_t \tilde\varphi-\hat w\partial_s \tilde\varphi)y^a y^a
\nonumber\\
&+&\!\!
2\tilde\phi^2 \partial_\a y^a\partial_\a y^a
+\ha \tilde\phi^2(\hat \nu^2+\hat w^2)y^2+8\tilde\phi^2|\partial_s \tilde{x}
-\ha \tilde{x}|^2+2\tilde\phi^2\partial_\a\tilde\varphi \partial_\a\tilde\varphi
-\ha y^2 \partial_\a y^a\partial_\a y^a\nonumber\\
&-&\!\!
\frac{1}{8}(\hat \nu^2+\hat w^2)y^4- y^a y^a \partial_\a\tilde\varphi \partial_\a\tilde\varphi\,.
\ee 
From the form of the interactions, we see that the possible structures for the loop integrals are now
\be
&&\int \frac{d^2p}{(2\pi)^2}\,\frac{d^2q}{(2\pi)^2} \frac{{\cal N}_1(p_0,p_1)}{p^2+m_y^2}  \frac{{\cal N}_2(q_0,q_1)}{q^2+m_y^2}\,,\label{b-bub-struc1}\\ 
&&\int \frac{d^2p}{(2\pi)^2}\,\frac{d^2q}{(2\pi)^2}  \frac{{\cal N}_1(p_0,p_1)}{p^2+m^2} 
\frac{{\cal N}_2(q_0,q_1)}{{\cal D}_B(q)}\,,~~~~m^2=m_{\tilde x}^2 ~{\rm or}~ m_y^2 \label{b-bub-struc2} \\
&&\int \frac{d^2p}{(2\pi)^2}\,\frac{d^2q}{(2\pi)^2}  \frac{{\cal N}_1(p_0,p_1)}{{\cal D}_B(p)}  \frac{{\cal N}_2(q_0,q_1)}{{\cal D}_B(q)}\,,
\label{b-bub-struc3}\ee
where  $p$ and $q$ are the momenta running in the two bubbles.
The two loop integrals are independent and this makes it possible to evaluate
 them exactly despite the presence of the ${\cal D}_B(p)$ denominator. 
We can therefore compute the bosonic double-bubble graphs
for any values of the $\hat \nu,\hat w$ parameters. 
%Of course, 
In view of combining the double-bubble contribution with the sunset diagram, 
we will  however keep only terms up to  fourth order
 in $\hat \nu$ and $\hat w$ in the final expression.

To 
%present our results in a compact way 
systematically organize the calculation it is convenient to introduce the following notation 
\be
\label{Fbos}
J_B(i,j)=\int \frac{d^2p}{(2\pi)^2}\frac{(p_0)^{i} (p_1)^{j}}{{\cal D}_B(p)}\,.
\ee 
These integrals can be evaluated exactly,  for example,  by first 
computing the angular integral in the $(p_0,p_1)$ plane and 
then performing an integral over the modulus $|p|$. 
In the process, one has to carefully separate a possible UV divergent part. The relevant
 explicit expressions for $J_B(i,j)$ are presented in  Appendix~\ref{sec:1loop_int}. 
 
%Now we come to discuss the various contributions. The simplest structure given in eq. (\ref{b-bub-struc1}) 
%comes from the quartic vertex in $y^a$ in ${\cal L}^{(4)}_b$
%\be
%{\rm Bos.Bub.}\,1=-\int \frac{d^2p\, d^2 q}{(2\pi)^4}\, \frac{3(\hat\nu^2+\hat w^2)+3p^2+5q^2-4 p\cdot q}{2(p^2+m_y^2)(q^2+m_y^2)}=-2 m_y^2\, {\rm I}\mbox{$\left({1\atop m_y^2}\right)$}^2\,.
%\ee

To give  an example, let us 
 present the contribution of the $x^2\tilde\phi^2$-interaction in the Lagrangian which is of the type (\ref{b-bub-struc2}), namely, 
% and reads
 \be
\int \frac{d^2p\, d^2 q}{(2\pi)^4} \frac{2 p^2 (1+4 q_1^2)}{{\cal D}_B(p)(q^2+m_{\tilde x}^2)}=
(\hat w^2-\hat \nu^2) \big[J_B(0,2)+J_B(2,0)\big]\,{\rm I}[m_{\tilde x}^2]\,.
\ee
The other contributions can be computed analogously.
%while the second one originates from the interactions $y^2\tilde\phi^2$,  $y^2\tilde\phi\tilde\varphi$ and $y^2\tilde\varphi^2$,  and yields
%\be
%&&{\rm Bos.Bub.}\,3=-\int \frac{d^2p\, d^2 q}{(2\pi)^4}  \frac{2 p^2 (1-\hat w^2+p^2)+4\,(\hat\nu p_0-\hat w p_1)^2
%-4 \,p^2(q^2+m_y^2)}{{\cal D}_B(p)(q^2+m_y^2)}\nonumber\\
%&&=-2\left(\hat \nu^2 J_B(2,0)+\hat w^2 J_B(0,2)-2\hat\nu\hat w J_B(1,1)\right)\,{\rm I}\mbox{$\left({1\atop m_y^2}\right)$}\,.
%\ee
%The last type of structure in (\ref{b-bub-struc3}) containing two ${\cal D}_B(p)$ factors comes from the 
%  $\tilde\phi^4$ and $\tilde \phi^2\tilde\varphi^2$ interactions and gives:
%\be
%&&{\rm Bos.Bub.}\,4=\int \frac{d^2p\, d^2 q}{(2\pi)^4} \left\{\frac{3 p^2 q^2(1 -\hat w^2+q^2)
%+2 q^2(\hat\nu p_0-\hat w p_1)^2+4p^2 (\hat\nu q_0-\hat w q_1)^2}{3{\cal D}_B(p){\cal D}_B(q)}\right. \nonumber \\
%&&~~~~~~~~~~~~~~~~~~~~~~~~~~~~~~\left.+\frac{p^2 q^2(1-\hat w^2-4 p \cdot q +2 q^2)-
%4  p \cdot q (\hat\nu p_0-\hat w p_1)(\hat\nu q_0-\hat w q_1)}{2{\cal D}_B(p){\cal D}_B(p)}\right\}\nonumber\\
%&&=-2 (J_B(2,0)+ J_B(0,2))\left(J_B(0,2)-2\hat\nu\hat w J_B(1,1)+(1+\hat \nu^2-\hat w^2)J_B(2,0)\right) -2(\hat\nu ^2+\hat w^2)J_B(1,1)^2 \nonumber\\
%&&-(\hat w^2-2\hat \nu^2-3)J_B(2,0)J_B(0,2)-\frac{3}{2}(\hat w^2-1)\left(J_B(2,0)^2+J_B(0,2)^2\right)\,.
%\ee
%The total bosonic double-bubble contribution to the effective action is then 
%\be
%W_{2~{\scriptsize \rm Bos.Bub.}}={\rm Bos.Bub.}\,1+{\rm Bos.Bub.}\,2+{\rm Bos.Bub.}\,3 +{\rm Bos.Bub.}\,4\,.
%\ee
After summing all the terms, using the explicit values of the integrals $J_B(i,j)$ 
and expanding up 
to fourth order in $\hat \nu, \hat w$, 
we obtain the following result for the bosonic double-bubble contribution 
\begin{equation}\la{dd} 
\begin{aligned}
&W_{2{\scriptstyle \rm B\ double-bubble}}=-\ha {\rm I}[1]^2-
\frac{1}{8\pi}{\rm I}[1] 
\left[\hat \nu^2 (2\ln 2-1+24 \pi \,{\rm I}[1])-\hat w^2 (2\ln 2-1-4 \pi\, {\rm I}[1])\right]\\
&-\frac{1}{32\pi^2}(\hat \nu^2+\hat w^2) \ln m_y^2 \left(\ln m_y^2-16 \pi\, {\rm I}[1]\right)
+\frac{1}{128\pi^2}\Big{[}\hat \nu^4 (4 \ln 2-1+74 \pi\, {\rm I}[1])\\
&+\hat w^4 (4 \ln 2-1+26 \pi\, {\rm I}[1])
-2 \hat \nu^2 \hat w^2 (4 \ln 2-1+22 \pi {\rm I}[1])-(6 \hat \nu^4+4 \hat \nu^2 \hat w^2-2 \hat w^4) \ln m_y^2)\Big{]}\,.
\end{aligned}
\end{equation}
%where as before we used the shorthand $m_y^2={\te {\te \frac{1}{4}}}(\hat\nu^2+\hat w^2)$. 
As above, we have used the 
identity ${\rm I}[m^2]= {\rm I}[1]-\frac{1}{4\pi} \ln m^2$ to keep for convenience only one representative of the UV divergent 1-loop integrals.

%We observe that in the sum of the 
Adding together  the 
sunset \rf{su} and double-bubble \rf{dd}   contributions we observe that
 all the UV  divergences cancel out  up to second order
in $\hat\nu$ and $\hat w$, i.e. 
 the bosonic 1PI 2-loop contribution  is finite  to this order.
 The remaining  UV divergences at fourth order in $\hat\nu$ and $\hat w$   will cancel
after we include the fermionic and tadpole contributions which we discuss next.

\subsection{Fermionic sunset}
The fermionic contribution of sunset topology 
arises from the bose-fermi-fermi interactions in the cubic Lagrangian
\begin{equation}
\begin{aligned}
{\cal L}^{(3)}_{BF^2}=&-2\ri \partial_t \tilde\varphi\, \tilde\eta_i (\rho^{56})^i_{\ j} \tilde \eta^j  
+2 \ri \partial_t y^a \tilde\eta_i (\rho^{a5})^i_{\ j} \tilde \eta^j+\hat \nu\, y^a \tilde\eta_i (\rho^{a6})^i_{\ j} 
\tilde \eta^j\\
&- \frac{\ri}{2} \delta \partial_t \tilde\varphi  
\left(\tilde\theta^k  (\rho^{56})_k^{\ l} \tilde\theta_l + \tilde\eta^k  (\rho^{56})_k^{\ l} \tilde\eta_l
+\tilde\theta_k  (\rho^{56})^k_{\ l} \tilde\theta^l + \tilde\eta_k  (\rho^{56})^k_{\ l} \tilde\eta^l \right)\\
& +2\ri \left(2\tilde\phi\, \rho^5_{kl} -(1-\delta) \tilde\varphi\, \rho^6_{kl} 
- y^a \rho^a_{kl}\right) \left(\tilde \eta^k(\partial_s \tilde\theta^l-\ha \tilde\theta^l) 
-\frac{\ri}{4} \hat w\, \tilde \eta^k (\rho^{56})^l_{\ u} \tilde\theta^u\right)\\
& +2\ri \left(2\tilde\phi\, (\rho^{\dagger}_5)^{kl} -(1-\delta) \tilde\varphi\, (\rho^{\dagger}_6)^{kl} 
- y^a(\rho^{\dagger}_a)^{kl}\right) \left(\tilde \eta_k(\partial_s\tilde\theta_l-\ha\tilde\theta_l) 
-\frac{\ri}{4} \hat w\, \tilde \eta_k (\rho^{56})_l^{\ u} \tilde\theta_u\right)\\
& -\ri \delta \,\partial_s \tilde\varphi \left(\rho^6_{kl} \tilde\eta^k \tilde \theta^l
+ (\rho^{\dagger}_6)^{kl} \tilde\eta_k \tilde \theta_l \right)
+2 (\partial_s \tilde{x}-\ha \tilde{x})\,\rho^5_{kl} \tilde\eta^k \tilde \eta^l
-2 (\partial_s \tilde{x}^*-\ha \tilde{x}^*)\,(\rho^{\dagger}_5)^{kl} \tilde\eta_k \tilde \eta_l\,.
\label{L3bff}
\end{aligned}
\end{equation}
Note that the vertices explicitly depend on the parameter $\delta$ introduced 
in the fermion rotation, see eq. (\ref{rot-par}). The computation of the
 corresponding Feynman diagrams in momentum space is straightforward although 
 somewhat lengthier than in the purely bosonic case. It is not difficult 
 to see that the possible structures of the loop integrals are
\begin{equation}
\begin{aligned}
&\int \frac{d^2p\, d^2q\, d^2r}{(2\pi)^4} \delta^{(2)}(p+q+r) \frac{{\cal N}(p,q,r)}{{\cal D}_F(p)
{\cal D}_F(q)(r^2+m^2)}+ {\rm c.c}\,,~~~~m^2=m_{\tilde x}^2 ~{\rm or}~  m_y^2 \\
&\int \frac{d^2p\, d^2q\, d^2r}{(2\pi)^4} \delta^{(2)}(p+q+r) \frac{{\cal
 N}(p,q,r)}{{\cal D}_F(p){\cal D}^*_F(q){\cal D}_B(r)}+ {\rm c.c}\,,
\label{fsun-struc}
\end{aligned}
\end{equation}
where ${\cal N}(p,q,r)$ is a sum of tensors of rank up to four, and
 ${\cal D}_F(p)$ is the characteristic denominator which appears 
 (together with its complex conjugate) in the fermion propagator (see \rf{fep} and  Appendix~\ref{FermProp} ) 
\begin{equation}\la{def}
{\cal D}_F(p) = p^2 -\frac{\ri}{2} (\hat \nu p_0-\hat w p_1)+\frac{1}{16}(3\hat \nu^2-\hat w^2+4)\,.
\end{equation} 
For example, a particularly simple contribution comes from the $\tilde x \tilde \eta \tilde \eta$-interaction in the last line of (\ref{L3bff}), which upon application of the Feynman rules yields
\begin{equation}
-\frac{1}{8}\int \frac{d^2p\, d^2q\, d^2r}{(2\pi)^4} \delta^{(2)}(p+q+r)
\frac{(4\ri p_0-1)(4\ri q_0-1)(1+4 r_1^2)}{{\cal D}_F(p){\cal D}_F(q)(r^2+m_{\tilde x}^2)}+{\rm c.c}\,.
\end{equation} 
Other contributions can be obtained analogously and we will omit the details here. 

In order to have Lorentz invariant denominators we  expand 
 the fermion denominators up to fourth order in $\hat \nu,\hat w$, 
 as we have done above in  the bosonic case. Then 
 the reduction to the basis of scalar integrals
  given in (\ref{int-basis}) proceeds in a straightforward fashion. The final expression in terms 
  of those integrals is,  however, 
   very lengthy, and so  we shall 
   limit ourselves to giving  the final answer
    obtained after using the explicit values
     of the integrals from  Appendix~\ref{sec:2loop_int}. Also, for the sake of brevity,
      we will only present the result for the simplest choice $\delta=1$
      of the fiducial  parameter  $\delta$. 
      We have explicitly
       checked, however, that all the $\delta$-dependence
        drops out as expected once we include the contribution
	 of the fermionic double-bubble topologies which will
	  be discussed in the next subsection. 
	  The total fermionic sunset contribution at $\delta=1$ is then 
\be\la{fsu}
%\begin{equation}\la{fsu}
%\begin{aligned}
W_{2{\scriptstyle \rm F\ sunset}}&=& -\frac{K}{8\pi^2} +\frac{1}{2\pi^2}
\left(2\pi\,{\rm I}[1]+\ln 2\right)^2+\frac{1}{4\pi^2}\left(2\pi\,{\rm I}[1]+\ln 2\right)\ln m_y^2\\
&&
-\frac{1}{64 \pi^2}\Big{[}\;\hat \nu^2 \Big{(}4K-13+96 \pi {\rm I}[1]
-8 \ln 2 (\ln 2-5+2 \pi {\rm I}[1])+19 \ln 4 m_y^2\Big{)}\cr
&&~~~~~~~
%\!\!\!\!\!\!\!\!\!\!\!
- \hat w^2 \Big{(}4K+9-4 \ln 2 (1+2 \ln 2)
-16 \pi {\rm I}[1] (5 \ln 2-1+8 \pi {\rm
I}[1]) 
\cr
&&~~~~~~~~~~~~~~~~~~~~~~~~~~~~~~~~~~~~~~~~~~~~~~~~~~~~
-(5-16 \ln 2-32 \pi {\rm I}[1])\ln 4 m_y^2\Big{)}\Big{]}\cr
&&
+\frac{1}{4608 \pi^2}\Big{[}\; \hat \nu^4 (126 K+1499+576 \ln 2+1872 \pi
{\rm I}[1]
+ 624 \ln 4 m_y^2 ) \cr
&&  ~~~~~~~~~~+
\hat w^4 (126 K-445-1008 \pi {\rm I}[1]+336 \ln 4 m_y^2 )\cr
&&~~~~~~~~~~+ 6 \hat \nu^2 \hat w^2 (18 K-47-96 \ln 2
-48 \pi {\rm I}[1]-288 \ln 4
m_y^2)\Big{]}\,.
\nonumber
%\end{aligned}
%\end{equation}
\ee
As before, here we have rewritten all UV divergent integrals ${\rm I}[m^2]$ 
in terms of ${\rm I}[1]$. 

\subsection{Fermionic double-bubble}

Finally, to conclude our analysis of 1PI diagrams, we have to include the fermionic
 contributions of  double-bubble topology.
 These come from the following bosonic-fermionic and fermionic quartic vertices
\footnote{Here we do not include vertices such as $\tilde\varphi y \tilde\eta \tilde\eta$ or 
$\tilde\varphi y \tilde\eta \tilde\theta$ 
which do not contribute to the double-bubble diagram as the corresponding propagator vanishes, 
$G_{\tilde\varphi y}$=0.} 
\be
%\begin{equation}
%\begin{aligned}
%\label{ferm-4-vert}
{\cal L}^{(4)}_{B^2F^2}&=&\!\!
\ri\left( (4\tilde\phi^2-(1-\delta)^2 \tilde\varphi^2-y^a y^a)\, \rho^5_{kl}-4
 (1-\delta) \tilde\phi \tilde\varphi\, \rho^6_{kl}\right) \left( \tilde\eta^k\,(\partial_s \tilde\theta^l-
 \ha \tilde\theta^l)-\frac{\ri\hat w}{4}\tilde\eta^k(\rho^{56})^l_{\, j}\,\tilde\theta^j\right)\cr
&+&\!\!
\ri\left( (4\tilde\phi^2-(1-\delta)^2 \tilde\varphi^2-y^a y^a)\, (\rho_5^{\dagger})^{kl}-4 (1-\delta) 
\tilde\phi \tilde\varphi\,(\rho_6^{\dagger})^{kl}\right) \left( \tilde\eta_k\,(\partial_s \tilde\theta_l-\ha \tilde\theta_l)
-\frac{\ri\hat w}{4}\tilde\eta_k(\rho^{56})_l^{\, j}\,\tilde\theta_j\right)\cr
&-&\!\!
\ri\delta\,\partial_t \tilde\varphi\,\left(2\tilde\phi\, \rho^5_{kl} +
(1-\delta) \tilde\varphi\, \rho^6_{kl}\right) \tilde\eta^k(\rho^{56})^l_{\, j}\,\tilde\theta^j
-\ri\delta\,\partial_t \tilde\varphi\,\left(2\tilde\phi\, (\rho_5^{\dagger})^{kl} +
(1-\delta) \tilde\varphi\, (\rho_6^{\dagger})^{kl}\right) \tilde\eta_k(\rho^{56})_l^{\, j}\,\tilde\theta_j
\cr
&-&\!\!
\hat \nu\, y^a y^a \tilde\eta_i (\rho^{56})^i_{\, j}\tilde\eta^j
+2\ri\, y^a \partial_t y^b \tilde\eta_i (\rho^{ab})^i_{\, j}\tilde\eta^j \ , \cr
{\cal L}^{(4)}_{F^4}&=&\!\!
-\tilde\eta_i (\rho^{56})^i_{\, j}\tilde\eta^j
\,\tilde\eta_k (\rho^{56})^k_{\,\, l}\tilde\eta^l- \sum_{a=1}^4\tilde\eta_i (\rho^{a 6})^i_{\, j}\tilde\eta^j
\,\tilde\eta_k (\rho^{a 6})^k_{\,\, l}\tilde\eta^l-\tilde\eta^i \tilde \eta_i \tilde\eta^j \tilde\eta_j
\label{ferm-4-vert}
%\end{aligned}
%\end{equation}
\ee
%\be\label{ferm 4-vertices}
%&&S^{(4)}_{y y\tilde\eta\tilde\theta}= \int dt ds\,\left(\frac{i}{2} y^a y^a %\tilde\eta^i\,\rho^5_{ij}(\partial_s\tilde\theta^j
%-\ha \tilde\theta^j)+\frac{\hat w}{8}y^a y^a \tilde\eta^i \rho^6_{ij}\tilde\theta^j \right)\,,\nonumber\\
%&&S^{(4)}_{y y\tilde\eta\tilde\eta}=\int dt ds\, \frac{\hat \nu}{2}y^a y^a \tilde\eta_i (\rho^{56})^i_{\, %j}\tilde\eta^j\,,\nonumber\\
%&&S^{(4)}_{\tilde \phi\tilde\phi\tilde\eta\tilde\theta}=
% -\int dt ds\,2i\tilde\phi^2\tilde\eta^i\,\rho^5_{ij}(\partial_s \tilde\theta^j-\ha \tilde\theta^j)
%+\frac{w}{2}\tilde\phi^2 \tilde\eta^i\,\rho^6_{ij}\tilde\theta^j\,,\\
%&& S^{(4)}_{\tilde{\phi}\tilde{\varphi}\tilde\eta\tilde\theta}=
%-\int dt ds\,  i\tilde\phi\partial_s\tilde\varphi \tilde\eta^i\rho^6_{ij}\tilde\theta^j\,,\nonumber\\
%&& S^{(4)}_{\tilde{\eta}\tilde{\eta}\tilde\eta\tilde\eta}=-\frac{1}{2} \int dt ds\,\tilde\eta_i %(\rho^{56})^i_{\, j}\tilde\eta^j
%\,\tilde\eta_k (\rho^{56})^k_{\,\, l}\tilde\eta^l+ \sum_{a=1}^4\tilde\eta_i (\rho^{a 6})^i_{\, j}\tilde\eta^j
%\,\tilde\eta_k (\rho^{a 6})^k_{\,\, l}\tilde\eta^l+(\eta^2)^2\,.\nonumber
%\ee
Note that, as for the cubic vertices involving fermions, these interactions depend 
non-trivially on the parameter $\delta$. While we have checked that $\delta$-dependence  drops out 
as a result of cancellations between the sunset and double-bubble topology, here we will only 
present  the result corresponding to the simplest choice of  $\delta=1$.

Since the quantity ${\cal D}_F(p)$ and its complex conjugate appear at the denominator of fermionic propagators,
 it is convenient to introduce the following integral
\be\label{KFerm}
J_F(i,j)=\int \frac{d^2 p}{(2\pi)^2} \frac{(p_0)^i (p_1)^j}{p^2-\frac{\ri}{2}(\hat\nu p_0-\hat w p_1)+\frac{1}{16}(3 \hat\nu^2-\hat w^2+4)
}\,,
\ee
as well as its complex conjugate $J_F^*(i,j)$.
As for the integral (\ref{Fbos}), it is possible to compute $J_F$ exactly. The expressions needed
for the present calculation are presented in   Appendix \ref{sec:1loop_int}. 
Similarly to the bosonic case, the fermionic double-bubble
contribution 
 may therefore be computed without the need to  expanding  in $\hat\nu$ and $\hat w$. 

As a  sample of the computation let us 
 consider the 4-point $y y\tilde\eta\tilde\theta$-interaction in the first two lines of (\ref{ferm-4-vert}), which yields
 (see Appendix~\ref{FermProp} for notation)
\be
&&\int \frac{d^2p\, d^2q}{(2\pi)^4}\,  \frac{4}{p^2+m_y^2} \nonumber\\
&& \times\Big[ \frac{\ri}{2}\left(\ri\, q_1+\ha\right)\left({\rm Tr}(G_{\eta\theta}(q)\rho_5^T)
+{\rm Tr}(G_{\eta^\dagger\theta^\dagger}(q)(\rho_5^{\dagger})^T)\right)-\frac{\hat w}{8}\left({\rm Tr}(G_{\eta\theta}(q)\rho_6^T)
+{\rm Tr}(G_{\eta^\dagger\theta^\dagger}(q)(\rho_6^{\dagger})^T)\right)\Big]\nonumber\\
&&=-\int \frac{d^2p\, d^2q}{(2\pi)^4}\,\frac{1}{2(p^2+m_y^2)}\nonumber\\
&&\times\Big[\frac{\hat w^2-4-16 q_1^2-8 \ri\hat w q_1}{q^2-\frac{\ri}{2}(\hat\nu q_0-\hat w q_1)+
\frac{1}{16}(3 \hat\nu^2-\hat w^2+4)
}
+\frac{\hat w^2-4-16 q_1^2+8 \ri \hat w q_1}{q^2+\frac{\ri}{2}(\hat\nu q_0-\hat w q_1)+\frac{1}{16}(3 
\hat\nu^2-\hat w^2+4)}\Big] \,.
\ee
Using (\ref{KFerm}), we can express the above momentum integral   as 
\be
%&&-\frac{1}{2}{\rm I}\mbox{$\left({1\atop m_y^2}\right)$}\left[(\hat w^2-4)\left( J_F(0,0)+J_F^*(0,0)\right)-16\left(J_F(0,2)+J_F^*(0,2)\right)
%-8 i \hat w\left(J_F(0,1)-J_F^*(0,1)\right)\right]\,,\nonumber\\
 -{\rm I}[m_y^2]\left[(\hat w^2-4) J_F(0,0)-16 J_F(0,2)
-8 \ri \hat w J_F(0,1)\right]
\ee
where we used the fact that $J_F(0,0)$, $J_F(0,2)$ and $J_F(2,0)$
are real while $J_F(1,0)$ and $J_F(0,1)$ are purely imaginary, see  Appendix~\ref{sec:1loop_int}.

We also observe that there is a contribution coming the 4-fermi vertex in the last line of (\ref{ferm-4-vert}). 
In the computation  of the 2-loop partition function of the standard null 
cusp surface ($\hat\nu=\hat w=0$) this contribution was absent \cite{grrtv}, while here it is non-vanishing. 
%and gives
%\be
%-\int \frac{d^2p\, d^2q}{(2\pi)^4}\,\frac{(4 p_0+i\hat \nu)(4 q_0-i\hat \nu)}{\left(p^2-\frac{i}{2}(\hat\nu p_0-\hat w p_1)+
%\frac{1}{16}(3 \hat\nu^2-\hat w^2+4)\right)\left(q^2+\frac{i}{2}(\hat\nu p_0-\hat w p_1)+\frac{1}{16}(3 \hat\nu^2-\hat w^2+4)\right)}\nonumber
%\ee
%For vanishing $\hat\nu$ and $\hat w$ this integral becomes proportional to
%\be
%\int \frac{d^2p\, d^2q}{(2\pi)^4} \frac{p_0 q_0}{(p^2+1/4)(q^2+1/4)}
%\ee
%and therefore vanishes by symmetric integration as expected. For non zero $\hat\nu$ and $\hat w$ the above integral yields instead
%\be
%-\left(\hat\nu J_F(0,0)-4 i J_F(1,0)\right)^2%\left(\hat\nu K^*(0,0)+4 i K^*(1,0)\right)
%\ee

%Summing up all contributions coming from the 4-vertices listed in (\ref{ferm-4-vert}) 
%we obtain our final expression for the fermionic double-bubble
%\be
%&&W_{2~{\rm Fer.Bub.}}=\\
%&&-\left((4-\hat w^2)J_B(2,0)-2\hat\nu \hat w J_B(1,1) \right)J_F(0,0)
%-\hat\nu^2 J_F(0,0)^2-8 J_B(2,0)\left(i \hat w J_F(0,1)+2 J_F(0,2)\right)\nonumber\\
%&&-J_B(0,2)\left((4+\hat w^2)J_F(0,0)+16 J_F(0,2)\right)+16 J_F(1,0)^2
%-8 i \hat\nu(J_B(1,1)J_F(0,1)-J_F(0,0)J_F(1,0))\nonumber\\
%&&-{\rm I}\mbox{$\left({1\atop m_y^2}\right)$}\left((\hat w^2-2(2+\hat\nu^2))J_F(0,0)+8 i\hat\nu J_F(1,0)-8i \hat w J_F(0,1)
%-16 J_F(0,2))\right)\nonumber\,.
%\ee
After combining everything together and expanding to fourth order in $\hat \nu,\hat w$, we obtain
\be
%\begin{equation}
%\begin{aligned}
\la{fdou}
W_{2{\rm F\ double-bubble}}\!\!&=&\!\!
-\frac{1}{4\pi^2}\left(2 \pi {\rm I}[1]+\ln 2\right)\ln m_y^2\\
&&\!\!
-\frac{1}{64\pi^2}\Big{[}\hat \nu^2 (4-48 \pi {\rm I}[1]
+8 \ln 2 (8 \ln 2-5+16 \pi {\rm I}[1]))
\cr
&&\!\!~~~~~~~~
+8 \hat w^2 (\ln 2+4 \pi {\rm I}[1])+
(\hat \nu^2 (16 \ln 2-15+32 \pi {\rm I}[1])-\hat w^2) \ln m_y^2\Big{]}
\cr
&&\!\!+\frac{1}{128\pi^2}\Big{[}\hat \nu^4 (106 \ln 2-39
+84 \pi {\rm I}[1])+8 \hat \nu^4 \ln m_y^2
+4 \hat \nu^2 \hat w^2 (5 \ln 2+3+10 \pi {\rm I}[1])\cr
&&\!\!~~~~~~~~~~
-\hat w^4 (1+6 \ln 2+12 \pi {\rm I}[1])
\Big{]}\,.
\nonumber
%\end{aligned}
%\end{equation}
\ee

%%%%%%%%%%%%%%%%%%%%%%%%%%%%%%%%%%%%%%%%%%%%%%%%%%%%%%%%%%%%%%%%%%%%%%%%%%%%%%%%%%%%%%%%%%%%%%%%

\subsection{Tadpole contributions}

%We conclude the analysis of the 2-loops
  Let us now  find the contribution of the non-1PI diagrams. 
 As it was already the case for the light-cone computation 
 for  the $\nu=0$  cusp anomaly \cite{grrtv}, these terms turn out to be non-zero and play an important role.
 % inclusion of these terms 
% will play an important role. 
 From the structure of the vertices and
  the propagators, it is not difficult to see that the only fluctuation 
  that can acquire a non-trivial one-point function is $\tilde\phi$. 
  Therefore the relevant non-1PI 2-loop diagrams are obtained by sewing 
  together two 1-loop tadpoles with a $\tilde\phi$ propagator at zero momentum, see  fig.\ref{non-1PI}.

From the  cubic part of the  bosonic Lagrangian \rf{cub}
 we find that the bosonic tadpoles give
\begin{equation}
\begin{aligned}
A^{\mbox{\tiny tadpole}}_B = \frac{1}{2}\int \frac{d^2p}{(2\pi)^2}&\Big{[}(\hat \nu^2-\hat w^2) \Big(\frac{1}
{p^2+{\te {\te \frac{1}{4}}}(1+\hat\kappa^2)} + \frac{2}{p^2+{\te {\te \frac{1}{4}}}(\hat \nu^2+\hat w^2)}\Big)\\
& + 2\frac{2 (p_0^4-p_1^4)+(\hat\kappa^2+\hat \nu^2)p_0^2 -(1+\hat w^2)p_1^2}{p^4 +
 \hat\kappa^2 p_0^2+p_1^2- 2 \hat \nu \hat w p_0 p_1}\Big{]}\,.
\end{aligned}
\end{equation}
The two terms in the first line come 
respectively from the  $\tilde{x}\tilde{x}^*$ and
  $y^a y^a$ loop (for the first term, a simplification of 
 the numerator was performed and a term proportional to $\int d^2p$ was discarded). The term in the second line comes
  from $\tilde\phi^2$, $\tilde\varphi^2$ and $\tilde\phi \tilde\varphi$ loops. 

The relevant fermionic tadpoles arise from the $\tilde\phi\tilde\eta\tilde\theta$-interactions in eq. (\ref{L3bff}). They give
\begin{equation}
\begin{aligned}
A^{\mbox{\tiny tadpole}}_F = \frac{1}{2}\int \frac{d^2p}{(2\pi)^2}\,
\frac{(1 + 4 p_1^2) (2 + \frac{3}{2} \hat \nu^2 + 8 p^2)-(1 + \frac{3}{8} \hat \nu^2 + 2 p_0^2- 4 p_1^2) \hat w^2 
- 8 \hat \nu \hat w p_0 p_1 + \frac{\hat w^4}{8}
}{\left(p^2+ \frac{4\hat\kappa^2-\hat\nu^2+3\hat w^2}{16}\right)^2 + 
 {\te {\te \frac{1}{4}}} (\hat\nu p_0 - \hat w p_1)^2}
\end{aligned}
\end{equation}
Performing the loop integration and separating a UV divergent part, we obtain 
for the bosonic  tadpole 
\begin{equation}
\label{boseTP}
\begin{aligned}
A^{\mbox{\tiny tadpole}}_B= &-\frac{1}{16\pi}\frac{\hat \nu^2 - \hat w^2}{\hat \nu^2+\hat w^2} 
\left( 4 + \hat \nu^2 - 3 \hat w^2 - 4 \sqrt{(1 + \hat \nu^2)(1-\hat w^2)}\right)\\
&~~~~+ \frac{1}{2}(\hat \nu^2 - \hat w^2) \left( \mbox{I}[{\te \frac{1}{4}} (1 + \hat \kappa^2)] + 
    2 \mbox{I}[{\te {\te \frac{1}{4}}} (\hat \nu^2 + \hat w^2)])+ \mbox{I}[{\te  \frac{1}{4}} (\sqrt{1 + \hat \nu^2} + \sqrt{1 - \hat w^2})^2]\right)\,,
\end{aligned}
\end{equation} 
and for the fermionic one 
\begin{equation}
\label{fermiTP}
\begin{aligned}
A^{\mbox{\tiny tadpole}}_F=\frac{\hat \nu^2-\hat w^2}{16\pi}+2 (1-\hat \nu^2) {\mbox I}[{\te {\te {\te \frac{1}{4}}}}(1+\hat\nu^2)]\,.
\end{aligned}
\end{equation}
Combining the two contributions, the total 1-loop tadpole for $\tilde\phi$ 
is found to be\footnote{The terms in the second line come from $ \mbox{I}[{\te {\te {\te \frac{1}{4}}}} (1 + \hat \kappa^2)] + 
    2 \mbox{I}[{\te {\te {\te \frac{1}{4}}}} (\hat \nu^2 + \hat w^2)])+ \mbox{I}[{\te {\te {\te \frac{1}{4}}}}
     (\sqrt{1 + \hat \nu^2} + \sqrt{1 - \hat w^2})^2]-4{\mbox I}[{\te {\te {\te \frac{1}{4}}}}(1+\hat\nu^2)]$, 
     which is finite.}
\begin{equation}
\begin{aligned}
&A^{\mbox{\tiny tadpole}}_B+A^{\mbox{\tiny tadpole}}_F = -\frac{1}{4\pi}(\hat \nu^2 - \hat w^2)\Big{[}\frac{1}{\hat \nu^2+\hat w^2} \left(1-\hat w^2-\sqrt{(1 + \hat \nu^2)(1-\hat w^2)} \right)\\
&~~~~~~~~~~~~~+\frac{1}{2} \left( 
\ln(1+\hat \kappa^2)-4 \ln (1+\hat\nu^2)+2\ln(\hat \nu^2+\hat w^2) 
+2 \ln \left(\sqrt{1 + \hat \nu^2} + \sqrt{1 - \hat w^2}\right)
\right)\Big{]}\\
&~~~~~~~~~~~~~+2 (1-\hat w^2) {\mbox{I}}[{\te {\te {\te \frac{1}{4}}}}(1+\hat\nu^2)]\,.
\end{aligned}
\end{equation}
For $\hat \nu=\hat w=0$, this reduces to the result we found
 for the ordinary cusp \cite{grrtv}, with
  the fermion tadpole being proportional 
  to ${\rm I}[\frac{1}{4}]$.

The total contribution of the 1-particle reducible diagrams
 to the 2-loop  partition function or $W=-\ln Z$  is then
\begin{equation}
W_{2~{\rm tadpoles}}= -\frac{1}{2} \left(A^{\mbox{\tiny tadpole}}_B+A^{\mbox{\tiny tadpole}}_F\right)^2\, G_{\tilde\phi\tilde\phi}(0)\,.
\label{W2tadpoles}
\end{equation} 
Using $G_{\tilde\phi\tilde\phi}(0)=\frac{1}{1-\hat w^2}$ and expanding up to fourth order in $\hat \nu,\hat w$, we obtain
\begin{equation}\la{taaa}
\begin{aligned}
&W_{2{\rm\  tadpoles}}= -\frac{1}{2\pi^2} \left(2\pi{\rm I}[1]
+\ln 2\right)^2\\
&~~~~
+\frac{1}{8\pi^2}(2 \pi {\rm I}[1]+\ln 2)\Big{[}
 (3+7 \ln 2) \hat \nu^2
+\hat w^2 (1-3 \ln 2+8 \pi {\rm I}[1])
+2 (\hat \nu^2-\hat w^2) \ln m_y^2\Big{]}\\
&~~~~
-\frac{1}{128 \pi^2}\Big{[}\; \hat \nu^4 (76 \ln 2+(3+7 \ln 2)^2+
152 \pi {\rm I}[1])
+\hat w^4 (1+\ln2 (-26+49 \ln 2)-24 \pi {\rm I}[1])\\
&~~~~~~~~~~~~~~
+4 (\hat \nu^2-\hat w^2) \ln m_y^2 \left((3+7 \ln 2) \hat \nu^2+(1-7 \ln 2) \hat w^2
+(\hat \nu^2-\hat w^2) \ln m_y^2\right)\\
&~~~~~~~~~~~~~~
+\hat \nu^2 \hat w^2 (6+4 \ln 2-98 \ln^2 2+64 \pi {\rm I}[1])\Big{]}\,.
\end{aligned}
\end{equation}
Let us briefly comment on the value $G_{\tilde{\phi} \tilde{\phi}}(0) =
\tfrac 1 {1 - \hat{w}^2}$ for the $\tilde{\phi}\tilde{\phi}$ propagator at zero momentum. 
As follows from~\eqref{Bo}, the $\tilde{\phi} \tilde{\phi}$
propagator at momentum $p$ is given by
\begin{equation}
  G_{\tilde{\phi} \tilde{\phi}}(p) = \frac {p^2}{\mathcal{D}_B(p)} =
  \frac {p^2}{p^4 + \hat{\kappa}^2 p_0^2 + p_1^2 - 2 \hat{\nu} \hat{w}
    p_0 p_1},
\end{equation} 
and it formally  does not have a well defined value when the
momentum  goes to  zero. 
% It doesn't even have a well defined limit when $p
%\to 0$.  %This poses a problem for defining $G_{\tilde{\phi}  \tilde{\phi}}(0)$.
To define the   propagator    in the $(\tilde \phi, \tilde \vp)$ sector of quadratic  action  \rf{qqw} 
one should  first  isolate the   constant (zero momentum)  mode 
of $\vp$. It must be projected out before computing the
propagator, because it corresponds to a reparameterization of the
classical background, not to a quantum fluctuation. 
Then  the propagator  is a $2 \times 2$ matrix at non-zero momentum, but at zero momentum   reduces to a
single term $G_{\tilde{\phi} \tilde{\phi}}(0) =
\tfrac 1 {\hat{\kappa}^2 - \hat{\nu}^2} = \tfrac 1 {1 - \hat{w}^2}$.
% If we first
%project out this zero mode and then invert the kinetic operator, we
%obtain the result $G_{\tilde{\phi} \tilde{\phi}}(0) =
%\tfrac 1 {\hat{\kappa}^2 - \hat{\nu}^2} = \tfrac 1 {1 - \hat{w}^2}$.
\foot{This follows from the second line in \rf{qqw} after  restricting the fields
to their  zero-momentum modes. The zero momentum propagator $G_{\tilde{\varphi} \tilde{\varphi}}(0)$
is also  not well defined, but it  never contributes as the shift symmetry of $\tilde\varphi$ implies that only its derivatives
can be generated quantum mechanically.}

%%%%%%%%%%%%%%%%%%%%%%%%%%%%%%%%%%%%%%%%%%%%%%%%%%%%%%%%%%%%%%%%%%%%%%%%%%%%%%%%%%%%%%%%%%%%%%%%%
\section{Two-loop partition function  and the generalized scaling 
function \label{generalized_scaling_function}}

We can now collect all the partial results listed in equations \rf{su}, \rf{dd}, \rf{fsu}, \rf{fdou}, \rf{taaa} 
to find  the full 2-loop contribution to the logarithm of the partition function\footnote{Here we 
 restored the overall factor $\frac{2\pi}{\sqrt{\lambda}}V_2= \frac{8\pi}{\sqrt{\lambda}}V$ in $W_2$ (cf. \rf{wv}).}
\begin{equation}
\begin{aligned}
W_2 &= W_{2{\rm B\ sunset}}+W_{2{\rm B\ double-bubble}}+W_{2{\rm F\  sunset
}}+W_{2{\rm F\ double-bubble}}+W_{2{\rm\ tadpoles}}\\
&\equiv \frac{V }{2\pi\sqrt{\lambda}} {\cal F}_2(\hat\nu,\hat w)\,.
\end{aligned}
\end{equation}
We will then be  ready to find the corresponding order $\ell^2$ and $\ell^4$ corrections to the generalized scaling function. 
Below we  shall also present the exact in $\ell$ expressions for the 
  coefficients of the two leading logarithms, 
$(\ln \ell)^2$ and $\ln \ell$ in the ${\cal O}({1 \ov \lambda})$ term in 
the scaling function.
    As we are interested in a comparison with the Bethe ansatz results 
of \cite{grom}, we will set ${\hat w}=0$ when constructing the generalized scaling function.

\subsection{Expansion to fourth order in $\ell$}

To this order, we find that ${\cal F}_2(\hat\nu,\hat w)$ is given by
\begin{equation}\la{F2}
\begin{aligned}
&{\cal F}_2(\hat\nu,\hat w)=-K
+{\te {\te {\te \frac{1}{4}}}} \Big{[}(9-2 K-6 \ln 2) \hat \nu^2
+(9+2 K-6 \ln 2) \hat w^2-4 (\hat \nu^2+\hat w^2) \ln(\hat\nu^2+\hat w^2)\Big{]}\\
&+\frac{1}{576} \Big{[}\big(126K-449+72 (17-9 \ln 2) \ln 2\big) \hat \nu^4
+6 \big(18 K-55+72 \ln 2 (-1+3 \ln 2)\big) \hat \nu^2 \hat w^2\\
&~~~~~~
+\big(126 K-1025+72 (17-9 \ln 2) \ln 2\big) \hat w^4
-48 \ln(\hat \nu^2+\hat w^2) \Big{(}(-17+18 \ln 2) \hat \nu^4\\
&~~~~~~
+6 (11-6 \ln 2) \hat \nu^2 \hat w^2+(-17+18 \ln 2) \hat w^4+
6 (\hat \nu^2-\hat w^2)^2 \ln(\hat \nu^2+\hat w^2)\Big{)}\Big{]}\,.
\end{aligned}
\end{equation}
Note that all UV divergences cancel out, i.e. 
the 2-loop partition function is finite.
 We 
 %can also check 
 have also checked that $W_2$ is invariant under 
 the simultaneous replacements $\nu \leftrightarrow w$, $\kappa \leftrightarrow \mu$ and $V_2 
 \leftrightarrow \hat \kappa^2 V_2$, as expected.\foot{To check this, one has to 
 notice that there is an interplay between different orders in the expansion in small $\hat\nu,\hat w$.}
The dependence  of $W_2$ on the physical winding number $m$ is  found 
%by the  doing  back  
through the replacement in \rf{esa}, i.e.  $\hat w = - \ri \hat m= - \ri  { m \ov \mu}$.  
%In what follows we shall consider the case of $w=0$. 

Setting $w=0$  in the 1-loop \rf{F1} and 2-loop \rf{F2}   expressions   we can 
 now use the relation  \rf{pp}  to compute  the 2-loop term in the  generalized  scaling function. 
As a result we get  (replacing $\hat \nu = \ell + O({1 \ov \sql})$) 
%Let us now use this result to obtain the string theory prediction for the generalized 
%scaling function corresponding to the twist operators ${\rm Tr} D_+^S Z^J$. For this 
%purpose, we set $\hat w=0$ and define ${\cal F}(\hat\nu) \equiv {\cal F}(\hat \nu,0)$. 
%Then we reconstruct the scaling function according to the recipe derived in \cite{rt2}
%\begin{equation}
%\begin{aligned}
%&E-S=\frac{\sqrt{\lambda}}{\pi}\ln S\left({\rm f}_0+\frac{1}{\sqrt{\lambda}}{\rm f}_1
%+\frac{1}{\lambda}{\rm f}_2+\ldots\right)\,,\\
%&{\rm f_0}={\cal F}_0(\hat \nu) \sqrt{1+\hat \nu^2}\\
%&{\rm f_1}=\frac{{\cal F}_1(\hat\nu)}{\sqrt{1+\hat \nu^2}}\\
%&{\rm f_2}=\frac{{\cal F}_2(\hat\nu)}{\sqrt{1+\hat \nu^2}}+\frac{1}{2}(1+\hat\nu^2)^{3/2} \left(
%\frac{d{\rm f_1}}{d\hat\nu}\right)^2\,.
%\end{aligned}
%\end{equation}
%The classical and 1-loop result are readily obtained from 
% eq.(\ref{F0})   and eq.(\ref{F1}). Finally, 
% using the result above for ${\cal F}_2$ and including the 1-loop shift, we find
\begin{equation}\la{f2f}
\begin{aligned}
&{\rm f}_2 = -K
\\
&~~~~~~
+ \ell ^2 \left(8\ln^2  \ell  -6 \ln  \ell-\frac{3}{2} \ln 2+\frac{11}{4}\right) 
\\
&~~~~~~
+  \ell^4 \left(-6 \ln^2 \ell -\frac{7}{6} \ln \ell +3 \ln 2 \ln \ell 
-\frac{9}{8} \ln^2 2+\frac{11}{8}\ln 2 +\frac{3}{32} K-\frac{233}{576}\right)+{\cal O}(\ell^6)\,.
\end{aligned}
\end{equation}
%Since $\nu hat = \ell + O({1 \ov \sql})$ the 
This is in  partial agreement   (for the 
  $-K+ \ell^2  (8\ln^2 \ell  -6 \ln \ell  -\frac{3}{2} \ln 2) $ terms)   with  an   earlier 
conformal gauge  computation \ci{rt2}
and also in 
complete agreement with the Bethe  ansatz  prediction \rf{tww}, \rf{qw}, \rf{ffq}  of \cite{grom}.

\subsection{Leading logarithms \label{sec:leadinglogs}}
In general, we may express the dependence of ${\rm f}_2$ on $\ell$ as
\be 
{\rm f}_2(\ell)= h_2(\ell) \ln^2\ell + h_1(\ell) \ln \ell + h_0(\ell)\ , 
\ee
where $h_2(\ell)$, $h_1(\ell)$ and $h_0(\ell)$ are expected to be analytic functions with a well-defined
Taylor expansion around $\ell=0$. The first few terms in this expansion can be read off from 
eq.~(\ref{f2f}) above. In this subsection we extract from our 2-loop superstring computation the 
exact expressions for the coefficient functions $h_2(\ell)$ and $h_1(\ell)$.
\iffalse
In this subsection we show that it is in fact possible to use our 2-loop superstring computation to extract an exact result for $h_2(\ell)$ and $h_1(\ell)$.
\fi

\iffalse
First, it is interesting to notice that the coefficient of the leading logarithm
 $\ln^2 \ell$ in (\ref{f2f}) receives contributions only from 
 the one-particle reducible diagrams
  and from the ``1-loop shift" induced by 
  ${\cal F}_1$ in eq. \rf{pp}. 
  \fi
  
 To find $h_2(\ell)$ we first notice that, to all orders in the small $\ell$ expansion, 
 the leading logarithm $\ln^2 \ell$ in (\ref{f2f}) receives contributions only from 
 the one-particle reducible diagrams
  and from the ``1-loop shift" induced by 
  ${\cal F}_1$ in eq. \rf{pp}.  
  The 1PI diagrams
%  on the other hand, 
 contribute only to $\ln \ell$ and other subleading terms. 
  %R
% The same pattern holds, in fact, to all orders in the $\ell$ expansion. 
 Indeed, even though 
 we have not computed exactly the sunset diagrams, it is possible to show that their complete contribution
 to the leading logarithms is $\ell^2 \ln^2 \ell$. These terms are already captured by our 
 small $\ell$ expansion and cancel against a similar double-bubble contribution.\footnote{The absence of 
 $\ln^2 \ell$ in the 1PI bosonic partition function can also be seen in the conformal gauge 
 calculation \cite{rt2}.} Moreover, it is clear that the 1PI diagrams involving fermions cannot
 yield $\ln^2\ell$ since they can only contain one propagator of the light $S^5$ fluctuations.
 %  
\iffalse
  Although we have not computed the
   sunset diagrams exactly, we expect this to remain
    true to all orders in $\ell$. \footnote{It is 
    clear that the 1PI diagrams involving fermions cannot
     yield $\ln^2\ell$ since they can only contain one 
     propagator of the light $S^5$ fluctuations. The bosonic
      1PI diagrams, on the other hand, do produce $\ln^2\ell$ 
      terms at order $\ell^2$, see eq. (\ref{su}) and (\ref{dd}),
       but they cancel between sunset and double-bubble diagram. 
      % At higher orders, we can directly see that the double-bubble 
      % does not produce $\ln^2\ell$ beyond order $\ell^2$, but what 
     %  about the sunset? Why can't ${\rm I}[1,m_y^2,m_y^2]$ in principle appear?
     }
\fi
%%%%
        Then, combining the exact expressions for $W_{2~{\rm tadpoles}}$ and ${\cal F}_1$, 
	we can deduce an all-order prediction for the coefficient of $\ln^2\ell$ in ${\rm
	 f}_2$. The relevant terms are (we use  eq. (\ref{F1}) and eq. (\ref{W2tadpoles}) and 
	 set $\hat w=0$)
\be
{\cal F}_{2~{\rm tadpoles}} = -2 \hat\nu^4 \ln^2 \hat\nu+\ldots \, 
\qquad \quad
{\cal F}_{1} = -2 \hat \nu^2 \ln\hat\nu+\ldots\ , 
\ee
which when inserted into (\ref{pp}) yield 
%the result
\begin{equation}
%({\rm f}_2)
% \Big{|}
%_{_{\ln^2 \ell}} 
h_2(\ell)= \frac{8 \ell^2 + 6 \ell^4}{(1+\ell^2)^{3/2}} = 
8\ell^2-6\ell^4+6\ell^6-\frac{25}{4}\ell^8+{\cal O}(\ell^{10})\,.
\end{equation}
This is in full agreement with the  Bethe ansatz 
%integrability
result of \cite{grom}, where an exact formula was given for the coefficient of the leading logarithm $\ln^{n}\ell/\lambda^{n/2}$ for all values of $n$.
%to all orders in perturbation theory.

With some effort, one may in fact extract also the exact contribution to $\ln\ell$ coming from the bosonic and fermionic sunset diagrams. It is clear that this can arise only from diagrams containing propagators of the $S^5$ fluctuations $y^a$, which become massless in the small $\ell$ limit. Isolating these contributions, setting ${\hat w}=0$ and keeping only the logarithmic terms we obtain the following %results (here we have set $\hat w=0$)
contributions from the bosonic and fermionic sunset diagrams:
\be 
&&{\cal F}_{2{\rm B~sunset}}=
2 \hat\nu^2 \ln^2 \frac{\hat\nu}{2} +4 \hat\nu^2 \ln \hat\nu\ 
\ln \frac{1+\sqrt{1+\hat\nu^2}}{2}+\ldots \cr
&&{\cal F}_{2{\rm F~sunset}}=
-\frac{\ln\hat\nu}{2\hat\nu^4}  \Bigg{(}\hat \nu^2 (24+3 \hat \nu^4-4 \hat \nu^2 (4\ln 2-5))
-8 (3+4 \hat \nu^2-\hat \nu^4) \ln(1+\hat\nu^2)\Bigg{)}+\ldots
\label{sunset-logs}
\ee
Notice that despite the presence of $\hat\nu^{-4}$ in the fermionic contribution, the small 
$\hat\nu$ expansion is perfectly regular. Expanding to fourth order, one recovers the logarithmic 
terms in the perturbative results  (\ref{su}) and (\ref{fsu}). 
%
%As for the double-bubble diagrams, as explained above they 
As explained in the previous section, the double-bubble diagrams
can be computed exactly in terms of the 1-loop integrals given in Appendix A; 
it is then straightforward to extract their logarithmic terms:
\be
 &&{\cal F}_{2{\rm B~double-bubble}}=-2 \hat\nu^2 \ln^2\frac{\hat\nu}{2} 
-2 \ln \hat \nu \left(2+\hat\nu^2-2\sqrt{1+\hat\nu^2}
+2\hat\nu^2 \ln\frac{1+\sqrt{1+\hat\nu^2}}{2} \right)+\ldots\cr
&&{\cal F}_{2{\rm F~double-bubble}}=\frac{1}{2} \ln \hat\nu \left(7 \hat\nu^2
+8 (1+\hat\nu^2) \ln \frac{1+\hat\nu^2}{4}\right)+\ldots
\label{bubbles-logs}
\ee
From these expressions one can see that, as claimed above, the $\ln^2\hat\nu$ terms cancel 
between bosonic sunset and double-bubble diagrams. 

To reconstruct the coefficient $h_1(\ell)$ of $\ln\ell$ in ${\rm f}_2$, we can now plug 
(\ref{sunset-logs}) and (\ref{bubbles-logs}), together with the exact ${\cal F}_{2~{\rm tadpoles}}$ 
and ${\cal F}_1$, into eq. (\ref{pp}). As a result we obtain the closed form expression
\be 
&&h_1(\ell)=\frac{2(1-\sqrt{1+\ell^2})^2}{\ell^8 (1+\ell^2)^{3/2}}\Bigg{\{}
-\ell^2 \sqrt{1+\ell^2} \left(12+22 \ell^2+12 \ell^4+\ell^6\right)
-\ell^2 \left(12+28 \ell^2+23 \ell^4+6 \ell^6\right)\cr
&& 
\!\!\!\!\!\!\!\!\!\!+\left(2+\ell^2+2 \sqrt{1+\ell^2}\right) 
\left[2(1+\ell^2) (3+4 \ell^2-2 \ell^6) \ln(1+\ell^2)
+\ell^8 \ln \left(\sqrt{2+\ell^2}(1+\sqrt{1+\ell^2})\right)\right]\Bigg{\}}
\ee
%As far as we know, this result has not appeared 
%elsewhere in the literature. 
This  all-order result is new, i.e. was not previously 
 derived directly from the Bethe  Ansatz. 
%was not previously derived from the 
The  small $\ell$ 
expansion  gives
\be
h_1(\ell) = -6 \ell^2+\left(-\frac{7}{6}+3\ln 2\right) \ell^4
+\left(\frac{26}{15}-\frac{9}{2}\ln 2\right) \ell^6
+\left(-\frac{181}{96}+\frac{45}{8}\ln 2\right) \ell^8+{\cal O}(\ell^{10})  
\ee
The first three terms can be seen to be in agreement with the analytic small $\ell$ expansion given 
in \cite{grom}. Remarkably, 
%it is possible to check that the 
higher order terms also agree, up to a considerably high power of $\ell$, with numerical results that 
can be obtained from the Bethe ansatz 
analysis of \cite{grom}.\footnote{We are very grateful to N. Gromov for 
sharing his numerical results and carefully checking them against ours.}

This provides convincing evidence  that the superstring and 
the Bethe ansatz expressions for the 2-loop term  ${\rm f}_2(\ell)$ 
in the generalized scaling function  are in full agreement. 

%ansatz result 
%This result is in complete agreement with the Bethe ansatz prediction of \cite{grom}.

Higher-loop calculations are in principle possible, but technically more involved. The leading 
logarithmic dependence on $\nuhat$, $(\nuhat^2 \ln \nuhat)^L$ at $L$-loops, is perhaps the 
most accessible. Based on the 2-loop results described in this section, one may expect that $n$-loop 
1PI diagrams can yield at most $(\ln \nuhat)^{n-1}$ beyond 1-loop. \footnote{This is trivial to see 
in conformal gauge. In light-cone gauge this is by no means obvious; however, based on the expected
gauge-independence of the 1PI part of the 
partition function, one may expect this to be generically true.}
Consequently, all 1PI graphs  as well as all non-1PI graphs containing a 1PI subgraph with more 
than one-loop should not contribute to the leading logarithmic terms. The only contributing graphs 
seem therefore to have at most 1-loop subgraphs; one might call them maximally-non-1PI graphs.

While the evaluation of the leading logarithmic terms is still nontrivial, the resulting partition 
function should take a simple form: indeed, the Bethe ansatz results for the leading logarithms 
are reproduced if the all-order ${\cal F}$ is given by
\be
&&{\cal F}^{\rm leading\;log}(\sql, \nuhat)=\sqrt{1+\frac{2}
{\sql}\ {\cal F}^{\rm leading\;log}_{\rm 1-loop}}
= \sqrt{1-\frac{2}{\sql} \nuhat^2\ln \nuhat^2 } \\
&&=1-\frac{1}{\sql}(\nuhat^2\ln \nuhat^2)-\frac{1}{2(\sql)^2}(\nuhat^2\ln \nuhat^2)^2
-\frac{1}{2(\sql)^3}(\nuhat^2\ln \nuhat^2)^3
-\frac{5}{8(\sql)^4}(\nuhat^2\ln \nuhat^2)^4+\dots
\nonumber
\ee
The first three terms on the second line reproduce the tree-level,  the 1- and the  2-loop terms
discussed earlier in this paper. The fourth term is generated at 3 loops and, as suggested above, 
appears to receive contributions only from the maximally-non-1PI Feynman diagrams. It would 
be very interesting to construct ${\cal F}^{\rm leading\;log}(\sql, \nuhat)$ through a direct field 
theory calculation, perhaps by reducing the partition function to a single integral over a constant 
(off-shell) mode.

\section{Concluding remarks   \label{remarks}}

In this paper 
 we   computed the first two nontrivial orders in the small $\nuhat$ ($S^5$ momentum density) 
 and $\hat m $ (winding number density)  expansion of
the 2-loop correction  to the partition function  of the 
generalized  null cusp surface or, equivalently, 
to  the  energy of a  generalization of the  large spin limit of the 
$(S,J)$ folded string  with extra winding in a  circle of $S^5$.

We have  found the corresponding correction to the generalized scaling function (for $\hat m=0$) 
and demonstrated   the complete agreement (which  was only partial    in the
previous string theory  computation \ci{rt2})    with the  result found 
\ci{grom} from the 
asymptotic Bethe ansatz (as well as  from $O(6)$ model combined with  BA information \ci{bass}).\foot{The
corresponding 
 $\hat m \not= 0$
expressions are still to be obtained from the Bethe ansatz as only the solution with $m=0$ 
was previously considered there.} 
This  provides a  highly non-trivial test of the  strong-coupling  asymptotic Bethe ansatz  proposal
beyond the 1-loop semiclassical level, thus 
extending earlier tests performed in \ci{rt1,bkk,rt2}.

%the most sophisticated   so far. 
Our  final 2-loop result is sensitive to 
all terms in the light-cone action \rf{la}   and thus also nontrivially checks its consistency,
%
%provides  a non-trivial consistency check of it, 
%
demonstrating its UV finiteness and, via the 
agreement with the Bethe ansatz result, 
the quantum integrability of the  corresponding world sheet theory.

The AdS light-cone gauge approach used in \ci{grrtv} and here 
 is substantially less complex than the conformal gauge
one used in \ci{rtt,rt1,rt2}.
As we demonstrated in  section \ref{sec:leadinglogs},  it is  also possible  
to extract analytically higher-order terms 
%also terms of higher orders  
in  the expansion in the parameters $\hat \nu,\,\hat m$ and thus provide 
further 2-loop tests of  the generalized scaling function
 (the $\nu^6$ terms were explicitly  worked out on the BA side in \ci{grom}).

In the   $\nuhat={\hat m}=0$  case  one may be able  also to carry out
%emphasized in \cite{grrtv} suggests that
higher-loop calculations of the cusp anomaly function $f(\l)$. 
Beyond the two-loop order, however, the  
momentum conservation  is  no longer sufficient to reduce all
 integrals to scalar integrals with constant
numerator factors. It seems likely that Lorentz-invariant integrals with nontrivial
momentum-dependent numerator factors will contribute to,  e.g.  the 3-loop partition function.

It may be of interest to study also  other generalizations of the 1-loop and 2-loop 
computations  of the partition function  for the  null cusp  surface by  including other 
(homogeneous?) profiles on $S^5$. For example, one may consider a string wrapped on a ``small''
circle  of a 2-sphere inside of $S^5$. 

An important extension of the calculations described here and in \ci{grrtv} is the evaluation
of finite size corrections and their comparison with L\"uscher term and TBA predictions.  A natural candidate  background 
is the spinning string. To leading order in the large spin 
expansion there are two types of
contributions to consider. On the one hand, the action is modified due to finite spin corrections 
to the classical background. On the other, the world sheet is no longer infinite, leading to momentum 
integration being replaced by summation over a discrete spectrum. As a step towards evaluating 
finite size corrections one may consider only the second type of contributions, i.e. simply use the 
spinning string solution in its asymptotic  scaling limit form 
 but assume that the world sheet circle is of finite radius.\footnote{There 
 is yet a third correction which may enter at sufficiently high loop order 
  due to 
finite size corrections to the thermodynamic argument in section \ref{E_vs_Z} as well as due to the
renormalization of the spin (having the same origin as  the renormalization of the orbital momentum $J$  discussed there).}

Another 
 potential future application of the AdS light-cone gauge action is  
the evaluation of energies of string states  with finite quantum numbers  
in a near-flat-space inverse tension expansion. There are, however, 
 various 
conceptual and technical complications along the way. Among the former is the realization of the superconformal algebra on excited string states. Among the latter, the light-cone 
expression for  the AdS energy is nonlocal, suggesting that the computation of its expectation 
value is not completely straightforward. However, the calculation in Appendix \ref{sec:one_loop_E_J} 
demonstrates that this conclusion may be premature. We hope to return to these issues in the future.

%While somewhat nontrivial from the closed string perspective, the inclusion of the "winding"
%$\what$ is quite trivial from the open string perspective. Similarly, other generalizations of
% the cusp Wilson line are possible,
%by including other (homogeneous?) profiles on $S^5$. Calculations appear difficult however due 
%to nontriviality of propagators of the $S^5$ excitations.

%%%%%%%%%%%%%%%%%%%%%%%%%%%%%%%%%%%%%%%%%%%%%%%%%
\subsection*{Acknowledgements}
%%%%%%%%%%%%%%%%%%%%%%%%%%%%%%%%%%%%%%%%%%%%%%%%%%
We are grateful to J. Bedford, N. Gromov, 
M. Kruczenski, A. Tirziu, I.~Tyutin and  D. Volin  for many  useful discussions. We are particularly grateful to N. Gromov for providing further terms in the Bethe ansatz result not explicitly included 
in \cite{grom}.
This work was supported in part by the US National Science Foundation under
DMS-0244464 (S.G.), PHY-0608114 and PHY-0855356 (R.Ro.) and PHY-0643150 (C.V.), 
  the US Department of Energy under contracts DE-FG02-201390ER40577 (OJI) (R.Ro.)
and DE-FG02-91ER40688 (C.V.), the Fundamental Laws Initiative Fund at 
Harvard University (S.G.) and the A. P. Sloan Foundation (R.Ro.).
It was also   supported by the EPSRC (R.Ri.). S.G. and R.Ri. 
would like to thank the Simons Center for Geometry and Physics for hospitality during 
the 7th Simons workshop on Physics and Mathematics.

%%%%%%%%%%%%%%%%%%%%%%%%%%%%%%%%%%%%%%%%%%%%%%%%%%%%%%%%%%%%%%%%%%%%%%%%%%%%%%%%%%%%%%%%%%%%%%%%%%%%%%%%%%%%%%%%%

\appendix

\section{Useful 1-loop integrals
\label{sec:1loop_int}}

We use the notation
\be
{\rm I}\mbox{$\left({a\atop m^2}\right)$}= \int \frac{d^2p}{(2\pi)^2}\frac{1}{(p^2+m^2)^a}
\ee
For $a > 1$ and non-zero $m$, this integral is convergent and equal to 
\be 
{\rm I}\mbox{$\left({a\atop m^2}\right)$}=\frac{m^{2-2a}}{4\pi(a-1)}\,.
\ee   
For $a=1$, on the other hand, the integral is logarithmically UV divergent. For convenience, in what follows and in the main text we use the notation
\be 
{\rm I}\mbox{$\left({1\atop m^2}\right)$} \equiv {\rm I}[m^2]\,.
\ee
The following identity will prove often useful
\be 
{\rm I}[m_1^2]-{\rm I}[m_2^2] = \int \frac{d^2p}{(2\pi)^2}\frac{m_2^2-m_1^2}{(p^2+m_1^2)(p^2+m_2^2)}=\frac{1}{4\pi}\left(\ln m_2^2-\ln m_1^2\right)\,.
\ee 

Define the integrals ($p^2= p_0^2 + p^2_1$) 
\begin{align}
  J_B(i,j) &= \int \frac{d^2 p}{(2 \pi)^2} \frac {p_0^i p_1^j}{p^4 +
    p^2 + (\hat{\nu}^2 - \hat{w}^2) p_0^2 - 2 \hat{w} \hat{\nu} p_0 p_1},\\
  J_F(i,j) &= \int \frac{d^2 p}{(2 \pi)^2} \frac {p_0^i p_1^j}{p^2 + \frac {4 \hat{\kappa}^2 - 
  \hat{\nu}^2 + 3 \hat{w}^2}{16} - \frac {\mathrm{i}} 2 (\hat{\nu} p_0 - \hat{w} p_1)}.
\end{align}
We get
{\allowdisplaybreaks
\begin{align}
  J_B(0,0) &= -\frac 1 {\sqrt{(1 + \hat{\nu}^2)(1 - \hat{w}^2)}}
    \Big(\frac 1 {4 \pi} \ln \frac {(\sqrt{1 + \hat{\nu}^2} +
          \sqrt{1 - \hat{w}^2})^2} 4  - {\rm I}[0] + {\rm I}[(1 +
      \hat{\nu}^2)(1 - \hat{w}^2)]\Big),\\
  J_B(1,0) &= J_B(0,1) = 0,\\
  J_B(2,0) &= - \frac {\hat{\nu}^2 - \hat{w}^2}{8 \pi (\hat{\nu}^2 +
    \hat{w}^2)^2} (\sqrt {1 + \hat{\nu}^2} - \sqrt{1 - \hat{w}^2})^2 +
  \frac 1 2 {\rm I}\Big[\frac {(\sqrt{1 + \hat{\nu}^2} +
          \sqrt{1 - \hat{w}^2})^2} 4\Big],\\
  J_B(0,2) &= \frac {\hat{\nu}^2 - \hat{w}^2}{8 \pi (\hat{\nu}^2 +
    \hat{w}^2)^2} (\sqrt {1 + \hat{\nu}^2} - \sqrt{1 - \hat{w}^2})^2 +
  \frac 1 2 {\rm I}\Big[\frac {(\sqrt{1 + \hat{\nu}^2} +
          \sqrt{1 - \hat{w}^2})^2} 4\Big],\\
  J_B(1,1) &= \frac {\hat{\nu} \hat{w}}{4 \pi (\hat{\nu}^2 +
    \hat{w}^2)^2} (\sqrt {1 + \hat{\nu}^2} - \sqrt{1 - \hat{w}^2})^2,\\
  J_F(0,0) &= {\rm I}\left[\frac{1+\hat{\nu}^2}4\right],\\
  J_F(1,0) &= \mathrm{i} \frac {\hat{\nu}}4 \left({\rm I}\left[\frac{1+\hat{\nu}^2}4\right] - \frac 1 {4 \pi}\right),\\
  J_F(0,1) &= -\mathrm{i} \frac {\hat{w}}4 \left({\rm I}\left[\frac{1+\hat{\nu}^2}4\right] - \frac 1 {4 \pi}\right),\\
  J_F(2,0) &= - \frac{2 + 3 \nu^2}{16} {\rm I}\left[\frac{1+\hat{\nu}^2}4\right] + \frac{7 \hat{\nu}^2 + \hat{w}^2}{256 \pi},\\
  J_F(0,2) &= - \frac{2 + 2 \hat{\nu}^2 + \hat{w}^2}{16} {\rm I}\left[\frac{1+\hat{\nu}^2}4\right] + \frac{\hat{\nu}^2 + 7 \hat{w}^2}{256 \pi},\\
  J_F(1,1) &= \frac{\hat{\nu} \hat{w}}{16} {\rm I}\left[\frac{1+\hat{\nu}^2}4\right] - \frac {3 \hat{\nu} \hat{w}}{128 \pi}.
\end{align}}

\section{Three-propagator integrals
\label{sec:int_computation}}

We want to compute the following integrals
\begin{equation}\la{ik}
  {\rm I} \Bigl(\begin{smallmatrix}
    \lambda_1 & \lambda_2 & \lambda_3\\
    m_1^2 & m_2^2 & m_3^2
  \end{smallmatrix}\Bigr) = \frac 1 {(2 \pi)^4} \int d^2 p_1 d^2 p_2 d^2 p_3 \frac {\delta^2 (p_1+p_2+p_3)}{(p_1^2 + m_1^2)^{\lambda_1} (p_2^2 + m_2^2)^{\lambda_2} (p_3^2 + m_3^2)^{\lambda_3}}.
\end{equation}
%Let us try to compute this  integral
% starting with the momentum space representation.  
 Let us use the following identity
\begin{equation}
  \frac 1 {(p^2 + m^2)^\lambda} = \frac 1 {\Gamma(\lambda)} \int_0^\infty d \alpha\, \alpha^{\lambda - 1} e^{-\alpha (p^2 + m^2)}.
\end{equation} 
 Using this $\alpha$ parameterization
 % in the expression for the integral,
  we get
\begin{equation}
  {\rm I} \Bigl(\begin{smallmatrix}
    \lambda_1 & \lambda_2 & \lambda_3\\
    m_1^2 & m_2^2 & m_3^2
  \end{smallmatrix}\Bigr) = \frac 1 {(2 \pi)^4} \int \prod_{i=1}^3 \frac{d^2 p_i d \alpha_i 
  \alpha_i^{\lambda_i-1}}{\Gamma(\lambda_i)} \delta^2 \big(\sum_{i=1}^3 p_i\big) \exp \Big(- \sum_{i=1}^3 
  \alpha_i (p_i^2 + m_i^2) \Big),
\end{equation} where the integrals over $\alpha_i$ run from zero to infinity.
Doing the gaussian integrals over $p_i$ gives 
\begin{equation}
  \frac 1 {16 \pi^2} \int \prod_{i=1}^3 \frac {d \alpha_i \, \alpha_i^{\lambda_i -1}}{\Gamma(\lambda_i)} \frac {\exp \left(- \sum_{i=1}^3 \alpha_i m_i^2 \right)}{\alpha_1 \alpha_2 + \alpha_1 \alpha_3 + \alpha_2 \alpha_3}.
\end{equation}
At this point, in general,  one changes the variables $\alpha_i$
 such that $\alpha_i = \alpha \xi_i$, with $\sum \xi_i = 1$ and then  performs the integral over $\alpha$
 % The result of this integration is given by
\begin{equation}
  \frac 1 {16 \pi^2} \Gamma\left(\sum_{i=1}^3 \lambda_i - 2\right) \int \prod_{i=1}^3 \frac {d \xi_i \xi_i^{\lambda_i-1}}{\Gamma(\lambda_i)} \frac {\delta\left(1 - \sum_{i=1}^3 \xi_i\right)}{(\xi_1 \xi_2 + \xi_1 \xi_3 + \xi_2 \xi_3) \left(\sum_{i=1}^3 m_i^2 \xi_i\right)^{\sum_{i=1}^3 \lambda_i -2}},
\end{equation} where the integrals over $\xi_i$ run from zero to infinity.

In this form, however, the integral is still hard to compute.  We will 
use a trick known as Cheng--Wu theorem (see refs.~\cite{Cheng:1987ga, Bjoerkevoll:1992cu, 
Smirnov:2004ym}).  This theorem states that the sum in the delta function in the above equation
 can be replaced by a sum over a restricted set of $\xi$ variables.  This can be proven by making
  a change of coordinates $\alpha_i = \alpha \xi_i$, with $\sum' \xi_i = 1$, where this time the 
  sum runs over the restricted set of $\xi_i$ variables.

In our case, we will choose the constraint to be $\xi_2 + \xi_3 = 1$ so that  we can then perform the 
integral over $\xi_1$ explicitly.  Let us specialize to the case of $\lambda_i=1$.  Then the integral
 to compute becomes
\begin{equation}
  \frac 1 {16 \pi^2} \int \prod_{i=1}^3 d \xi_i \frac{\delta(1 - \xi_2 - \xi_3)}{(\xi_1 + \xi_2 \xi_3) 
  (m_1^2 \xi_1 + m_2^2 \xi_2 + m_3^2 \xi_3)}.
\end{equation}  The integral over $\xi_1$ runs from $0$ to $\infty$ and can be done trivially, 
and then another integral can be done using the delta function constraint.  As a result 
\begin{equation}
  \frac 1 {16 \pi^2} \int_0^1 d x \frac {\ln \frac {m_1^2 x (1-x)}{m_2^2 x + m_3^2 (1-x)} }{m_1^2 x (1-x)
   - m_2^2 x - m_3^2 (1-x)}.
\end{equation}  It may seem 
that the integral becomes divergent when the denominator vanishes.
  However, this singularity is cancelled by the numerator 
  so that the integral is convergent for all values of the masses.

Let us study the case of  two equal masses, $m_2 = m_3 = \tfrac 1 {\b} m_1$. 
 In that case the integral simplifies to
\begin{equation}
  \frac 1 {16 \pi^2} \frac 1 {m_1^2} \int_0^1 d x \frac {\ln(\b^2 x (1-x))} {x(1-x) - \frac 1 {\b^2}}.
\end{equation}
This integral can be computed in terms of logarithms and di-logarithms:
\begin{equation}
\int_0^1 \frac {d x \ln(\b^2 x (1-x))} {x(1-x) - \frac 1 {\b^2}} = \frac 1 {x_1-x_2} \left[\ln 
\b^2 \ \ln \left(\frac {x_2^2}{x_1^2}\right) + 2 \Li_2 \left(\frac 1 {x_1}\right) - 2 \Li_2 \left(\frac 
1 {x_2}\right) \right],
\end{equation} where $x_1$, $x_2$ are the solutions of the equation $x(1-x) - \tfrac 1 {\beta^2} = 0$.

\section{Values of  the integrals
\label{sec:2loop_int}}

{\allowdisplaybreaks
\begin{alignat*}{2}
%%%%
{\rm I} \Bigl(\begin{smallmatrix}1&1&1 \\ \tfrac 1 2 & \tfrac 1 4 & \tfrac 1 4\end{smallmatrix}\Bigr) &= \frac{K}{2 \pi^2}, \quad &
{\rm I} \Bigl(\begin{smallmatrix}1&1&2 \\ \tfrac 1 2 & \tfrac 1 4 & \tfrac 1 4 \end{smallmatrix}\Bigr) &= \frac {2 K - \ln 2}{2 \pi^2}, \\
{\rm I} \Bigl(\begin{smallmatrix}1&1&3 \\ \tfrac 1 2 & \tfrac 1 4 & \tfrac 1 4 \end{smallmatrix}\Bigr) &= \frac {-1 + 4 K - 2 \ln 2}{\pi^2}, \quad &
{\rm I} \Bigl(\begin{smallmatrix}1&1&4 \\ \tfrac 1 2 & \tfrac 1 4 & \tfrac 1
  4 \end{smallmatrix}\Bigr) &= \frac {2 (24 K - 7 (1 + 2 \ln 2))}{3 \pi^2}, \\
{\rm I} \Bigl(\begin{smallmatrix}1&1&5 \\ \tfrac 1 2 & \tfrac 1 4 & \tfrac 1
  4 \end{smallmatrix}\Bigr) &= \frac {2 (-33 + 102 K - 64 \ln 2)}{3 \pi^2},\quad &
{\rm I} \Bigl(\begin{smallmatrix}1&2&2 \\ \tfrac 1 2 & \tfrac 1 4 & \tfrac 1 4 \end{smallmatrix}\Bigr) &= \frac {4 (K - \ln 2)}{\pi^2}, \\
{\rm I} \Bigl(\begin{smallmatrix}1&2&3 \\ \tfrac 1 2 & \tfrac 1 4 & \tfrac 1 4 \end{smallmatrix}\Bigr) &= -\frac {2 (3 - 12 K + 10 \ln 2)}{\pi^2}, \quad &
{\rm I} \Bigl(\begin{smallmatrix}1&3&3 \\ \tfrac 1 2 & \tfrac 1 4 & \tfrac 1 4 \end{smallmatrix}\Bigr) &= \frac {4 (-15 + 42 K - 32 \ln 2)}{\pi^2}, \\
{\rm I} \Bigl(\begin{smallmatrix}2&1&1 \\ \tfrac 1 2 & \tfrac 1 4 & \tfrac 1 4 \end{smallmatrix}\Bigr) &= \frac {\ln 2}{2 \pi^2}, \quad &
{\rm I} \Bigl(\begin{smallmatrix}2&1&2 \\ \tfrac 1 2 & \tfrac 1 4 & \tfrac 1 4 \end{smallmatrix}\Bigr) &= \frac {1 - 2 K + 2 \ln 2}{\pi^2}, \\
{\rm I} \Bigl(\begin{smallmatrix}2&1&3 \\ \tfrac 1 2 & \tfrac 1 4 & \tfrac 1 4 \end{smallmatrix}\Bigr) &= \frac {4 (1 - 3 K + 3 \ln 2)}{\pi^2}, \quad &
{\rm I} \Bigl(\begin{smallmatrix}3&1&1 \\ \tfrac 1 2 & \tfrac 1 4 & \tfrac 1
  4 \end{smallmatrix}\Bigr) &= \frac {2 K - 1}{2 \pi^2}, \\
%%%%
{\rm I} \Bigl(\begin{smallmatrix}1&1&1 \\ 1 & \tfrac 1 4 & \tfrac 1 4 \end{smallmatrix}\Bigr) &= \frac {\ln 2}{2 \pi^2}, \quad &
{\rm I} \Bigl(\begin{smallmatrix}1&1&2 \\ 1 & \tfrac 1 4 & \tfrac 1 4 \end{smallmatrix}\Bigr) &= \frac {1 + 2 \ln 2}{6 \pi^2}, \\
{\rm I} \Bigl(\begin{smallmatrix}1&1&3 \\ 1 & \tfrac 1 4 & \tfrac 1 4 \end{smallmatrix}\Bigr) &= \frac {2 (1 + 8 \ln 2)}{15 \pi^2}, \quad &
{\rm I} \Bigl(\begin{smallmatrix}1&2&2 \\ 1 & \tfrac 1 4 & \tfrac 1 4 \end{smallmatrix}\Bigr) &= \frac {2 (7 - 4 \ln 2)}{15 \pi^2},\\
{\rm I} \Bigl(\begin{smallmatrix}2&1&1 \\ 1 & \tfrac 1 4 & \tfrac 1 4 \end{smallmatrix}\Bigr) &= \frac {-1 + 4 \ln 2}{12 \pi^2}, \quad &
%%%%
{\rm I} \Bigl(\begin{smallmatrix}1&1&1 \\ 1 & \tfrac 1 2 & \tfrac 1 2 \end{smallmatrix}\Bigr) &= \frac {K}{4 \pi^2}, \\
{\rm I} \Bigl(\begin{smallmatrix}1&1&2 \\ 1 & \tfrac 1 2 & \tfrac 1 2 \end{smallmatrix}\Bigr) &= \frac {2 K - \ln 2}{8 \pi^2}, \quad &
{\rm I} \Bigl(\begin{smallmatrix}1&1&3 \\ 1 & \tfrac 1 2 & \tfrac 1 2 \end{smallmatrix}\Bigr) &= \frac {-1 + 4 K - 2 \ln 2}{8 \pi^2}, \\
{\rm I} \Bigl(\begin{smallmatrix}1&2&2 \\ 1 & \tfrac 1 2 & \tfrac 1 2 \end{smallmatrix}\Bigr) &= \frac {K - \ln 2}{2 \pi^2}, \quad &
{\rm I} \Bigl(\begin{smallmatrix}2&1&1 \\ 1 & \tfrac 1 2 & \tfrac 1 2 \end{smallmatrix}\Bigr) &= \frac {\ln 2}{8 \pi^2}, \\
{\rm I} \Bigl(\begin{smallmatrix}2&1&2 \\ 1 & \tfrac 1 2 & \tfrac 1 2 \end{smallmatrix}\Bigr) &= \frac {1 - 2 K + 2 \ln 2}{8 \pi^2}, \quad &
{\rm I} \Bigl(\begin{smallmatrix}3&1&1 \\ 1 & \tfrac 1 2 & \tfrac 1 2 \end{smallmatrix}\Bigr) &= \frac {2 K - 1}{16 \pi^2}.
\end{alignat*}}

We will also need some integrals of the type ${\rm I}\Bigl(\begin{smallmatrix}1&1&k \\ \mu^2& \tfrac
 1 4 & \tfrac 1 4 \end{smallmatrix}\Bigr)$
 %.  The necessary integrals are
\begin{align}
  {\rm I} \Bigl(\begin{smallmatrix} 1 & 1 & 2 \\ \mu^2 & \tfrac 1 4 & \tfrac 1 4 \end{smallmatrix}\Bigr) =& \frac 1 {1 - \mu^2} {\rm I} \Bigl(\begin{smallmatrix} 1 & 1 & 1 \\ \mu^2 & \tfrac 1 4 & \tfrac 1 4 \end{smallmatrix}\Bigr) - \frac {\ln (4 \mu^2)}{4 (1 - \mu^2) \pi^2}, \\
  {\rm I} \Bigl(\begin{smallmatrix} 1 & 2 & 2 \\ \mu^2 & \tfrac 1 4 & \tfrac 1
    4 \end{smallmatrix}\Bigr) =& \frac {4 \mu^2 - 1}{(\mu^2 - 1)^2 \mu^2}
  {\rm I} \Bigl(\begin{smallmatrix} 1 & 1 & 1 \\ \mu^2 & \tfrac 1 4 & \tfrac 1
    4 \end{smallmatrix}\Bigr) - \frac{\ln (4 \mu^2)-2}{4 \mu^2 \pi^2} + \frac{\ln (4 \mu^2)+2}{4 (\mu^2-1) \pi ^2} - \frac{3 \ln (4 \mu^2)}{4 (\mu^2-1)^2 \pi^2},\\
  {\rm I} \Bigl(\begin{smallmatrix} 1 & 1 & 3 \\ \mu^2 & \tfrac 1 4 & \tfrac 1
    4 \end{smallmatrix}\Bigr) =& \frac{2 \mu^2+1}{2 (\mu^2-1)^2 \mu^2}
  {\rm I} \Bigl(\begin{smallmatrix} 1 & 1 & 1 \\ \mu^2 & \tfrac 1 4 & \tfrac 1
    4 \end{smallmatrix}\Bigr) + \frac{\ln (4 \mu^2)-2}{8 \mu^2 \pi^2} + \frac{3
    \ln (4 \mu^2)+2}{8 (\mu^2-1) \pi^2} -\frac{3 \ln (4 \mu^2)}{8
    (\mu^2-1)^2 \pi^2}.
\end{align}

\section{Computation of tensor integrals
\label{sec:tensor_ints}}

In the computation of the partition function we encounter vacuum tensor
integrals, i.e. integrals of the form
\begin{equation}
  \label{eq:tensor_integral}
  I^{\mu_1 \cdots \mu_n} = \int \prod_{i=1}^L \frac {d^2 k_i}{(2 \pi)^2} 
  \frac {p_1^{\mu_1} \cdots p_n^{\mu_n}} {\mathcal{D}},
\end{equation} 
where $L$ is the number of loops, $k_i$ are the loop
momenta, $ {\mathcal{D}}    $ is a Lorentz invariant denominator arising from the
product of propagators and the momenta $p_j$ can be expressed in terms
of the loop momenta $k_i$.  As throughout the paper, we are using a regularization in which
the dimension of space is two.

Lorentz invariance implies that the tensor $I^{\mu_1 \cdots \mu_n}$ should be
expressible in terms of invariant tensors $\delta^{\mu \nu}$ and
$\epsilon^{\mu \nu}= -\epsilon^{\nu \mu} $. 
%where $\epsilon$ is the antisymmetric tensor.
Furthermore, parity invariance implies that  $I^{\mu_1
  \cdots \mu_n}$ cannot actually depend on the $\epsilon$ tensor. 
  % It follows
Thus  the result should be expressible solely in terms of $\delta$
tensors.  In particular, tensors of odd rank should vanish automatically
because there are no odd rank invariant tensors.

The strategy for reduction is to first find a basis of Lorentz
invariant tensors and then to compute the decomposition of the tensor
integral in this basis.  To find the coefficients of the
decomposition we contract the indices of the decomposition and of the
integrals in all possible ways and then solve the resulting linear
system.

%What is a basis for rank $2 r$ Lorentz invariant tensors?  
The construction of a basis of  rank $2 r$ Lorentz invariant tensors is 
somewhat subtle due to nontrivial relations existing between tensors of 
sufficiently high rank.
At rank two
there is only one possibility, $\delta^{\alpha \beta}$ and at rank
four there are three possibilities $\delta^{\alpha \beta}
\delta^{\gamma \delta}$, $\delta^{\alpha \gamma} \delta^{\beta
  \delta}$, $\delta^{\alpha \delta} \delta^{\beta \gamma}$.  If this
pattern continued to higher rank then there would be
\begin{equation}
  \frac 1 {r!} \binom{2 r}{r} \binom{2 r-2}{2} \cdots \binom{2}{2} = (2 r - 1)!!
\end{equation} possibilities at rank $2 r$.  However, it turns out
that starting at rank six there are some linear relations between the
elements of this naive basis.  To give  an example of such relations
consider the rank six tensor $\delta^{\alpha \beta} \delta^{\gamma
  \delta} \delta^{\epsilon \zeta}$ and antisymmetrize in the three
indices $\alpha$, $\gamma$ and $\epsilon$: since  the 
antisymmetrization of three two-dimensional indices yields a vanishing result,
this construction generates a nontrivial relation between rank six tensors.

One simple way to solve the constraints arising from
antisymmetrization in the indices is to consider invariant tensors of
type $(p,q)$, with $p+q=2 r$ which are completely symmetric in a group
of $p$ indices and also in the remaining group of $q$ indices.  For
this kind of tensors the antisymmetrization constraint is empty,
because one ends up antisymmetrizing in two symmetric indices.

It is easy to see that one can use the tensors of type $(p,q)$ to
build tensors of type $(p+1,q-1)$ by symmetrizing one of the $q$
indices together with the $p$ completely symmetric indices.  It
follows that the tensors of type $(r,r)$ can be used to build all the
allowed symmetries.  There are $\tfrac 1 2 \binom{2 r}{r}$ such
tensors (3 for rank four, 10 for rank six and 35 for rank eight).  A
careful counting of the independent constraints confirms that the
number of independent invariant tensors is given by $\tfrac 1 2
\binom{2 r}{r}$.

In this paper we will need to reduce tensor integrals of rank up to
eight.  The high rank integrals arise from the expansion of the
denominators of some propagators.  In the cases of rank two and four
we can replace the numerators in eq.~\eqref{eq:tensor_integral} as
follows
\begin{align}
  p_1^{\mu} p_2^{\nu} \to & \ \  \frac 1 2 \delta^{\mu \nu} p_1 \cdot p_2,\\
  p_1^\alpha p_2^\beta p_3^\gamma p_4^\delta \to & \begin{aligned}[t]
&\hphantom{+}\left(\frac 3 8 p_1 \cdot p_4 p_2 \cdot p_3 -
      \frac 1 8 p_1 \cdot p_3 p_2 \cdot p_4 -
      \frac 1 8 p_1 \cdot p_2 p_3 \cdot p_4\right)
\delta^{\alpha \delta} \delta^{\beta \gamma} \\
&+\left(- \frac 1 8 p_1 \cdot p_4 p_2 \cdot p_3 +
        \frac 3 8 p_1 \cdot p_3 p_2 \cdot p_4 -
        \frac 1 8 p_1 \cdot p_2 p_3 \cdot p_4\right)
  \delta^{\alpha \gamma} \delta^{\beta \delta} \\
&+\left(- \frac 1 8 p_1 \cdot p_4 p_2 \cdot p_3 -
        \frac 1 8 p_1 \cdot p_3 p_2 \cdot p_4 +
        \frac 3 8 p_1 \cdot p_2 p_3 \cdot p_4\right)
  \delta^{\alpha \beta} \delta^{\gamma \delta} ~~.
\end{aligned}
\end{align}  
We do not include the reduction formulae for higher rank
tensors because they are rather  lengthy.  As explained above, tensor
integrals of odd rank vanish.

We have therefore reduced the problem of computing tensor integrals to
the simpler problem of computing scalar integrals, with numerators
containing scalar products of loop momenta.  In the cases we encounter at
two loops, these  products can be simplified further by using
momentum conservation and finally the resulting expressions can be
simplified by partial fractioning.  In the end only a relatively small
number of master integrals need to be computed.  The values of these
integrals are tabulated in Appendices~\ref{sec:1loop_int}
and~\ref{sec:2loop_int}.

%%%%%%%%%%%%%%%%%%%%%%%%%%%%%%%%%%%%%

So far we have been discussing the tensor reduction of vacuum integrals.
However, for some purposes it turns out to be useful to reduce tensor
integrals depending on external momenta.  In these cases the integrals
are no longer Lorentz invariant and apart from dependence on constant invariant tensors, 
their expression includes dependence on the components of external momenta.
 Below we will only study
the case with one external momentum (propagator integrals) and will present reduction
formulas only for rank one and two; the method we use can be
extended to higher rank tensors and to more external momenta.

Let us denote by $I^\mu(q)$ an tensor integral dependent on an
external momentum $q$.  This integral is going to be of the form
\begin{equation}
 I^\mu = \int \prod_{i=1}^L \frac {d^2 k_i}{(2 \pi)^2} \frac
 {p_1^\mu}{\mathcal{D}},
\end{equation} where $\mathcal{D}$ is a Lorentz invariant denominator
and $p_1$ is a momentum which can be written in terms of the loop
momenta and the external momentum $q$.

Lorentz invariance requires that the integral be proportional to the
external momentum $q$ and the proportionality constant can be found by
contracting with $q$.  In the end, we find that the numerator
$p_1^\mu$ can be replaced by 
\be
p_1^\mu \mapsto \frac {p_1 \cdot q}{q^2} q^\mu~~.
\ee  
In some cases this new integrand can be simplified further.

At rank two the same procedure can be followed.  In this case we will
study integrals of type
\begin{equation}
 I^{\mu \nu} = \int \prod_{i=1}^L \frac {d^2 k_i}{(2 \pi)^2} \frac
 {p_1^\mu p_2^\nu}{\mathcal{D}}.
\end{equation}  For vacuum integrals the only available rank two
tensor was $\delta^{\mu \nu}$.  In the case of an integral with one
external momentum $q$ there is another rank two tensor, $q^\mu q^\nu$,
on which the integral can depend.  So the most general ansatz for a
rank two integral with one external momentum is a linear combination
of the two possible rank two tensor structures, $\delta^{\mu \nu}$ and
$q^\mu q^\nu$.  By solving the associated linear system, we find that
the rank two numerator $p_1^\mu p_2^\nu$ can be replaced by
\begin{equation}
p_1^\mu p_2^\nu\mapsto  \left(-\frac {p_1 \cdot p_2}{q^2} + 2 \frac {(p_1 \cdot q) (p_2
     \cdot q)}{(q^2)^2}\right) \delta^{\mu \nu} + \left(p_1 \cdot p_2
   - \frac {(p_1 \cdot q) (p_2 \cdot q)}{q^2}\right) q^\mu q^\nu.
\end{equation}  
In this case also the numerator can be further simplified, by using momentum 
conservation and partial fractioning.
%%%%%%%%%%%%%%%%%%%%%%%%%%%%%%%%%%%%%

\section{Fermionic propagators
\label{FermProp}}

The  elements of the fermionic  propagator $G_F= K^{-1}_F$ following from \rf{fw} are 
\be
G_{\theta\theta^\dagger}&=&\left(\begin{array}{cccc} 
  \frac{-p_0+\frac{3 \ri\nu}{4}}{{\cal D}_F(p)} & 0  &0 &0\\
0& \frac{-p_0-\frac{3 \ri\nu}{4}}{{\cal D}_F^*(p)}  & 0& 0\\
0&0&\frac{-p_0+\frac{3 \ri\nu}{4}}{{\cal D}_F(p)} &0\\
0&0 &0&\frac{-p_0-\frac{3 \ri\nu}{4}}{{\cal D}_F^*(p)} 
\end{array}\right) 
\,,\ee
\be G_{\eta\eta^\dagger}&=&\left(\begin{array}{cccc} 
  \frac{-p_0+\frac{ \ri\nu}{4}}{{\cal D}_F^*(p)} & 0  &0 &0\\
0& \frac{-p_0-\frac{ \ri\nu}{4}}{{\cal D}_F(p)}  & 0& 0\\
0&0&\frac{-p_0+\frac{ \ri\nu}{4}}{{\cal D}_F^*(p)} &0\\
0&0 &0&\frac{-p_0-\frac{ \ri\nu}{4}}{{\cal D}_F(p)} 
\end{array}\right)
\,,\ee
\be
G_{\theta\eta}&=&\left(\begin{array}{cccc} 
 0&0&\frac{\ri p_1+\frac{2-\hat w}{4}}{{\cal D}_F(p)} &0\\
0&0  & 0& \frac{\ri p_1+\frac{2+\hat w}{4}}{{\cal D}_F^*(p)} \\
\frac{-\ri p_1-\frac{2-\hat w}{4}}{{\cal D}_F(p)} &0&0&0\\
0&\frac{-\ri p_1-\frac{2+\hat w}{4}}{{\cal D}_F^*(p)}  &0&0
\end{array}\right) \,,
\ee
\be
G_{\theta^\dagger\eta^\dagger}&=&\left(\begin{array}{cccc} 
 0&0&\frac{\ri p_1+\frac{2+\hat w}{4}}{{\cal D}_F^*(p)} &0\\
0&0  & 0& \frac{\ri p_1+\frac{2-\hat w}{4}}{{\cal D}_F(p)} \\
\frac{-\ri p_1-\frac{2+\hat w}{4}}{{\cal D}_F^*(p)} &0&0&0\\
0&\frac{-\ri p_1-\frac{2-\hat w}{4}}{{\cal D}_F(p)}  &0&0
\end{array}\right)
\,,
\ee
They may be decomposed in terms of the $\rho$ matrices as:
\be
G_{\theta\theta^\dagger}\!\!&=&\!\!\left[\frac{ 3{\rm i}{\hat\nu}-4p_0}{8{\cal D}_F(p)   }
	+\frac{-3{\rm i}{\hat\nu}-4p_0}{8{\cal D}^*_F(p)}\right]{\bf 1}
         -\left[\frac{3 {\hat\nu}+4{\rm i}p_0}{8{\cal D}_F(p)  }
         +\frac{3 {\hat\nu}-4{\rm i}p_0}{8{\cal D}^*_F(p)}
                 \right]\rho^{\dagger[5}\rho^{6]}
\,, \\[5pt]
G_{\eta\eta^\dagger}\!\!&=&\!\!\left[\frac{-{\rm i}{\hat\nu}-4p_0}{8{\cal D}_F(p) }
         +\frac{ {\rm i}{\hat\nu}-4p_0}{8{\cal D}^*_F(p)}\right]{\bf 1}
         -\left[\frac{ {\hat\nu}-4{\rm i}p_0}{8{\cal D}_F(p)}
                 +\frac{ {\hat\nu}+4{\rm i}p_0}{8{\cal D}^*_F(p)}\right]\rho^{\dagger[5}\rho^{6]}
\,,\\[5pt]
G_{\theta\eta}\!\!&=&\!\!\left[\frac{4p_1-{\rm i}(2-{\hat w})}{8{\cal D}_F(p) }
         +\frac{4p_1-{\rm i}(2+{\hat w})}{8{\cal D}^*_F(p)}\right]\rho^{\dagger 5}
         +\left[\frac{ -4{\rm i}p_1-(2-{\hat w})}{8{\cal D}_F(p)}
                 +\frac{  4{\rm i}p_1+(2+{\hat w})}{8{\cal D}^*_F(p)}\right]\rho^{\dagger 6}
\,, \\[5pt]
G_{\theta^\dagger\eta^\dagger}\!\!&=&\!\!\left[\frac{4p_1-{\rm i}(2-{\hat w})}{8{\cal D}_F(p) }
         +\frac{4p_1-{\rm i}(2+{\hat w})}{8{\cal D}^*_F(p)}\right]\rho^{ 5}
         +\left[\frac{ -4{\rm i}p_1-(2-{\hat w})}{8{\cal D}_F(p)}
                 +\frac{  4{\rm i}p_1+(2+{\hat w})}{8{\cal D}^*_F(p)}\right]\rho^{ 6} 
\,,
\ee
where
\be
{\cal D}_F(p)=p^2-\frac{\ri}{2}(\hat\nu p_0-\hat w p_1)+\frac{1}{16}(3\hat \nu^2-\hat w^2+4)\,.
\ee

\section{ A comment on a thermodynamic relation
\label{sec:thermodynamics}}

A derivation of the relation (\ref{UFrel}) in section 2  for general (potentially related) chemical potentials $h_i$
proceeds as follows. One starts with the infinitesimal variation of the logarithm of the partition function
under the variation of the temperature and  the independent chemical potentials $h_s$:
\be
d\ln Z= \frac{\partial\ln Z}{\partial\beta}d\beta+\sum_i\frac{\partial \ln Z}{\partial h_i} dh_i
= -\langle {\widetilde H}_{2d}\rangle d\beta
                    -\beta\sum_{i,s} \langle Q_i\rangle \frac{\partial h_i}{\partial h_s}dh_s
\ee
This may be reorganized as
\be
\b^{-1}d\big[-\ln Z-\langle {\widetilde H}_{2d}\rangle \beta\big]&=&
\sum_{i,s} \langle Q_i\rangle \frac{\partial h_i}{\partial h_s}dh_s-d\langle {\widetilde H}_{2d}\rangle
\ee
In the thermodynamic limit the differential of the two-dimensional energy is
% nothing but (accounting for potential
%relations between chemical potentials)
\footnote{This is the analog of the general thermodynamic relation
$dU=TdS+\sum_i F_i dX^i$ where $X_i$ are extensive parameters and $F^i$ are the 
conjugate intensive quantities (e.g. charges and external potentials).}
\be
d\langle {\widetilde H}_{2d}\rangle = TdS +
\sum_{i,s} \langle Q_i\rangle \frac{\partial h_i}{\partial h_s}dh_s \ . 
\ee
Here the  last  term arises from the differentiation of the chemical potentials in the 
${\widetilde H}_{2d}$ prefactor 
 and the first term  is due to  the differentiation of the chemical potentials 
in the probability measure $e^{-\beta {\widetilde H}_{2d}}$, which amounts to changing the 
density of states which
%, for lack of a better name,
 we call (the infinitesimal change of the) entropy   ($T= \beta^{-1}$). 
Putting this  all together we find
\be
d\big[-\ln Z-\langle {\widetilde H}_{2d}\rangle \beta \big]  = \beta T dS = dS   \ . 
\ee
Integrating this relation gives 
\be
\Sigma=\langle {\widetilde H}_{2d}\rangle -\b^{-1}S-\b^{-1}{\cal C}
= \langle { H}_{2d}\rangle+\sum_i h_i \langle Q_i\rangle 
                - \b^{-1}S-\b^{-1}{\cal C} \ , 
\ee
where ${\cal C}$ is a constant that is independent of the chemical potentials and  $\beta$. In our case, 
$\beta\rightarrow \infty$ and thus we recover the equation (\ref{UFrel}).

%%%%%%%%%%%%%%%%%%%%%%%%%%%%%%%%%%%%%%%%%%%%%%%%%%%%%%%%%%%%%%%%%%%%%%%%%%%%%%%%

\section{Direct computation of the one-loop expectation values of  $J $ and  $E-S$
\label{sec:one_loop_E_J}}

To test the general arguments in section \ref{E_vs_Z}, in this appendix we evaluate directly 
the one-loop expectation values of $J\equiv J^{56}$   and   $E-S$. We will see explicitly that, in the 
presence of the chemical potential ${\hat\nu}$ and of the parameter ${\hat w}$, they take 
the form following from the equations (\ref{ej}) and (\ref{pq}) where 
$\hat\kappa=\sqrt{1+\hat\nu^2-\hat w^2}$:
\be
\langle E-S\rangle&=& \ \frac{\sqrt{\lambda}}{\pi}\,\ln S\;
\frac{\sqrt{1+{\hat\nu}{}^2-{\hat w}{}^2}}{1-{\hat w}{}^2}\ 
\bigg[ {\cal F}({\hat\nu},{\hat w})-{\hat\nu}\frac{\del {\cal F}({\hat\nu},{\hat w})}{\del {\hat\nu}}\bigg] \ , 
\la{qqq} \\
\langle J \rangle 
% \frac{\pi\langle J \rangle }{\sqrt{\lambda}\ln S} 
&=&  \  \frac{\sqrt{\lambda}}{\pi}\,\ln S\; \frac{1}{1-{\hat w}{}^2}\ \bigg[ {\hat\nu}{\cal F}({\hat\nu},{\hat w})-
(1+{\hat\nu}{}^2-{\hat w}{}^2) \frac{\del {\cal F}({\hat\nu}, {\hat w})}{\del {\hat\nu}}\bigg] \ .
\la{ouu}
\ee
In the absence of any background, the current $J^{56}$ and the corresponding charge 
were constructed in \cite{mtt}. It is not hard to expand $J^{56}$  around the generalized null cusp 
solution discussed in section 3. Since this current is nothing but the momentum conjugate to the field 
$\varphi$ in \rf{nee}, it is much simpler to extract it  from the already expanded 
action (\ref{Action}). We find 
\be
J \equiv J^{56}&=&\frac{\sqrt{\lambda}}{2\pi}\int ds\  ({\widetilde{\cal J}}){}^{56} \ , 
\cr
 ({\widetilde{\cal J}}){}^{56} &=& 
 {-2i}\Big[({R{\tilde z}})^5\Big( 
\partial_t (R{\tilde z})^6 -[((\partial_t R)R^{-1})(R{\tilde z})]^6  +
{{\rm i} }
{\teta}_i  (\rho^{6M}){}^i{}_j {\teta}^j  \frac{(R{\tilde Z})_M}{{\tilde z}^2}  \Big)
     - (5\leftrightarrow 6)\Big]
     \cr
&& \  - \ \ttheta_i(\rho^{\dagger[5}\rho^{6]}){}^i{}_j\ttheta^j 
      - \ \teta_i(\rho^{\dagger[5}\rho^{6]}){}^i{}_j\teta^j~~.
\ee
It is interesting to note that this is also the derivative of the expanded Lagrangian
with respect to the chemical potential ${\hat\nu}$ for the charge $J^{56}$ {\it before} the 
relation $\hat\kappa=\sqrt{1+\hat\nu^2-\hat w^2}$ is used.

The classical contribution and the terms relevant for a one-loop computation are:
\be\label{terms_exp_J}
{\widetilde{\cal J}} ^{56}
={\hat\nu}+\left(2{\hat\nu}\,{\tilde\phi}-2\ri \,\partial_t{\tilde\varphi}\right)
+
\left(2{\hat\nu} \, {\tilde \phi}^2 - {\hat\nu} {\tilde y}^a {\tilde y}^a 
-4 \ri {\tilde \phi} \partial_t {\tilde \varphi}
-
\ttheta_i(\rho^{\dagger [5}\rho^{6]}){}^i{}_j\ttheta^j
-
3\teta_i(\rho^{\dagger [5}\rho^{6]}){}^i{}_j\teta^j 
\right) \ .  
\ee
The first term leads to the classical expectation value of $J$
\be
\langle \frac{\pi J }{\sqrt{\lambda}\ln S}\rangle_0= {\hat\nu}~~,
\ee
which reproduces the tree-level 
component of the equation (\ref{ouu}) for all ${\hat \nu}$ and ${\hat w}$.

The one-loop expectation value\footnote{Here we anticipate that the expectation 
value is constant, so the $s$ integral is trivial, giving a length factor, cf. \rf{my}.} 
of $\langle \frac{J }{2\ln S}\rangle_1$ is given 
by a sum of two terms, 
\be\la{kp}
\langle \frac{J}{2\ln S}\rangle_1=\langle {\widetilde{\cal J}}  ^{56}\rangle_1=
\langle {\widetilde{\cal J}}  ^{56}\rangle^{(1)}+\langle {\widetilde{\cal J}}  ^{56}\rangle^{(2)}
~,
\ee
\begin{figure}
\begin{center}
\includegraphics[width=.5\textwidth]{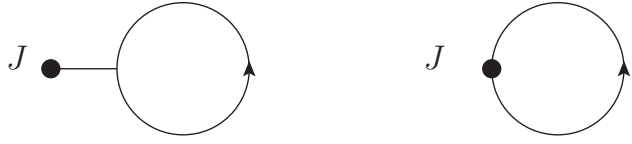}
\parbox{13cm}{\caption{The two contributions to the 1-loop expectation value of $J^{56}$.}
\label{Jvev-fig}}
\end{center}
\end{figure}
corresponding to the second and third parenthesis 
in the equation (\ref{terms_exp_J}), respectively. The two contributions are depicted diagrammatically in fig. \ref{Jvev-fig}. The first one is simply proportional to 
the one-loop tadpole for the field ${\tilde \phi}$ (the expectation value of $\partial_t{\tilde\varphi}$ vanishes because of the homogeneity of the background) 
\be
\langle {\widetilde{\cal J}}^{56}  \rangle^{(1)}=2{\hat \nu}\langle{\tilde \phi}\rangle=
-\frac{2{\hat \nu}}{1-{\hat w}^2}(A^{\mbox{\tiny tadpole}}_B+A^{\mbox{\tiny tadpole}}_F) \ , 
\label{vevJ1}
\ee
with $A^{\mbox{\tiny tadpole}}_B$ and $A^{\mbox{\tiny tadpole}}_F$ given in 
equations (\ref{boseTP}) and (\ref{fermiTP}). 
The second contribution to the expectation value of ${\widetilde{\cal J}}  ^{56}$ comes 
from one-loop diagram with one vertex from the second term in (\ref{terms_exp_J}):
\be
%\label{vevJ2}
\langle {\widetilde{\cal J}}  ^{56}\rangle^{(2)}&=&
8{\hat \nu}\   {\mbox{I}}[{\te {\frac{1}{4}}}(1+\hat\nu^2)]
-4{\hat \nu}\   {\mbox{I}}[{\te {\frac{1}{4}}}(\hat\nu^2+\hat w^2)]\\
&+&
2\int\frac{d^2p}{(2\pi)^2}
\frac{{\hat\nu}\,(p_1^2-p_0^2)+2{\hat w}\,p_0\, p_1}
{p^2(p^2+1) + ({\hat \nu}^2-{\hat w}^2) \,p_0^2
-2{\hat\nu}{\hat w} \,p_0\, p_1}
\cr
&+&2\int\frac{d^2p}{(2\pi)^2}
\frac{  {\hat\nu}(p_0^2-p_1^2) -
{\hat\nu}(\frac{1}{4} + \frac{3}{16} {\hat\nu}^2-\frac{1}{16}{\hat w}^2)
-2{\hat w}\,p_0\,p_1 }
        {(p^2+\frac{1}{4} + \frac{3}{16} {\hat\nu}^2 -\frac{1}{16}{\hat w}^2)^2 +
        \frac{1}{4} ({\hat\nu}\, p_0 -{\hat w}\, p_1)^2}~~.
\label{vevJ2}
\ee
%The divergent part of the fermionic contribution was explicitly separated.
These integrals may be evaluated in terms of the basic integrals listed in Appendix \ref{sec:1loop_int}. 
Putting together (\ref{vevJ1}) and (\ref{vevJ2}) we find that
\be
\langle {\widetilde{\cal J}}^{56} \rangle_1 &=&
\frac{{\hat\nu}}{2\pi} \left(2\ln({\hat\nu}^2+{\hat w}^2)-2\ln(1+{\hat\nu}^2)\right)
-\frac{\hat\nu}{2\pi}
+\frac{{\hat\nu}}{2\pi}\,\frac{1}{\hat\nu^2+\hat w^2}(\sqrt{1+\hat\nu^2}-\sqrt{1-\hat w^2})^2
\cr
&+&\frac{\hat\nu}{2\pi}\frac{(\hat \nu^2 - \hat w^2)}{1-{\hat w}^2}\Big{[}
\frac{1}{\hat \nu^2+\hat w^2} \left(1-\hat w^2-\sqrt{(1 + \hat \nu^2)(1-\hat w^2)} \right)\\
&+&\frac{1}{2} \left( 
\ln(1+\hat \kappa^2)-4 \ln (1+\hat\nu^2)+2\ln(\hat \nu^2+\hat w^2) 
+2 \ln \left(\sqrt{1 + \hat \nu^2} + \sqrt{1 - \hat w^2}\right)
\right)\Big{]}\nonumber
\ee
One may check that this may be rewritten  (using \rf{F1}) as
\be
\langle \frac{\pi J}{\sqrt{\lambda}\ln S}\rangle_1=\frac{2\pi}{\sqrt{\lambda}}
\langle {\widetilde{\cal J}}  ^{56} \rangle_1=
\frac{1}{\sqrt{\lambda}}\; \frac{1}{1-{\hat w}{}^2}\ \bigg[ {\hat\nu}{\cal F}_1({\hat\nu},{\hat w})-
(1+{\hat\nu}{}^2-{\hat w}{}^2) \frac{d{\cal F}_1({\hat\nu}, {\hat w})}{d{\hat\nu}}\bigg] 
\ee
i.e. we recover the one-loop component of the equation (\ref{ouu}) for all ${\hat \nu}$ and ${\hat w}$.

One may also use  a  similar approach  to evaluate $\langle E-S\rangle_1$. A direct construction 
of $E-S$ in the light-cone gauge is complicated as at first sight it is to involve the field $x^-$. A simpler approach 
is to extract $E-S$ as the derivative of the expanded Lagrangian with respect to ${\hat \kappa}$, again 
{\it before} using the relation $\hat\kappa=\sqrt{1+\hat\nu^2-\hat w^2}$. We find that
the classical contribution and the terms relevant for a one-loop computation are
\be
E-S=\frac{\sqrt{\lambda}}{2\pi}\int ds \;\left[{\hat\kappa}+2{\hat\kappa}{\tilde\phi}
+{\hat\kappa}(2{\tilde\phi}^2+|{\tilde x}|^2)\right]~~.
\label{derLka}
\ee
Interestingly, to this order there is no fermionic contribution.
The classical value of the energy is just
\be
\langle E-S\rangle_0=\frac{\sqrt{\lambda}}{\pi}\ln S\; {\hat \kappa}
\ee
this reproduces the tree-level 
component of the equation (\ref{qqq}) for all ${\hat \nu}$ and ${\hat w}$.

Similarly to the expectation value of $J$,  
As in the case  of $\langle J\rangle_1$  in  \rf{kp} the one-loop expectation value
$\langle E-S\rangle_1$  is also a sum of two terms:
\be
\frac{1}{2\ln S}\langle E-S\rangle_1=\langle {\cal E} \rangle^{(1)}+\langle {\cal E}\rangle^{(2)} \ , 
\ee
corresponding to the second and third terms in equation (\ref{derLka}). The first one is again 
proportional to the one-loop tadpole
\be
\langle {\cal E}\rangle^{(1)}&=&{2}{\hat\kappa}\langle{\tilde\phi}\rangle=
-\frac{2{\hat\kappa}}{1-{\hat w}^2}(A^{\mbox{\tiny tadpole}}_B+A^{\mbox{\tiny tadpole}}_F) \ . 
\ee
The second contribution  comes from one-loop diagram with one vertex from the third 
term in (\ref{derLka}):
\be
\langle{\cal E}\rangle^{(2)}&=&\frac{1}{2}{\hat\kappa}\,{\mbox{I}}[{\te {\frac{1}{4}}}(1+\hat\kappa^2)]
+\frac{1}{2}{\hat\kappa}\int\frac{d^2p}{(2\pi)^2}\frac{p^2}
{p^2(p^2+1) + {\hat \nu}^2 \,p_0^2-2{\hat\nu}{\hat w} \,p_0\, p_1}  \ . 
\ee
The remaining integrals may be evaluated using the basic integrals in Appendix \ref{sec:1loop_int}. 
As a result, we find 
%Putting together the components we find that
\be
&&
\langle {\cal E} \rangle^{(1)}+\langle {\cal E}\rangle^{(2)}=
\frac{\hat\kappa}{2\pi}\big[\ln(1+\hat\nu^2)-\ln(1+\hat\kappa^2)  \big]
\no \\
&&\ \ \ \ \ \ \  - \frac{\hat\kappa}{2\pi}\Big[
\ln(2+\hat\nu^2-{\hat w}^2+2\sqrt{(1+\hat\nu^2)(1-{\hat w}^2)})-\ln(1+\hat\nu^2)\Big]
\cr
&& \ \ \ \ \ \ \ +\frac{\hat\kappa}{2\pi}
\frac{\hat \nu^2 - \hat w^2}{1-{\hat w}^2}\Big{[}\frac{1}{\hat \nu^2+\hat w^2} \left(1-\hat w^2
-\sqrt{(1 + \hat \nu^2)(1-\hat w^2)} \right)\la{oyt} \\ 
&& \ \ +\frac{1}{2} \left( 
\ln(1+\hat \kappa^2)-4 \ln (1+\hat\nu^2)+2\ln(\hat \nu^2+\hat w^2) 
+2 \ln \left(\sqrt{1 + \hat \nu^2} + \sqrt{1 - \hat w^2}\right)
\right)\Big{]}
\nonumber
\ee
which may  be written as
\be
\langle E-S\rangle_1= {1 \ov \pi} 
{\ln S}\,\frac{\hat\kappa}{1-{\hat w}^2}\Big[{\cal F}_1(\hat \nu,\hat w)
-{\hat\nu}\frac{\del{\cal F}_1(\hat \nu,\hat w) }{\del\hat\nu}\Big] \ , 
\ee
i.e. we recover the one-loop component of the equation (\ref{qqq}) for all ${\hat \nu}$ and ${\hat w}$.

Since the equations (\ref{qqq}) and (\ref{oou}) hold for the nonvanishing $\hw=\ri \hat w$, we
 may then
eliminate $\hn$ between them and derive the expressions analogous to the  equations (\ref{gg}) and
(\ref{pp}) in the presence of nontrivial winding ${\hat m}$.:
\be
E-S&=&\frac{\sqrt{\lambda}}{\pi} \ff(\ell, \hw, \sql)\ln S \  , \ \ \   \ \ 
\ff(\ell, \hw, \sql)=\ff_0(\ell, \hw)+\frac{1}{\sqrt{\lambda}}\ff_1(\ell, \hw)
+\frac{1}{\lambda}\ff_2(\ell, \hw)+\dots \no 
\\
\ff_0(\ell, \hw)&=&\sqrt{1+\ell^2 + \hw^2} \ , \ \ \ \ \ \ \ \ \ \ \ \ 
\ff_1(\ell, \hw)=\frac{{\cal F}_1(\ell, \hw)}{\sqrt{1+\ell^2+\hw^2}} \ , 
\\
\ff_2(\ell, \hw)&=&\frac{{\cal F}_2(\ell, \hw)}{\sqrt{1+\ell^2+\hw^2}}
           +\frac{(1+\ell^2+\hw^2)^{3/2}}{2(1+\hw^2)}\Big(\frac{d\ff_1(\ell, \hw)}{d\ell}\Big)^2 \ . 
\ee
%
%R
It is in principle possible that, depending on the precise definition  of the winding and 
similarly to the relation between $J$ and ${\hat \nu}$, 
the relation between the physical winding number and the parameter $\hat m$ also receives 
quantum corrections.
Assuming that such corrections do not occur (e.g.  by defining the 
winding number as $\int ds\,\partial_s\varphi$, whose expectation 
value vanishes at the quantum level due to the homogeneity of the classical solution) and
using the 2-loop result for $\cal F$ (\ref{F2}), it is then straightforward to 
extract $\ff_2(\ell,\hw)$ up to fourth order in small $\ell,\hw$
\be
\ff_2(\ell,\hw)&=& -K +
\left(2 \ell^2 \ln^2(\ell^2-\hw^2)-(3\ell^2-\hw^2)\ln(\ell^2-\hw^2)-\frac{3}{2}(\ell^2-\hw^2)\ln 2+\frac{11}{4}\ell^2-\frac{9}{4}\hw^2 \right)\cr
&&-\frac{1}{2}(3\ell^4+\hw^4)\ln^2(\ell^2-\hw^2)
-\left((\frac{7}{12}-\frac{3}{2}\ln 2)\ell^4-(\frac{11}{12}-\frac{3}{2}\ln 2)\hw^4
- 4 \ell^2 \hw^2\right)\ln(\ell^2-\hw^2)\cr
&&-\frac{9}{8}(\ell^2+\hw^2)^2\ln^2 2
+\frac{1}{8}(11\ell^4+11 \hw^4-6 \ell^2 \hw^2)\ln 2
+\frac{1}{32} (3 \ell^4+3 \hw^4-14 \ell^2 \hw^2)K \no \\
&&-\frac{1}{576}(233 \ell^4+377 \hw^4-618 \ell^2 \hw^2) +{\cal O} (\ell^6,\ell^4 \hw^2, ...)  \ . 
\label{f2ell-m}
\ee
Setting $\hw=0$, one recovers the result (\ref{f2f}) given in the main text.

%%\
%%{\bf

%% why should not we get more serious modification ?? 
 
 %% where is the difference between $\n$ and $w$  ??  
%%}

 \end{document}